\documentclass[10pt, a4paper]{article}
\usepackage[utf8]{inputenc}
\usepackage{mlmodern}
\usepackage[T1]{fontenc}
\usepackage{color}
\usepackage{mathtools}
\usepackage{amsmath}
\usepackage{amssymb}
\usepackage{setspace}
\usepackage{float}
\usepackage{subfig}
\usepackage{xcolor}
\usepackage{braket}
\usepackage{mathrsfs}
\usepackage{fullpage}
\usepackage{cite}
\allowdisplaybreaks

\global\long\def\ty#1{\mathbf{T}_{#1}^{y}}%
\global\long\def\tz#1{\mathbf{T}_{#1}^{z}}
\global\long\def\tr#1{\mathbf{T}_{#1}^{r}}%
\global\long\def\tk#1{\mathbf{T}_{#1}^{k}}%
\global\long\def\ta#1{\mathbf{T}_{#1}^{a}}%
\global\long\def\tb#1{\mathbf{T}_{#1}^{b}}%
\global\long\def\tc#1{\mathbf{T}_{#1}^{c}}%
\global\long\def\td#1{\mathbf{T}_{#1}^{d}}%
\global\long\def\tg#1{\mathbf{T}_{#1}^{g}}%
\global\long\def\te#1{\mathbf{T}_{#1}^{e}}%
\global\long\def\tkk#1{\mathbf{T}_{#1}^{k}}%
\global\long\def\tm#1{\mathbf{T}_{#1}^{m}}%
\global\long\def\tn#1{\mathbf{T}_{#1}^{n}}%

\newcommand{\D}[1]{(\mathbf{1}_{#1},0)}
\newcommand{\report}[1]{\textcolor{black}{#1}}

\newcommand{\e}{\epsilon}

\newcommand{\as}{\alpha_s}

\definecolor{indigo}{rgb}{0.0, 0.25, 0.42}
\usepackage[colorlinks,allcolors=indigo]{hyperref}
\newcommand\scalemath[2]{\scalebox{#1}{\mbox{\ensuremath{\displaystyle #2}}}}
\newenvironment{Repo}{\color{black}}

\begin{document}

\begin{titlepage}
	\hfill \today 
\vspace*{\fill}
	\begin{flushleft}
		{\Large \sc Correlator webs of massive multiparton amplitudes at four loops: A study of boomerang webs} \bigskip \bigskip \\
		{Neelima Agarwal$\,^a$, 
			Sourav Pal$\,^b$, Aditya Srivastav$\,^c$, Anurag Tripathi$\,^c$ }
	\end{flushleft}
\vspace{0.5cm} 
{\it
$\,^a$Department of Physics, Chaitanya Bharathi Institute of Technology, \\ Gandipet, Hyderabad, Telangana State 500075, India  \medskip \\
$\,^b$Theoretical Physics Division, Physical Research Laboratory, \\ Navrangpura, Ahmedabad 380009, India \medskip \\
$\,^c$Department of Physics, Indian Institute of Technology Hyderabad, \\ Kandi, Sangareddy, Telangana State 502284, India}
\vspace{0.5cm}

\noindent {\it E-mail}: \href{mailto:dragarwalphysics@gmail.com}{dragarwalphysics@gmail.com}, \href{mailto:sourav@prl.res.in}{sourav@prl.res.in}, \href{mailto:shrivastavadi333@gmail.com}{shrivastavadi333@gmail.com}, \\ \href{mailto:tripathi@phy.iith.ac.in}{tripathi@phy.iith.ac.in}
\vspace{0.4cm}
\hrule     
\vspace{2.0cm}

\noindent {\sc Abstract}: Logarithm of the soft function \report{can be organized into} sets of Feynman diagrams known as Cwebs. We introduced a new formalism in~\cite{Agarwal:2022wyk}, that allows to determine several of the building blocks of Cweb mixing matrices without explicit computations. \report{ In~\cite{Agarwal:2022xec} we used this formalism to obtain the diagonal blocks of four general classes of Cwebs to all orders in perturbation theory which also covered all the four loop Boomerang Cwebs connecting four Wilson lines.
In this work we present complete mixing matrices and exponentiated colour factors for Boomerang Cwebs at four loops that connect three and four Wilson lines. Also, we present a more efficient version of the algorithm of generating Cwebs that was presented in~\cite{Agarwal:2020nyc}. This new algorithm has been used 
to generate the Cwebs in the present work.} 
\vspace*{\fill}

\end{titlepage}	
	
\hrule
\tableofcontents
\vspace{0.4cm}
\hrule 	

\bigskip 

\section{Introduction}

\report{The study of infrared (IR) singularity structure of scattering amplitudes in non-abelian gauge theories has a rich and long history~\cite{Bloch:1937pw,Sudakov:1954sw,Yennie:1961ad,Kinoshita:1962ur,Lee:1964is,Grammer:1973db,Mueller:1979ih,Collins:1980ih,Sen:1981sd,
	Sen:1982bt,Korchemsky:1987wg,Korchemsky:1988hd,Magnea:1990zb,
	Dixon:2008gr,Gardi:2009qi,Becher:2009qa,Feige:2014wja}}. A comprehensive review of the IR singularities of gauge theories is elucidated in a recent review article~\cite{Agarwal:2021ais}. The universal structure of IR singularities of gauge theories, that is their independence of relevant hard scattering process helps one to study their all order structure in perturbation theory. The structure of IR singularities not only helps us in understanding gauge theories to all orders in the perturbation theory, but also remains important for phenomenological applications. These studies are relevant to the high energy scattering experiments at different colliders. The calculations of physical observables such as scattering cross section and decay rates suffer from these IR divergences which eventually cancel upon adding the real emission contribution to the virtual correction at a given perturbative order. This cancellation results in large logarithms of kinematic invariants that damage the predictive power of fixed order calculations in certain kinematical regions. Again the universality property of IR singularity provides a way of summing these logarithms to all order in perturbation and helps in recovering the predictions in those regions. The intricate cancellation of IR poles between real and virtual contributions for observables relevant to colliders is a non-trivial task for which several subtraction procedures~\cite{GehrmannDeRidder:2005cm,Somogyi:2005xz,Catani:2007vq,
	Czakon:2010td,Boughezal:2015dva,Sborlini:2016hat,Caola:2017dug,
	Herzog:2018ily,Magnea:2018hab,Magnea:2018ebr,Capatti:2020xjc,Magnea:2022twu,Bertolotti:2022aih,Bertolotti:2022lcj} are developed.  

The QCD factorization theorem enables us in studying the singular parts of scattering amplitudes involving massless gauge bosons, without calculating the  complicated hard part. The soft function, which captures the IR singular parts of a scattering amplitude, is expressed as the correlators of Wilson line operators. The usual Wilson-line operators $\Phi ( \zeta )$  evaluated on smooth space-time contours $\zeta$ are defined as,
\begin{align}
	\Phi \left(  \zeta \right) \, \equiv \, \mathcal{P} \exp \left[ {\rm i} g \!
	\int_\zeta d x \cdot {\bf A} (x) \right] \, .
	\label{genWL}
\end{align}
Here ${\bf A}^\mu (x) = A^\mu_a (x) \, {\bf T}^a$ is a non-abelian gauge field, 
and ${\bf T}^a$ is a generator of the gauge algebra, which can be taken to belong
to any desired representation, and $ \mathcal{P} $ denotes path ordering of the gauge fields. The soft function in general can be expressed as, 
\begin{align}
	{\cal S}_n \left( \zeta_i \right) \, \equiv \, \bra{0} \prod_{k = 1}^n
	\Phi \left(  \zeta_k \right) \ket{0} \, .
	\label{genWLC}
\end{align}
These Wilson lines are semi-infinite Wilson lines along the direction of the hard particle. Thus, one can write the soft function as, 
\begin{align}
	{\cal S}_n \Big( \beta_i \cdot \beta_j, \as (\mu^2), \e \Big) \, \equiv \, 
	\bra{0} \prod_{k = 1}^n \Phi_{\beta_k} \left( \infty, 0 \right) \ket{0} , \quad
	\Phi_\beta \left( \infty, 0 \right) \, \equiv \, \mathcal{P} \exp \left[ {\rm i} g \!
	\int_0^\infty d \lambda \, \beta \cdot {\bf A} (\lambda \beta) \right]\,.
	\label{softWLC}
\end{align} 
The object $ \mathcal{S}_n $ suffers from both UV and IR (soft) singularities, and thus needs renormalization. In dimensional regularization $ d=4-2\e $, \report{ higher order terms of unrenormalized soft function $ \mathcal{S}_n $ vanish, giving $ \mathcal{S}_{n}=1 $. The simple argument for this is that the soft function is made out of Wilson line correlators, which involve only scaleless integrals. Thus, the renormalized soft function $ \mathcal{S}_n^{(\text{ren})} $ contains pure UV counterterms.} 

The renormalized soft function obey a renormalization group equation and solving this equation results in an exponentiation of the form, 
\begin{align}
	\mathcal{S}_n^{(\text{ren})} \Big( \beta_i \cdot \beta_j, \as (\mu^2), \e \Big) \, = \, 
	\mathcal{P} \exp \left[ - \frac{1}{2} \int_{0}^{\mu^2} \frac{d \lambda^2}  
	{\lambda^2} \, {\bf \Gamma}_n \Big( \beta_i \cdot \beta_j, \alpha_s (\lambda^2), 
	\e \Big) \right]  \, ,
	\label{softmatr}
\end{align}    
where $ {\bf \Gamma}_n $ is known as the soft anomalous dimension. In case of processes involving multi-parton scatterings, the soft anomalous dimension is a matrix, which is an interesting theoretical object to  study, and is our main focus in this article. The renormalization group approach has been used for more than two decades to calculate the soft-anomalous dimension. One-loop calculations for ${\bf \Gamma}_n$ were performed in \cite{Kidonakis:1996aq,Kidonakis:1997gm,Kidonakis-1998,Korchemskaya:1994qp}, while two-loop calculations were done for both the massless case in \cite{Aybat:2006wq,Aybat:2006mz} and the massive case in \cite{Mitov:2009sv, Ferroglia:2009ep, Ferroglia:2009ii, Kidonakis:2009ev,Kidonakis:2010tc,Kidonakis:2010ux,Kidonakis:2011wy, Chien:2011wz}. The three-loop calculations were finally carried out for the massless case in \cite{Almelid:2015jia, Almelid:2017qju}. The calculation of soft anomalous dimension at four loops is an ongoing effort. Several studies in this direction are available in the literature in~\cite{Becher:2019avh,Falcioni:2020lvv,Falcioni:2021buo,Falcioni:2021ymu,Catani:2019nqv,Moch:2017uml,Ahrens:2012qz,
	Moch:2018wjh,Chetyrkin:2017bjc,
	Das:2019btv,Das:2020adl,vonManteuffel:2020vjv,Henn:2019swt,Duhr:2022cob,Kidonakis:2023lgc}. The kinematic dependence of the soft anomalous dimension for scatterings involving only massless lines is restricted due to the constraints discussed in~\cite{Gardi:2009qi,Gardi-Magnea,Becher:2009cu,Becher:2009qa}. However, these constraints do not hold true for scatterings involving massive particles. The state-of-the-art knowledge for soft anomalous dimension is known upto two loops for scatterings involving massive particle~\cite{Mitov:2009sv, Ferroglia:2009ep, Ferroglia:2009ii, Kidonakis:2009ev, Kidonakis:2010tc, Kidonakis:2010ux, Kidonakis:2011wy,  Chien:2011wz}, and at three-loop for one massive Wilson line~\cite{Liu:2022elt}. 

An alternative approach to determine the exponent of the soft function is through diagrammatic exponentiation. In terms of Feynman diagrams, the soft function has the form, 
\begin{align}
	{\cal S}_n \left( \gamma_i \right) \, = \, \exp \Big[ {\cal W}_n \left( \gamma_i \right) 
	\Big]  \, ,
	\label{diaxp}
\end{align}
where $ {\cal W}_n \left( \gamma_i \right) $ is known as \textit{webs}, and can be directly computed using Feynman diagrams.  Webs were first defined as two-line irreducible diagrams for scattering involving two Wilson lines~\cite{Sterman-1981,
	Gatheral,Frenkel-1984}.  In case of multiparton scattering process, webs in non-abelian gauge theory~\cite{Mitov:2010rp,Gardi:2010rn} are defined as the sets of diagrams that differ among each other by the order of gluon attachments on each Wilson lines.
\report{A generalization of a web called Cweb was introduced in~\cite{Agarwal:2020nyc,Agarwal:2021him}}.

The state-of-the-art studies for massive multiparton webs at three loops~\cite{Gardi:2021gzz}, for massless webs upto three loops~\cite{Gardi:2010rn,Gardi:2011yz,Gardi:2013ita,Gardi:2013saa}, at four loops~\cite{Agarwal:2020nyc,Agarwal:2021him}, and partially at five loops~\cite{Mishra:2023acr} advance significantly.
The kinematics and the colour factors of diagrams in a Cweb mix among themselves, through a web mixing matrix. 

These web mixing matrices are crucial objects in the study of non-abelian exponentiation, and the general method to calculate them is by applying a well-known replica trick algorithm~\cite{Gardi:2010rn,Laenen:2008gt}; an alternative approach of generating functionals was developed in \cite{Vladimirov:2015fea,Vladimirov:2014wga,Vladimirov:2017ksc}. The web mixing matrices are combinatorial objects and have also been studied from the viewpoint of combinatorial mathematics using posets~\cite{Dukes:2013gea,Dukes:2013wa,Dukes:2016ger}. Further, a novel method of calculating the diagonal blocks --- using several new concepts such as Normal Ordering, Uniqueness theorem and Fused-Webs --- of the mixing matrices has been developed in~\cite{Agarwal:2022wyk} and has been applied for a certain classes of webs in~\cite{Agarwal:2022xec}.  

Recently, Boomerang webs are introduced in \cite{Gardi:2021gzz} to calculate the soft anomalous dimension at three loops for scattering processes involving massive particles, such as top quark, whose mass cannot be ignored, in several QCD processes. Boomerang webs are defined as webs, which contain at least one gluon propagator whose both ends attach to the same massive Wilson line. \report{Note that, Boomerang webs vanish identically in the case of massless Wilson lines.}  
Following the definition of Cwebs, Boomerang Cwebs are defined in~\cite{Agarwal:2022xec} as the Cwebs that contain at least one two-point gluon correlator whose both ends are attached to the same massive Wilson line. In \cite{Gardi:2021gzz}, the authors have computed the web mixing matrices, their exponentiated colour factors, and kinematics for all three-loop Boomerang webs. Recently, the diagonal blocks of mixing matrices for four-loop Boomerang Cwebs connecting four Wilson lines are presented in~\cite{Agarwal:2022xec} using the concept of Fused-Webs~\cite{Agarwal:2022wyk} and combinatorial properties of the Cwebs. However, to determine the exponentiated colour factors and the kinematic contribution of these Cwebs one needs to calculate the explicit form of the mixing matrices.    

In this article we present the explicit results of the mixing matrices for all the Boomerang Cwebs that connect three and four massive Wilson lines. To enumerate these Cwebs uniquely, we have modified the algorithm of enumerating Cwebs~\cite{Agarwal:2020nyc,Agarwal:2021him} and implemented it in a Mathematica code. Further, we have modified the older version of an in-house Mathematica code used in~\cite{Agarwal:2020nyc,Agarwal:2021him} to calculate the mixing matrices for Cwebs using the replica-trick algorithm. These modifications are incorporated in our in-house Mathematica Code  \texttt{CwebGen 2.0}.   

The rest of the paper is structured as follows. In section~\ref{sec:CwebProper}, we review Cwebs and the properties of web mixing matrices. In the section~\ref{sec:Algo} we give the details of the modified algorithm to generate Cwebs at a given perturbative order and its comparison with the older versions. Section~\ref{sec:repl} discusses the working of \texttt{CwebGen 2.0}.  In section \ref{sec:BoomCwebsEx}, we describe the calculation of mixing matrices for two Boomerang Cwebs at four loops. Finally in section \ref{sec:summary} we conclude our findings and give a future outlook on our results. In ancillary file \textit{Boomerang.nb}, we provide the explicit form of the mixing matrices and the corresponding column weight vector for all Boomerang Cwebs that are considered in this article.


\begin{figure}[b]
	\centering
	\includegraphics[scale=0.28]{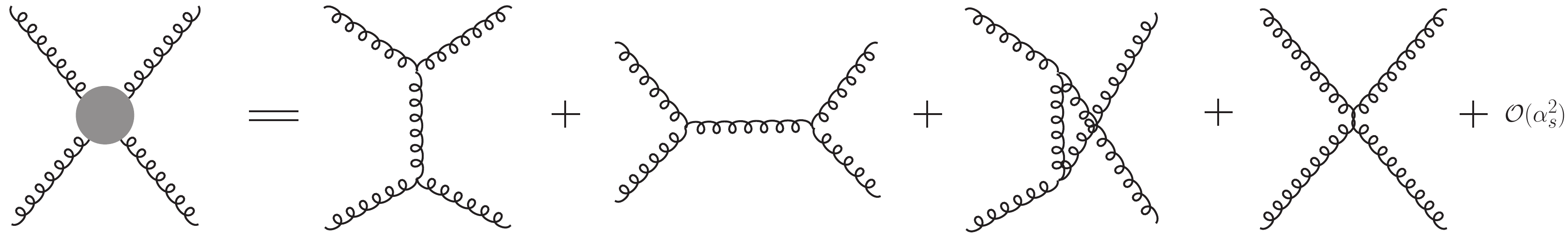}
	\caption{An all order four-point connected gluon correlator is shown. The blob on the left hand side contains all higher order corrections an expansion of which is given on the right hand side.}
	\label{fig:correlator}
\end{figure}

\section{Cwebs and web mixing matrices}\label{sec:CwebProper}

For the convenience of the reader, we collect here some of the definitions that are used in this article.  

\subsubsection*{Definitions}
\begin{itemize}
	\item [] \textcolor{black!70}{\textbf{Web}}: \textcolor{black!70}{\report{A set of diagrams closed under shuffles of the gluon attachments --- originating from different gluon propagators or three-point or four-point gluon vertices --- on each Wilson line.}}
	
	\item [] \textcolor{black!70}{\textbf{Cweb}}:  \textcolor{black!70}{\report{A set of diagrams closed under shuffles of the gluon attachments --- originating from different gluon correlators (see fig.~\eqref{fig:correlator} for an example) --- on each Wilson line.}}

	\item [] \report{\textbf{Multi-pronged-boomerang correlator:} An $ m $-point gluon correlator ($ m \geq 2 $) whose all ends attach to one Wilson line. The contributions of these for massless Wilson lines vanish identically.}

	\item [] \textcolor{black!70}{\textbf{Boomerang Cweb}}:  \textcolor{black!70}{A Cweb that contains at least one  \report{multi-pronged-boomerang correlator} with $ m=2 $ and none with $ m > 2 $. }

	\item [] \textcolor{black!70}{\textbf{Weight factor ($s$-factor)}}:  \textcolor{black!70}{The weight factor $s(d)$ for a diagram $ d $ is defined as the number of  ways in which the gluon correlators 
	can be {\it sequentially} shrunk to their common origin. }
	
	\item [] \textcolor{black!70}{\textbf{Column weight vector}}: \textcolor{black!70}{
	We can construct a column weight vector out of the $ s $-factors for a Cweb with $ n $ diagrams as}
	\begin{align} 
		\textcolor{black!70}{S=\{s(d_1),s(d_2), \ldots,s(d_n)\}}.
	\end{align}	
	
	\item[] $\textcolor{black!70}{\mathbf{W_n^{(c_2, \ldots , c_p)} (k_1, \ldots  , k_n)}}$:  \textcolor{black!70}{
	This is how we denote a Cweb constructed out of $c_m$ $m$-point connected gluon correlators 
	($m = 2, \ldots, p$), where $k_{i}$ denotes number of attachments on \report{$ i^{\text{th}} $ Wilson line.}}

\end{itemize}
  
  Note that the 
perturbative expansion for an $m$-point connected gluon correlator starts 
at ${\cal O} (g^{m - 2})$, while each attachment to a Wilson line carries a 
further power of $g$, the perturbative expansion for a Cweb can be written as
\begin{equation}
	W_n^{(c_2, \ldots , c_p)} (k_1, \ldots  , k_n)  \, = \, 
	g^{\, \sum_{i = 1}^n k_i \, + \,  \sum_{r = 2}^p c_r (r - 2)} \, \sum_{j = 0}^\infty \,
	W_{n, \, j}^{(c_2, \ldots , c_p)} (k_1, \ldots  , k_n) \, g^{2 j} \, ,
	\label{pertCweb}
\end{equation}
which defines the perturbative coefficients $W_{n, \, j}^{(c_2, \ldots , c_p)} 
(k_1, \ldots  , k_n)$. Here the perturbative order of this Cweb is $ g^{\, \sum_{i = 1}^n k_i \, + \,  \sum_{r = 2}^p c_r (r - 2)} $.

 Cwebs are the proper building blocks of the logarithm of the Soft function --- they are useful in the organization and counting of diagrammatic contributions at higher perturbative orders ~\cite{Agarwal:2020nyc, Agarwal:2021him}. The logarithm of the Soft function is a sum over all the Cwebs at each perturbative order:
\begin{align}
	{\cal S} \, = \, \exp \left[ \sum_W
	\sum_{d,d' \in  W} {K} (d) \, R_W (d, d') \, C (d')
	\right] \, .
	\label{Snwebs}
\end{align}
The $d$ here denotes a diagram in a  Cweb $W$ and its corresponding kinematic and colour factor are denoted by $ {K} (d) $ and $C(d)$. The action of web mixing matrix $R_W$ on the colour of a diagram $ C(d) $ generates its exponentiated colour factor $\widetilde{C}(d)$,  
\begin{align}
	\widetilde{C} (d)  \, = \, \sum_{d'\in W} R_W (d, d') \, C(d') \, .
	\label{eq:ecf}
\end{align}
The contribution of colour and kinematic factors to a web $ W $ can be arranged in a more transparent manner if we diagonalize the mixing matrix $ R $:\begin{align}\label{eq:YC}
	W  =   \left( K^T Y^{-1} \right) Y R Y^{-1}  
	\left( Y C \right) \,=\,  \sum_{j =  1}^{r} \left(  K^T Y^{-1}  \right)_j 
	\left( Y C \right)_j  \, ,
\end{align}
where  $ Y $ is the diagonalizing matrix  and $ Y R Y^{-1} \equiv \mathcal{D} $ is the diagonal matrix that we get; furthermore we have arranged $\mathcal{D}_{jj} =1$ for $1 \leq j \leq r$.
$ \left( Y C \right)_j $ are also referred to as  exponentiated colour factors and the corresponding kinematic factors 
are $ \left(  K^T Y^{-1}  \right)_j $. In this article we will present the exponentiated colour factors in terms of $ \left( Y C \right)_j $. 

To understand the structure of the soft anomalous dimension, the study of the web mixing matrices is crucial.  We will 
begin this study by listing down the properties that they are known to obey~\cite{Mitov:2010rp,
	Gardi:2010rn,Gardi:2011wa,Gardi:2011yz,Dukes:2013wa,Dukes:2016ger,
	Dukes:2013gea,Agarwal:2022wyk} which we list down below. 

\subsubsection* {Properties of mixing matrices}
\begin{enumerate}
	\item  \textit{Idempotence}: A Cweb mixing matrix is idempotent, that is, $ R^2=R $. This implies that: (a) the eigenvalues of $R$ can either be  0 or 1, (b) trace  $\textrm{tr}(R)$ is equal to its rank, and (c) it acts as a projection operator; acting on its right, it give completely connected exponentiated colour factors.  
	\item  \textit{Zero row sum rule}: The entries of $R$ obey the zero row sum rule $\sum_{d'} 
	R(d, d') = 0$. This ensures that while acting on the vector of kinematic factors, the web mixing matrix implements cancellation of the leading divergences among the diagrams of the Cweb.
	\item \textit{Column sum rule}: Along with these general properties, the mixing matrices also obey a conjectured column-sum rule of the form
	\begin{align}
		\sum_d s(d) R(d, d')=0\,.
	\end{align}	  
	\item  \textit{Uniqueness}: For a given column weight vector $S=\{s(d_1),s(d_2), \ldots,s(d_k)\}$ 
	with all $s(d_{i}) \neq 0$, the mixing matrix is unique.
\end{enumerate}
In section~\ref{sec:BoomCwebsEx}, we will verify these properties for four loop Boomerang Cwebs. 

In the next section, we describe a recursive algorithm that generates Cwebs present at $ \mathcal{O}(g^{2l+2}) $ using Cwebs at $ \mathcal{O}(g^{2l}) $. 

\section{An improved algorithm to generate Cwebs}\label{sec:Algo}
An algorithm to generate Cwebs at $ l+1 $ loops using  Cwebs at $ l $ loops was presented in~\cite{Agarwal:2020nyc} in which some of the present authors were also involved.
This was later used in~\cite{Agarwal:2021him} to obtain the Cwebs that are present at four loops.
Now we will  present (for ease of reading) the original algorithm below and then spell out a modified and more efficient version of the algorithm, which is implemented in a subroutine of in-house Mathematica code \texttt{CwebGen \,2.0}.

\subsubsection*{Original algorithm}
The original algorithm presented in~\cite{Agarwal:2020nyc,Agarwal:2021him}  to generate Cwebs at $ l+1 $ loops using Cwebs at $ l $ loops is  reproduced below:
\begin{enumerate}
	\item Add a two-gluon connected correlator  connecting any two Wilson lines
	(including Wilson lines that  had no attachments at lower orders).
	\item Connect an existing $m$-point correlator to any Wilson line (again, 
	including Wilson lines with no attachments at lower orders), turning  it into
	an  $(m+1)$-point correlator.
	\item Connect an existing $m$-point correlator to an existing $n$-point 
	correlator, resulting in an $(n+m)$-point correlator. 
	\item Discard the duplicate Cwebs. 
\end{enumerate}
\begin{figure}[b]
	\centering
	\subfloat[][$ \text{W}_2^{(2)}(2,2) $]{\includegraphics[height=3.5cm,width=3.5cm]{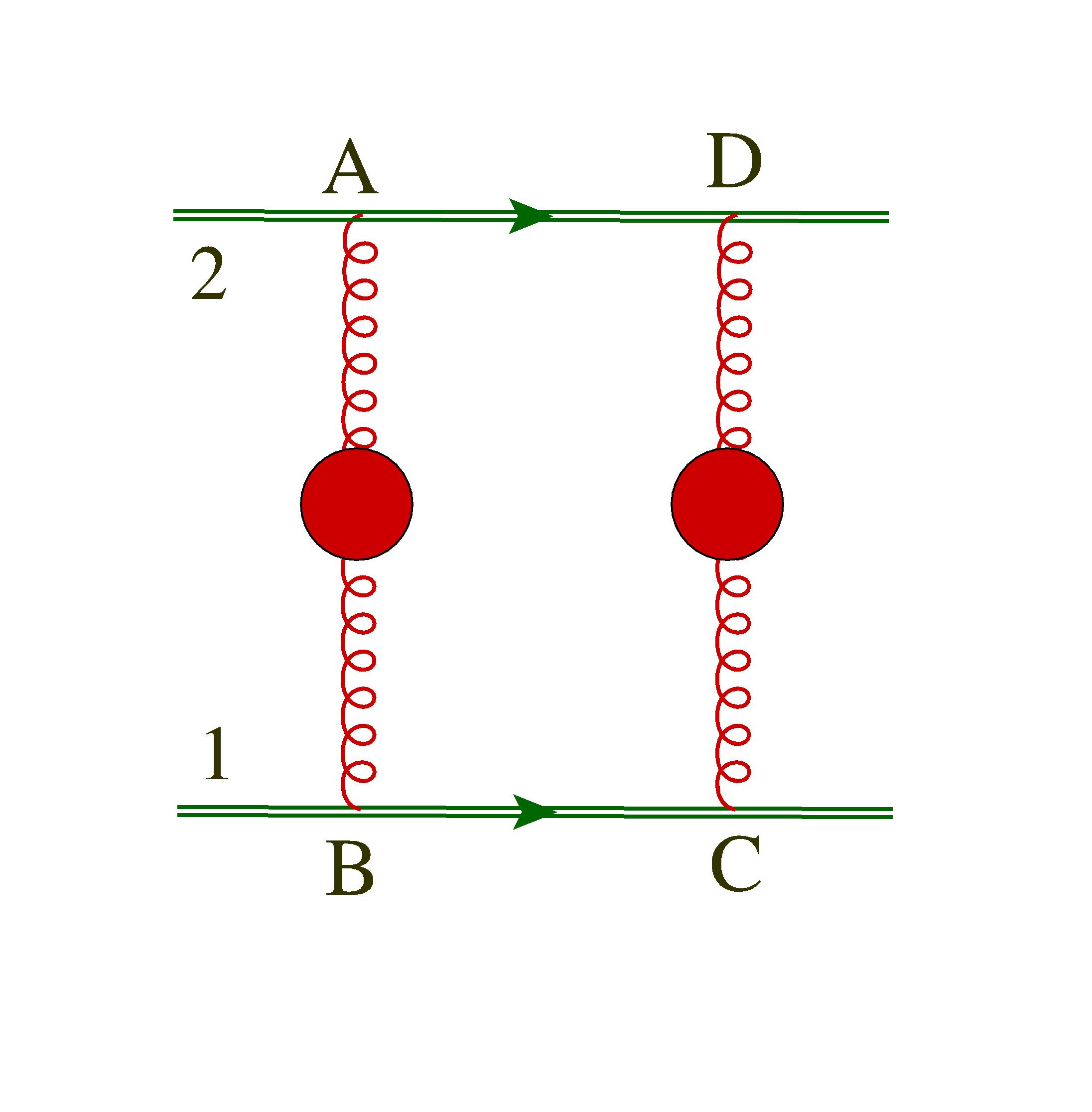} }
	\subfloat[][$ \text{W}_2^{(0,1)}(2,1) $]{\includegraphics[height=3.5cm,width=3.5cm]{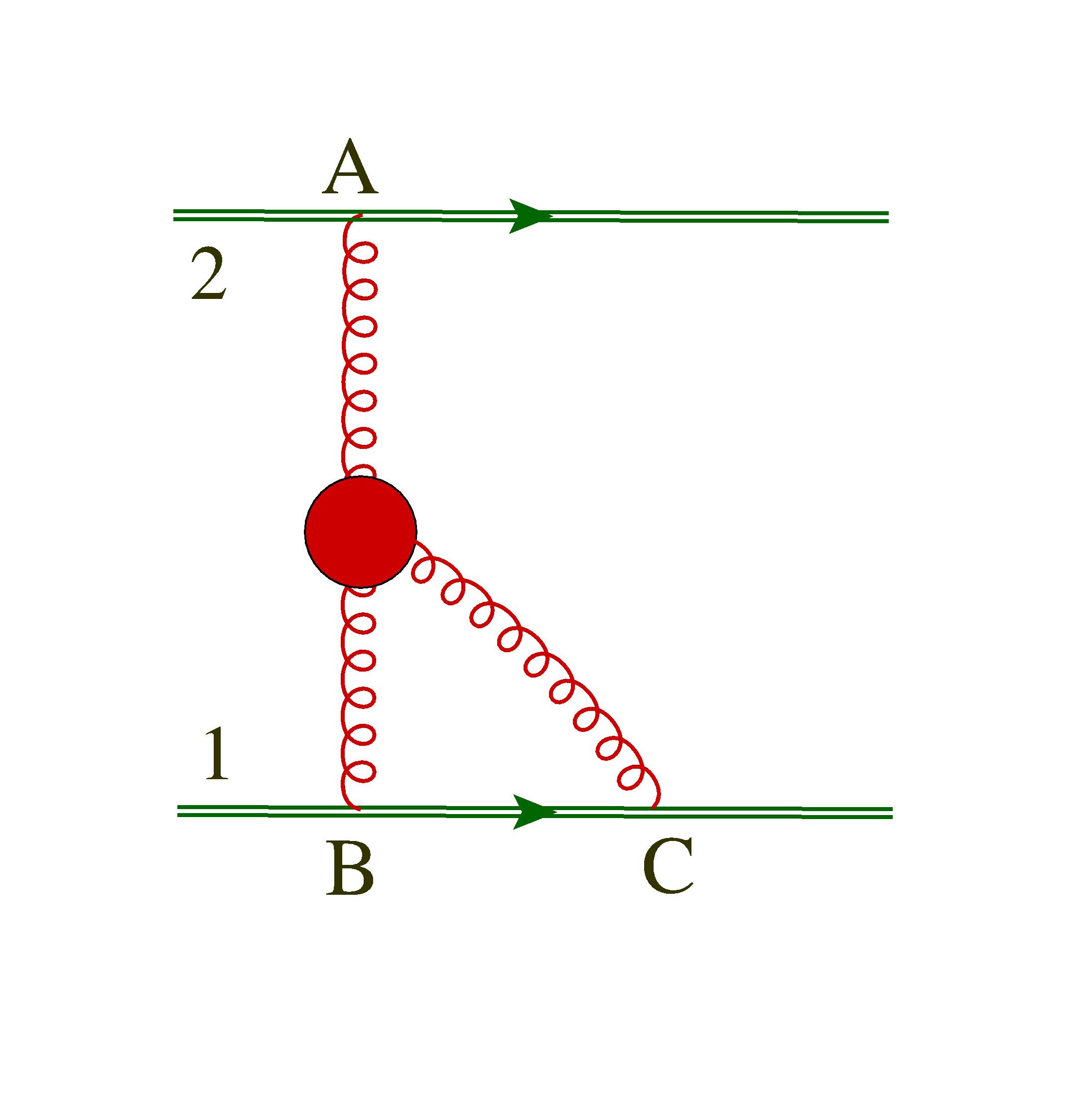} }
	\subfloat[][$ \text{W}_3^{(2)}(2,1,1) $]{\includegraphics[height=3.5cm,width=3.5cm]{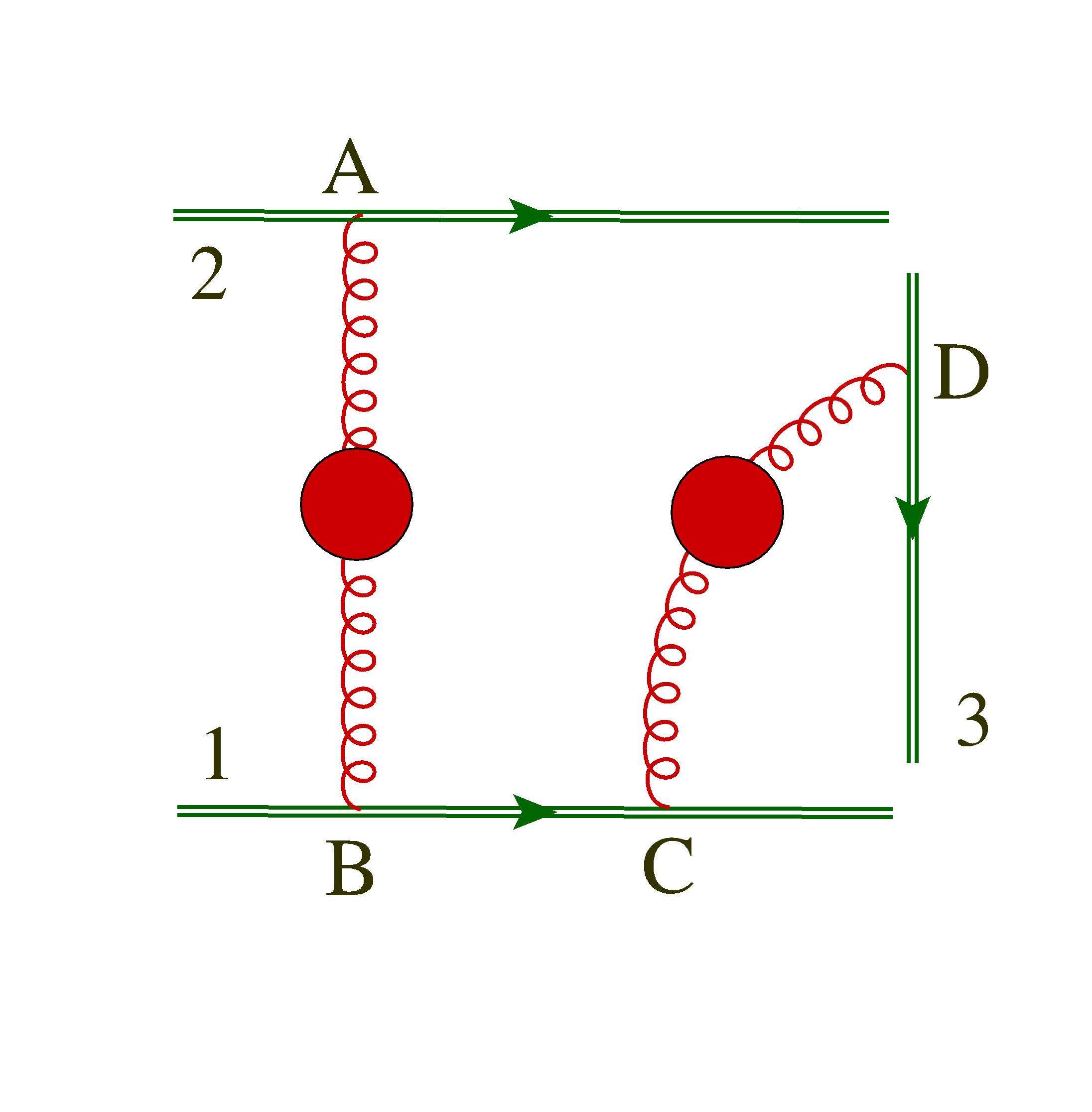} }
	\subfloat[][$ \text{W}_3^{(0,1)}(1,1,1) $]{\includegraphics[height=3.5cm,width=3.5cm]{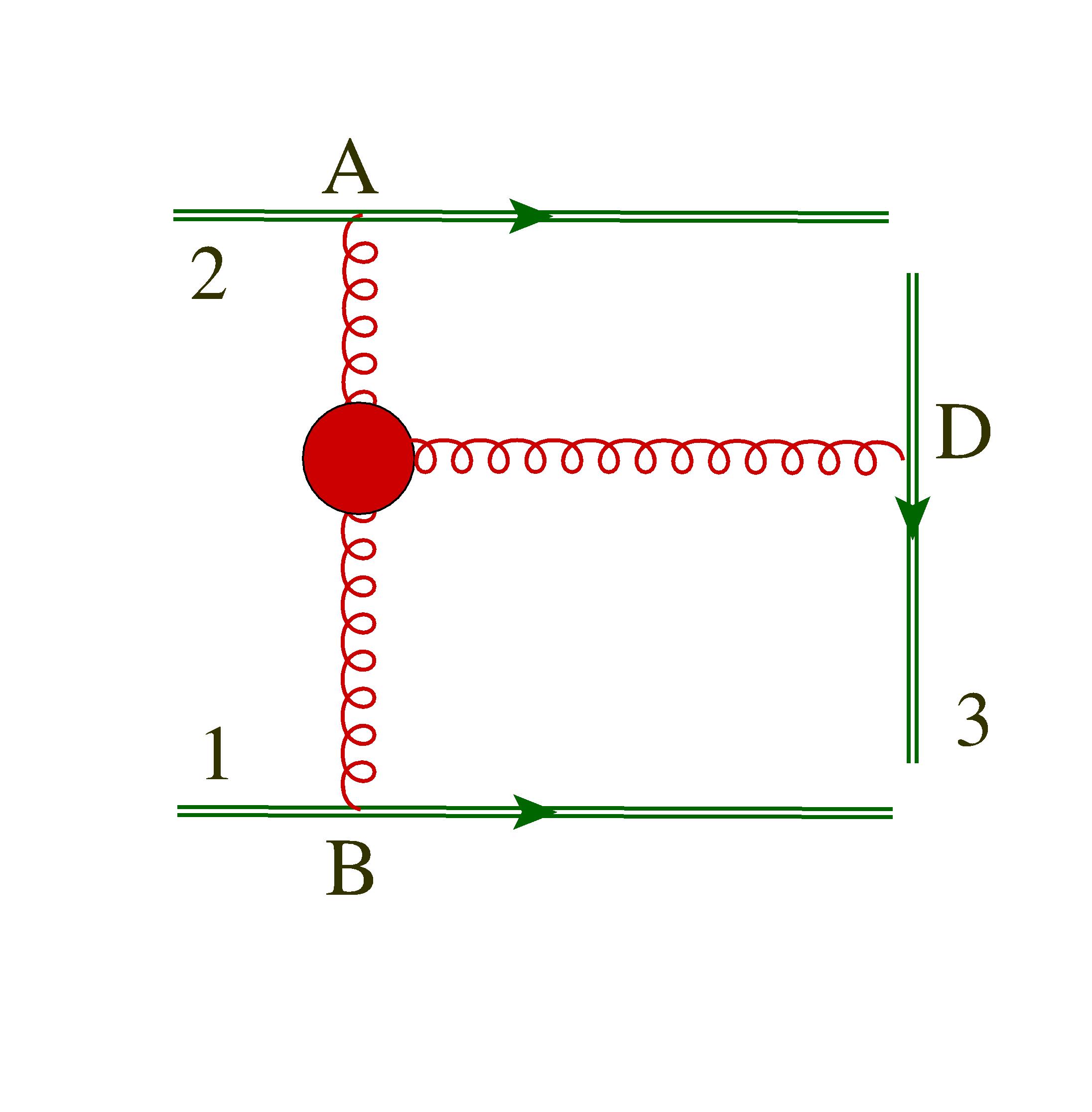} }
	\caption{ Representative diagrams of Cwebs appearing at two loops connecting light like Wilson lines. Cwebs (\textcolor{blue}{a}), and~(\textcolor{blue}{b}) belong to $\{W_{2}^{2}\}$, and $\{W_{3}^{2}\}$ has Cwebs~(\textcolor{blue}{c}), and~(\textcolor{blue}{d})}
	\label{fig:Two-loop-Cwebs-Lightlike}.
\end{figure}
As discussed in~\cite{Agarwal:2020nyc,Agarwal:2021ais}, one of the major drawback of the above algorithm is that it generates multiple copies of a Cweb and they need to be discarded before starting the calculation of mixing matrices at a given perturbative order. 

\subsubsection*{New algorithm}
We find that it is sufficient to use Cwebs at $ l $ loops connecting $ m $ Wilson lines to generate all Cwebs at $ l+1 $ loops connecting the same number of Wilson lines;  also
the third step of the original algorithm generates the Cwebs that are already generated using the first and second steps. Based on these, an improved and efficient algorithm to generate Cwebs recursively is 
given below:

\begin{enumerate}	
	\item To generate Cwebs at $ l+1 $ loops connecting at most $ l+1 $ Wilson lines starting from a Cweb at $ l $ loops connecting $ m $ ($ 1\leq m \leq l+1 $) lines
	\begin{enumerate}
		\item Connect any two existing Wilson lines by introducing a two-point gluon correlator (for massive Wilson lines a two point correlator can also be attached to a single line),
		\item Connect any existing $k$-point gluon correlator to an existing Wilson line.
	\end{enumerate}
	\item To generate Cwebs at $ l+1 $ loops connecting the highest number ($ l+2 $) of Wilson lines, the following steps need to be applied to 
	Cwebs at $l$ loops connecting the highest number $(l+1)$ of lines allowed at this order: 
	\begin{enumerate}
		\item Connect a new Wilson line to any of the existing lines by introducing a two-point gluon correlator.
		\item Connect any existing $k$-point gluon correlator to a new Wilson line.
	\end{enumerate}
	
	\item Discard the duplicate Cwebs.
\end{enumerate}
Let us now explain how the above algorithm works and also contrast it with the older version of the algorithm. Let us denote the set of all Cwebs, $\{W_n^{(c_2, \ldots , c_p)} (k_1, \ldots  , k_n) \} $ that are present $l$ loops connecting $ n $ lines by $\{W_{n}^{l}\}$.
To generate  $\{W_{n}^{l+1}\}$ the earlier version of the algorithm required implementation of operations on both  $\{W_{n}^{l}\}$ and $\{W_{n-1}^{l}\}$. In contrast, the new algorithm requires 
operations only on $\{W_{n}^{l}\}$ --- the Cwebs in $\{W_{n}^{l+1}\}$ generated from $\{W_{n-1}^{l}\}$ need an extra gluon that can connect the existing lines or \textit{blobs} with a new Wilson line, however, we can simply start with  $\{W_{n}^{l}\}$ and attach a gluon to the existing lines or the blobs of the correlators in all possible ways to generate $\{W_{n}^{l+1}\}$.

\begin{figure}[h]
	\centering
	\centering
	\subfloat[][$ \text{W}_3^{(3)}(3,2,1) $]{\includegraphics[height=4cm,width=4cm]{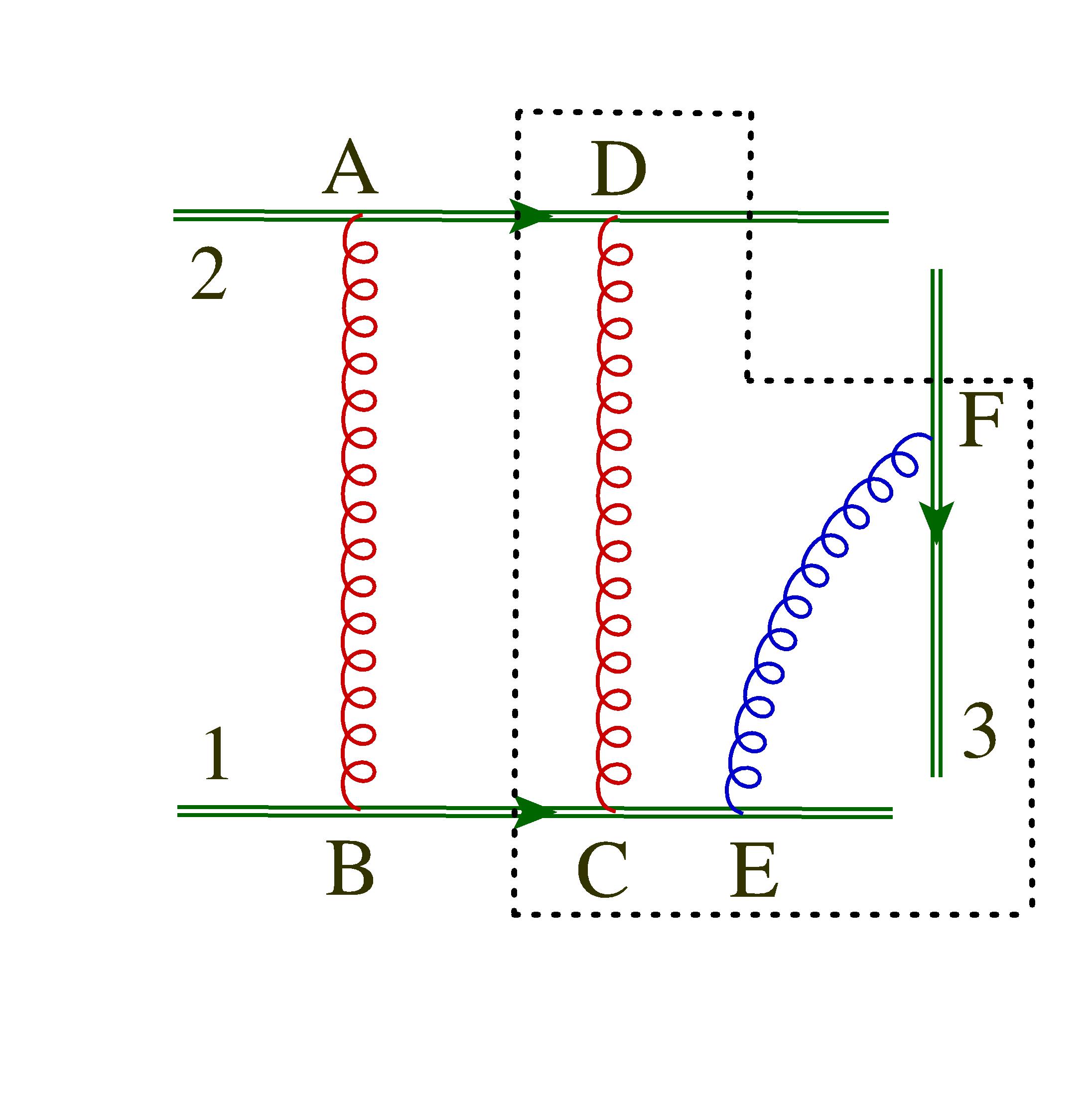} }
	\quad 	
	\subfloat[][$ \text{W}_3^{(3)}(2,2,2) $]{\includegraphics[height=4cm,width=4cm]{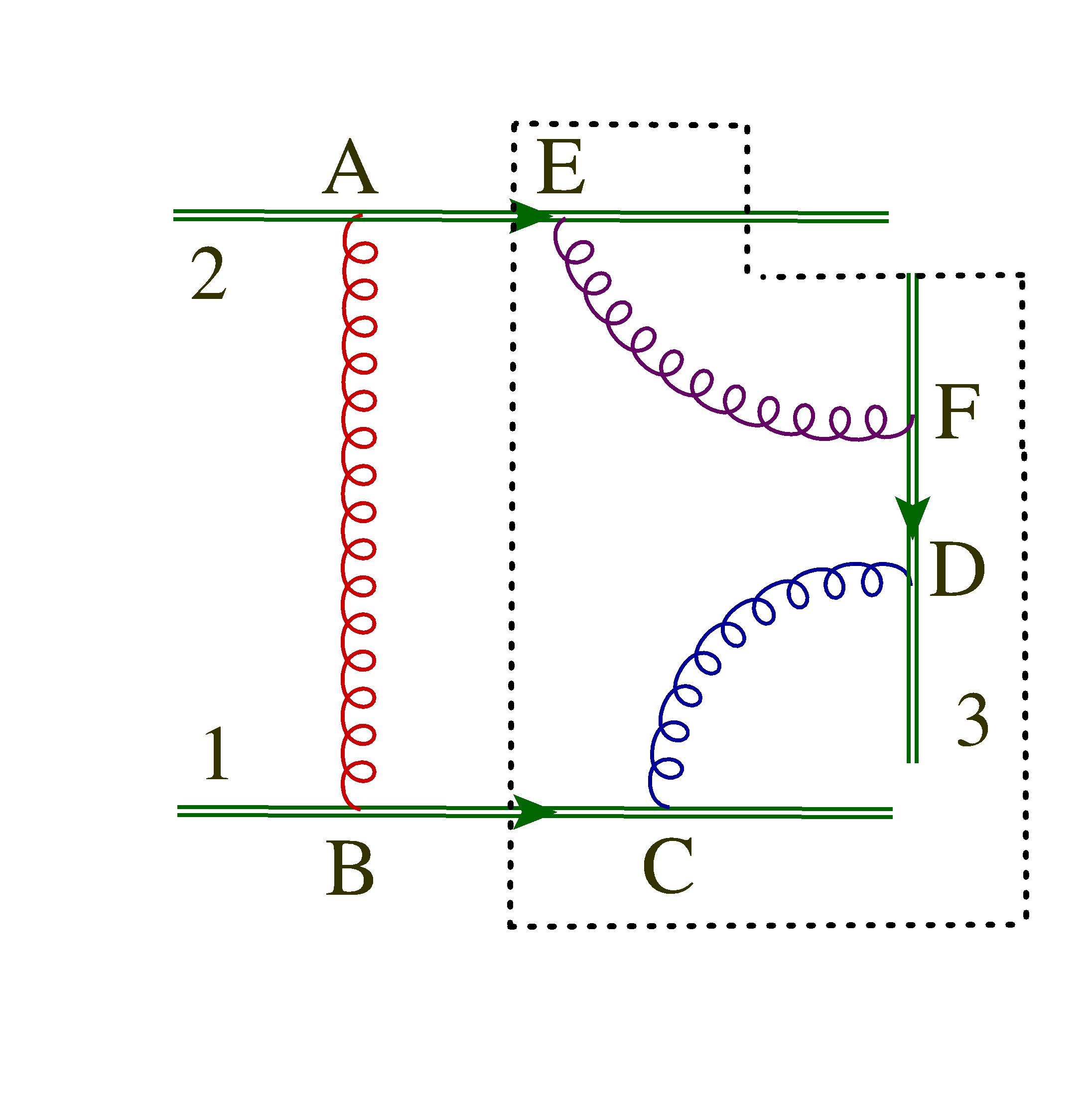} }
	\quad 	
	\subfloat[][$ \text{W}_3^{(1,1)}(3,1,1) $]{\includegraphics[height=4cm,width=4cm]{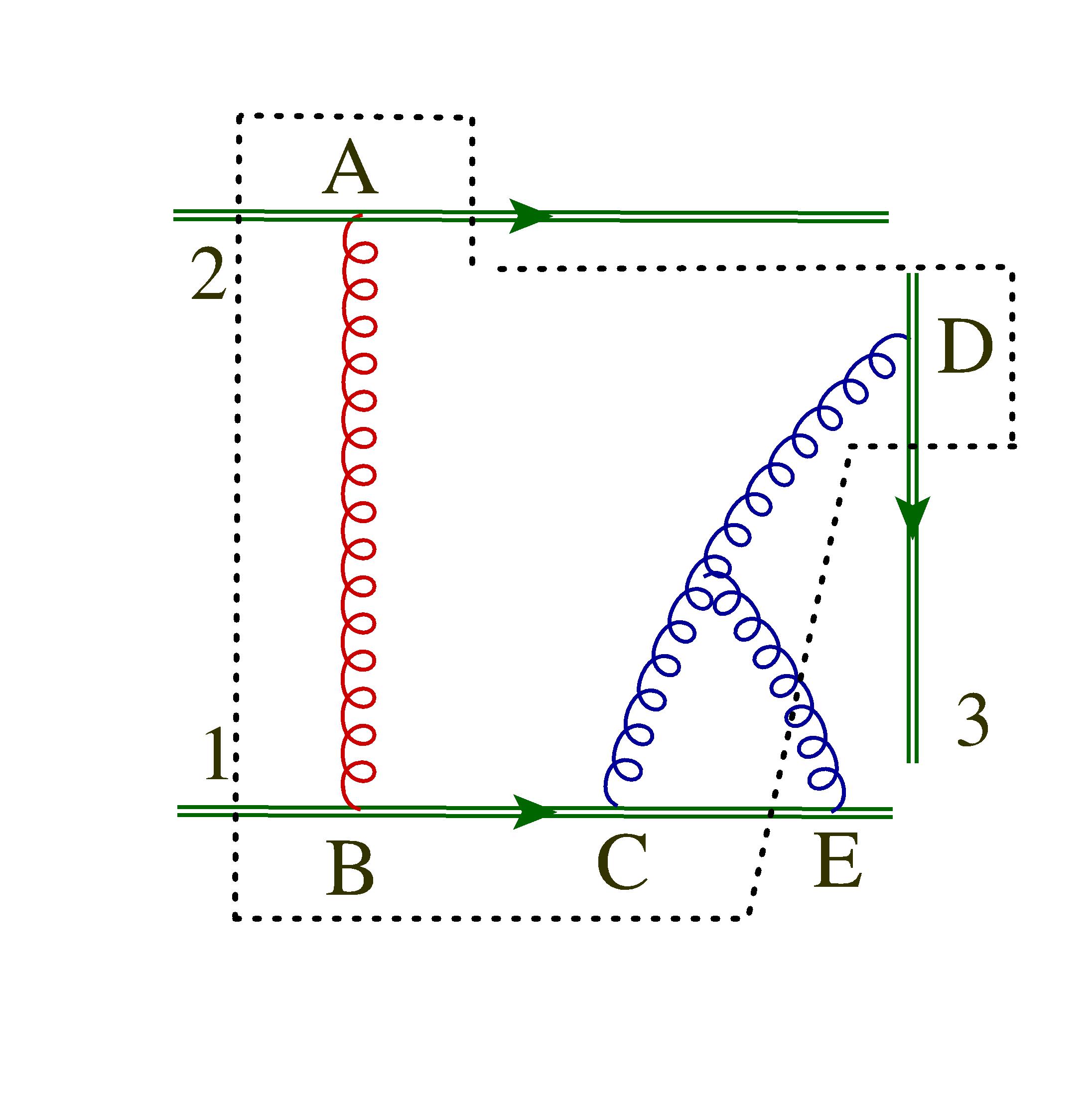} }
	\quad 	
	\subfloat[][$ \text{W}_{3,\,\text{I}}^{(1,1)}(2,1,2) $]{\includegraphics[height=4cm,width=4cm]{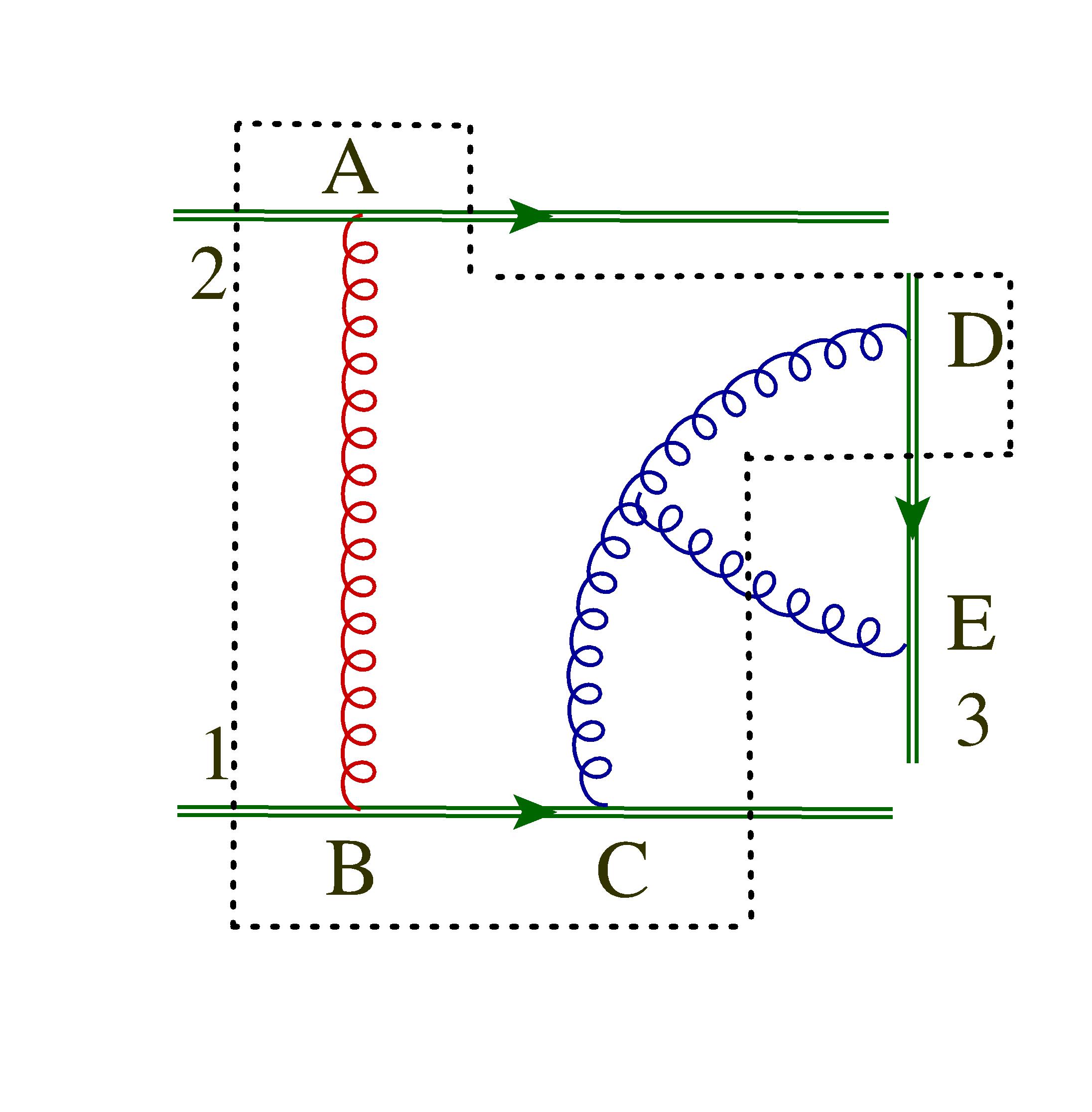} }
	\quad
	\subfloat[][$ \text{W}_{3,\,\text{II}}^{(1,1)}(2,1,2) $]{\includegraphics[height=4cm,width=4cm]{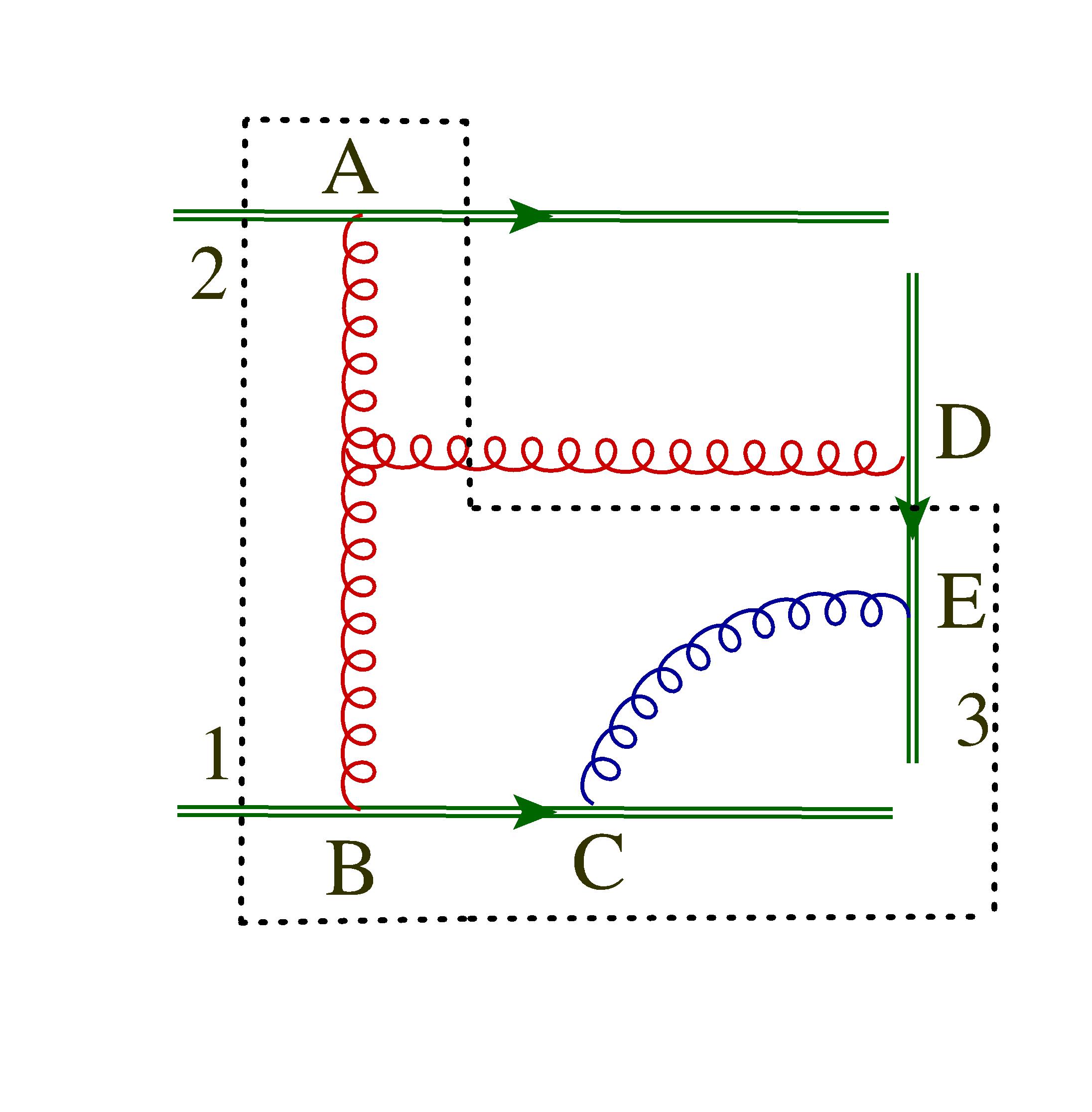} }
	\quad 	
	\subfloat[][$ \text{W}_3^{(0,0,1)}(2,1,1) $]{\includegraphics[height=4cm,width=4cm]{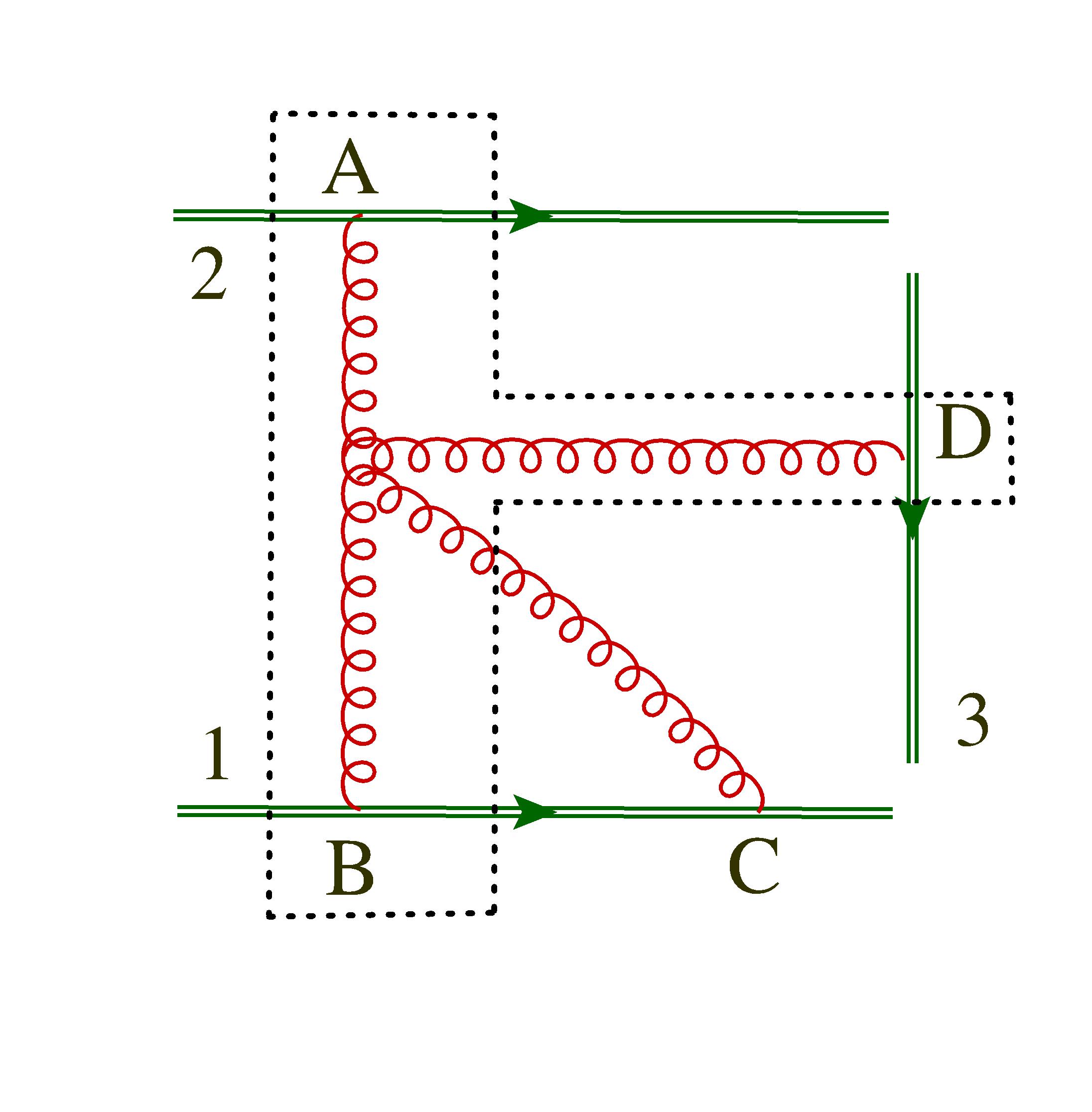} }
	\caption{Cwebs (webs) of $\{W_{3}^{3}\}$ generated using the non recursive algorithm}
	\label{fig:3loop-CWeb-Light-AL2-}
\end{figure}
We apply above reasoning to generate three loop Cwebs $\{W_{3}^{3}\}$ using two loop Cwebs $\{W_{3}^{2}\}$. Cwebs at two loops, as shown in fig.~\eqref{fig:Two-loop-Cwebs-Lightlike} fall into two sets: $\{W_{2}^{2}\}$ containing Cwebs (a) and (b), and $\{W_{3}^{2}\}$ containing Cwebs (c) and (d). 
All the Cwebs at three-loops connecting three lines~\cite{Gardi:2013ita,Agarwal:2020nyc} are shown in the fig.~\eqref{fig:3loop-CWeb-Light-AL2-}. 
The same Cwebs are generated if we use the algorithm on $\{W_{3}^{2}\}$.
In the fig.~\eqref{fig:3loop-CWeb-Light-AL2-} we have indicated by dashed {\it boxes} the Cwebs that are contained in $\{W_{3}^{2}\}$ in fig.~\eqref{fig:Two-loop-Cwebs-Lightlike}; note that {\it each} of them 
has a box that contains one of the Cwebs in  $\{W_{3}^{2}\}$.

In a similar fashion Cwebs in the set $\{W_{2}^{2}\}$ are present in $\{W_{3}^{3}\}$ except those in figs.~(\ref{fig:3loop-CWeb-Light-AL2-}\textcolor{blue}{b}), and~(\ref{fig:3loop-CWeb-Light-AL2-}\textcolor{blue}{e}). This example, thus, provides a justification for our above algorithm.

\subsubsection*{Comparison of older and newer Mathematica implementations}
\label{sec:Codes}

 The in-house Mathematica code \texttt{CwebGen 2.0} that we have used to generate the results reported in this paper
incorporates a new subroutine that is based on the more efficient algorithm of generating Cwebs presented above.
With this new algorithm we have reproduced the known results of Cwebs 
at three~\cite{Gardi:2013ita} and four loops~\cite{Agarwal:2020nyc,Agarwal:2021him} where multi-pronged-boomerang correlators do not appear, which provides a check on our implementation. This algorithm is completely general and applicable at any perturbative order. As mentioned in the algorithm, certain Cwebs (colour factors) are generated multiple times and we discard the duplicates to obtain the unique ones. 
At four loops, \report{for Cwebs without multi-pronged-boomerang correlator}, the new algorithm generates $ 150 $ Cwebs as compared to $ 226 $ Cwebs that are generated by original algorithm; after discarding the duplicates only $ 60 $ remain \cite{Agarwal:2020nyc,Agarwal:2021him}.
The new algorithm generates $ 2.5 $ times of the unique Cwebs, whereas the previous algorithm generates $ 3.8 $ times. Thus, this new algorithm increases the efficiency by around $ 35\% $.

In this work, we have used the subroutine where the colour factors of one diagram from each three-loop Boomerang Cwebs are given as inputs to generate colour factors of Boomerang Cwebs at four loops. The older algorithm when applied on each of the $ 9 $ Boomerang Cwebs present at three loops, generates $ 95 $ Boomerang Cwebs, whereas, the new algorithm generates $ 71 $. After discarding the duplicates only $ 45 $ remain at four loops. Thus the new algorithm increases the efficiency in finding unique Cwebs by $ 25\% $.  
As we go beyond four loops, the number of Cwebs at lower orders increase, thus we expect that new algorithm will become more useful in finding Cwebs at higher orders. 


\section{A brief description of the code \texttt{CwebGen \,2.0} }\label{sec:repl}
One of the most powerful techniques in the combinatorial problems in physics, which involves exponentiation, is the replica trick \cite{MezaPariVira}. For Wilson line correlators, the replica trick algorithm was developed in \cite{Gardi:2010rn,Laenen:2008gt}. The same replica trick was adopted in \cite{Agarwal:2020nyc,Agarwal:2021him} for the calculation of the mixing matrices for four-loop Cwebs. The details of replica trick --- discussed in appendix~\ref{sec:replAppen} --- will be used in the calculation of mixing matrices for Boomerang Cwebs at four loops.  

We present below the algorithm of current version of the in-house Mathematica code \texttt{CwebGen \,2.0}, that incorporates replica trick algorithm and is a significantly improved version of the codes that were used in 
\cite{Agarwal:2020nyc,Agarwal:2021him}. This code has been used in this work to obtain the Cweb mixing matrices.
\begin{itemize}
	
	\item The code begins by generating the colour of one of the Cweb diagrams; however, it is different from the usual colour assignment as we also 
	assign  replica variables $\{i,j,k, \ldots\}$  to each of the gluon correlators that are present in the diagram.
	
	\item A subroutine then shuffles the attachments on each of the Wilson lines belonging to different correlators, and correspondingly different replica variables,  to generate all the diagrams of the Cweb.  
	\item Next hierarchies between the replica variables are generated. For example, for a Cweb with two correlators, with replica variables $i$ and $j$, the code generates hierarchies
	$\{h\} =\{ i=j, i>j,i<j \} $. Then number of distinct replica variables $ N_r $ for each hierarchy is calculated, for example, for hierarchy $ i=j $, $ N_r=1 $, whereas, for $ i>j $, and $ i<j $, $ N_r=2 $. 
	Using this, $ M_{N_r}(h) $  is obtained using eq. (\ref{expocolf}).
	\item[] Note that previous versions of the code could generate all the possible hierarchies only if the number of replica variables was less than or equal to 4.  \texttt{CwebGen \,2.0} can generate
	the hierarchies for any number of replica variables.
	\item Another  subroutine then generates replica ordered colour factors $  \textbf{R} \big[ C(d) \big| h \big] $ for each of the hierarchies (see appendix~\ref{sec:replAppen}) for every diagram of the Cweb.  
	The code then generates the data for each diagram in the Cweb (as given in the Table 1 of~\cite{Gardi:2010rn}) and gives the mixing matrix $ R_W $. 
	A diagonalizing matrix $ Y_W $ is constructed to diagonalize $ R_W $ using its right eigenvectors. $ Y_W $ then acts on the column vector containing the colour factors of each of the diagrams of the Cweb, and then gives the corresponding independent exponentiate colour factors. 
\end{itemize}

The runtime of the code depends on the number of correlators or equivalently number of replica variables. The computation time for four loops is larger as compared to three loops, since the maximum number of hierarchies at four loops is 75 as opposed to 13 at three-loops. In addition to large number of hierarchies, the  total number of diagrams in  any Boomerang Cwebs is larger than that for Cwebs involving \report{no multi-pronged-boomerang correlator}, at a given perturbative order. 

The algorithm and its implementation in two older versions of the same code can be found in~\cite{Agarwal:2020nyc,Agarwal:2021him}. 
To make a comparison of the present version with the previous versions, we have calculated the mixing matrix of $ \text{W}_5^{(4)}{(1,1,1,1,4)} $ appearing at four loops~\cite{Agarwal:2020nyc}. The largest dimension  mixing matrix is $ 24 \times 24 $ and it took $ 7 $ days in the first version~\cite{Agarwal:2020nyc}, upon subsequent improvement it took $ 6.4 $ hours ~\cite{Agarwal:2021him}. The same calculation
by \texttt{CwebGen \,2.0} takes only $ 1.54 $ seconds. Thus, the latest version of code is almost $ 15000 $ times faster as compared to the previous versions.

\begin{Repo}
To test the efficiency of our code, we choose a class of Cwebs $ W^n_{n+1} (1,1,\ldots n) $ whose mixing matrices have dimension $ n! $, where $ n $ is the perturbative order. Further, the Cwebs of this class have the largest dimension of mixing matrices at a given perturbative order connecting maximum number of lines. We run a benchmark test on the computation time for the two loop Cweb that belongs to this category. We also provide the computation times for the Cwebs of this category for three and four loops, which are already known previously in the literature. To test the capability of our code \texttt{CwebGen 2.0} in handling higher orders, we have computed mixing matrices of this category appearing at five and six loops. The computation times of all these mixing matrices are provided in table~\eqref{tab:benchmark}. 
It is evident that \texttt{CwebGen 2.0} is not only well-suited for four-loop but is also capable of handling higher loops.
\begin{table}[H]
\begin{center}
	\begin{tabular}{|c|c|c|}
		\hline 
	loops &	Dimension & Computation time  \\
	\hline  
	2 & 2 & 0.02 s\\
	\hline 
	3 & 6 & 0.07 s\\
	\hline 
	4     & 24        &  1.07 s \\ 
	\hline 
	5     & 120       &   6.05 m\\
	\hline
	6     & 720       &  2.79 h \\
	\hline
	\end{tabular}
\end{center}
\caption{Benchmark tests of \texttt{CwebGen 2.0.} All these tests are performed in a computer with Intel (R) Xeon (R) CPU E5-2697 v4 @ 2.30GHz and 64GB RAM.\label{tab:benchmark}}
\end{table} 	
\end{Repo}

In the next section, we provide the results for two Boomerang Cwebs that are present at four loops and the results for the remaining Cwebs will be presented in the appendix~\ref{sec:result}.


\section{Boomerang Cwebs at Four loops}
\label{sec:BoomCwebsEx}

Using  \texttt{CwebGen \,2.0}  and discarding the duplicates we get 8 and 20 Boomerang Cwebs connecting four and three Wilson lines respectively. In~\cite{Agarwal:2022xec}, we had obtained  the diagonal blocks of mixing matrices for all the 8 Boomerang Cwebs that are present at four loops using the formalism introduced in~\cite{Agarwal:2022wyk}.
Additionally, we were also able to make predictions for the diagonal blocks of 4 classes of Cwebs to all orders in $\alpha_{s}$.
Of course, we could not obtain the complete mixing matrices and exponentiated colour factors in that work which is the subject matter of this paper. 

\begin{Repo}
Before going into the details of the calculation, let us briefly review the concepts which will be useful in understanding the structure of the mixing matrices presented in this paper. The mixing matrices of Cwebs have a general structure, which is manifest,
when the diagrams of these Cwebs are put in a \textit{Normal order}. This structure was first discovered in~\cite{Agarwal:2022wyk}. 

\paragraph*{Normal Order:}
First the diagrams of a Cweb are classified according to their $ s $-factors:
a diagram $d$ is  \textit{reducible} if $s(d) \neq 0$, whereas it is \textit{irreducible} if $s(d) = 0$.  Further irreducible diagrams are classified into the following two categories.

\noindent {\textit{Completely entangled diagram}}: An irreducible diagram in which all the gluon correlators are entangled and thus none of them can be independently shrunk to the origin.

\noindent {\textit{Partially entangled diagram}}: An irreducible diagram which has at least one gluon correlator which is not entangled with the other correlators. \\
\noindent Normal Ordering of diagrams in a Cweb with $ n $ diagrams, is then, defined as the order in which first $k$ diagrams are completely entangled, followed by $(l-k)$
partially entangled diagrams, then followed by reducible diagrams in ascending order of their $s$-factors. The mixing matrix $ R $ then takes the following general form~\cite{Agarwal:2022wyk},
\begin{align}
	R=\left(\begin{array}{cc}
		A_{l\times l}  & B_{l\times m} \\
		O_{m \times l} & D_{m\times m}
	\end{array}\right)\,=\left(\begin{array}{c|c}
	\begin{array}{cc}
		\hspace*{-0.5cm}{I}_{k \times k} & (A_U)_{k\times (l-k)} \\
		{O}_{(l-k)\times k} & \quad \, \, (A_L)_{(l-k) \times (l-k)}
	\end{array} & B_{l\times m} \\ 
	\hline
	{O}_{m\times l } & D_{m\times m}
\end{array}\right)\,,
	\label{eq:R-gen-big-form}
\end{align}
where $ l+m=n $. Here A and D are square matrices of order $ l $, and $ m  $ corresponding to irreducible and
reducible diagrams, respectively; O is null matrix of order $ m \times l $. Additionally $ D $ itself is a mixing matrix as proved in~\cite{Agarwal:2022wyk}. Further $ A $ has a sub-structure --- it contains an identity matrix $ I $ of order $ k $ that corresponds to completely entangled diagrams, a square matrix $ A_L $ of order $ (l-k) $ corresponding to partially entangled diagrams, a null matrix $ O $  of order $ (l-k)\times k $, and a matrix $ A_U $  of order $ k \times (l-k) $.
  
Above structure follows from the replica trick algorithm which can always disentangle two correlators, however can never entangle them. Therefore, the exponentiated colour factors of reducible diagrams do not contain the colour of the irreducible diagrams; and those of completely entangled diagrams do not contain the colour of the partially entangled diagrams. The diagrams of all the Cwebs considered in this paper are Normal Ordered and thus the mixing matrices of them have the form given in eq.~(\ref{eq:R-gen-big-form}).  

\end{Repo}

We now present the results for two Cwebs:  one connecting four Wilson lines, and  another connecting three Wilson lines. In each case we will present 
(i) Column weight vector $ S $, 
(ii) Mixing Matrix $R$,
(iii) Exponentiated colour factors $(YC)$.

\subsubsection*{$\text{W}\,_{4}^{(1,0,1)}\,(3,1,1,1)$: a four-line Boomerang Cweb}

\begin{figure}[t]
	\captionsetup[subfloat]{labelformat=empty}
	\centering	
	\subfloat[][$ C_1 $]{\includegraphics[scale=0.05]{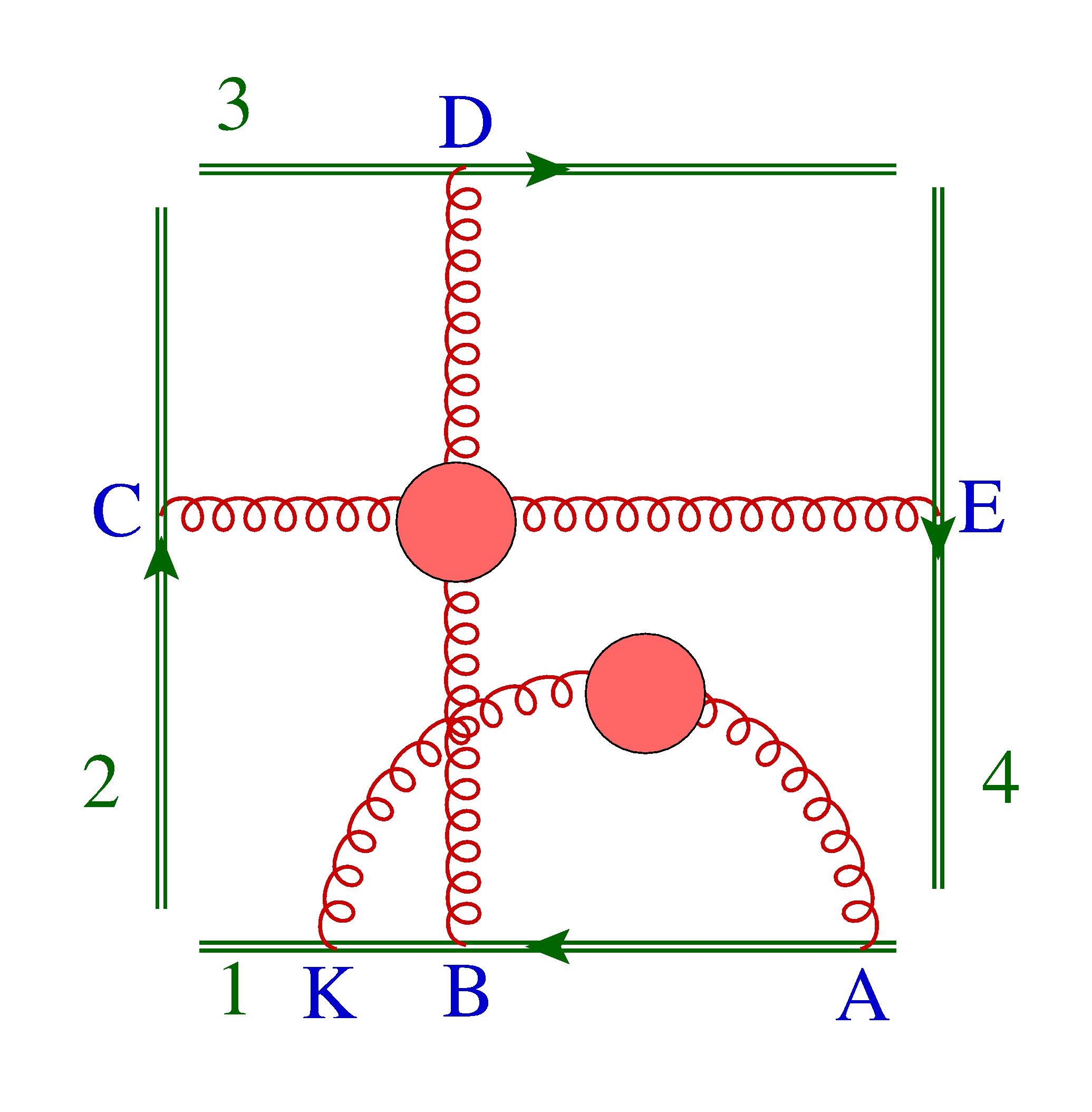} }
	\qquad
	\subfloat[][$ C_2 $]{\includegraphics[scale=0.05]{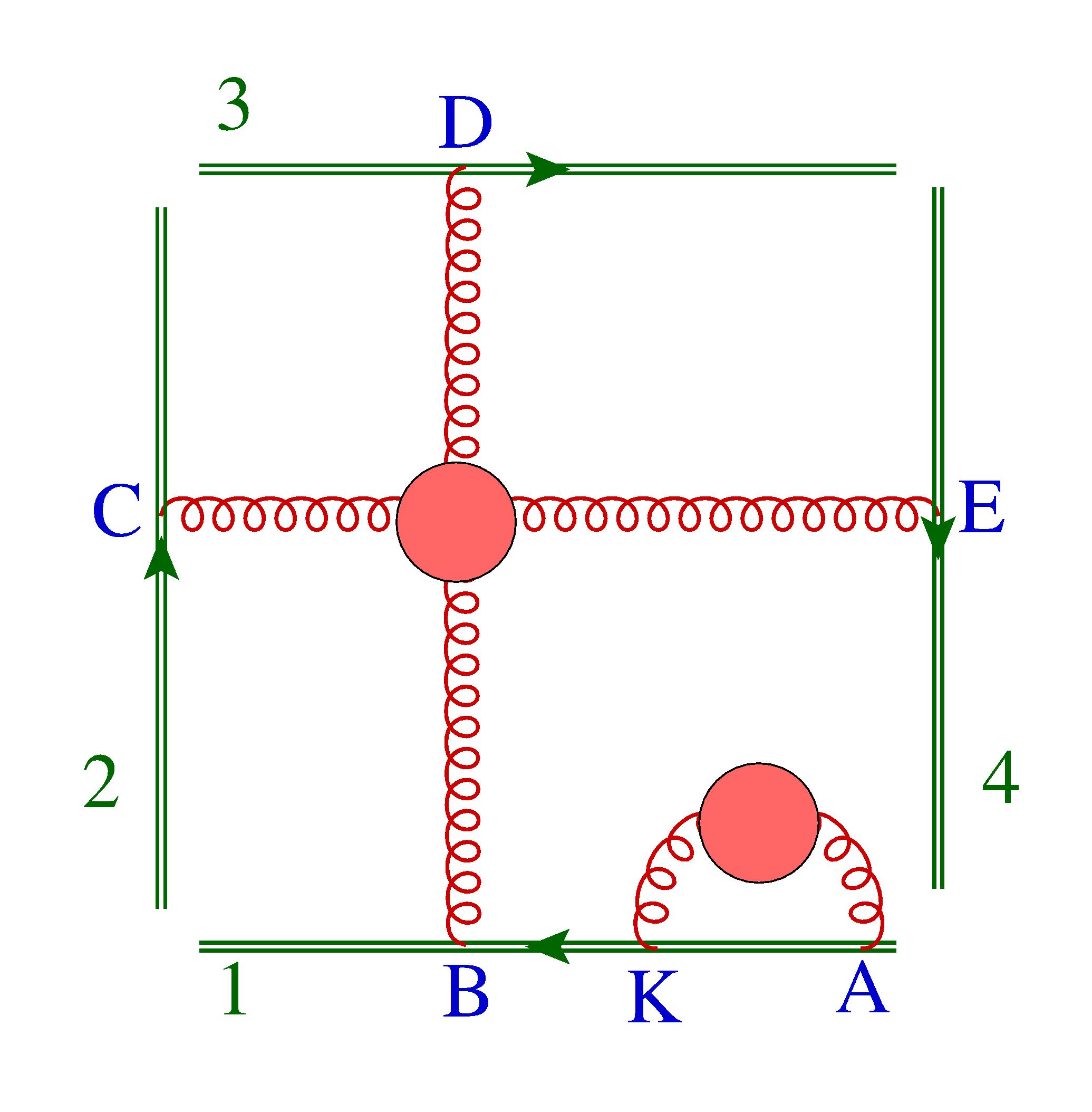} }
	\qquad
	\subfloat[][$ C_3 $]{\includegraphics[scale=0.05]{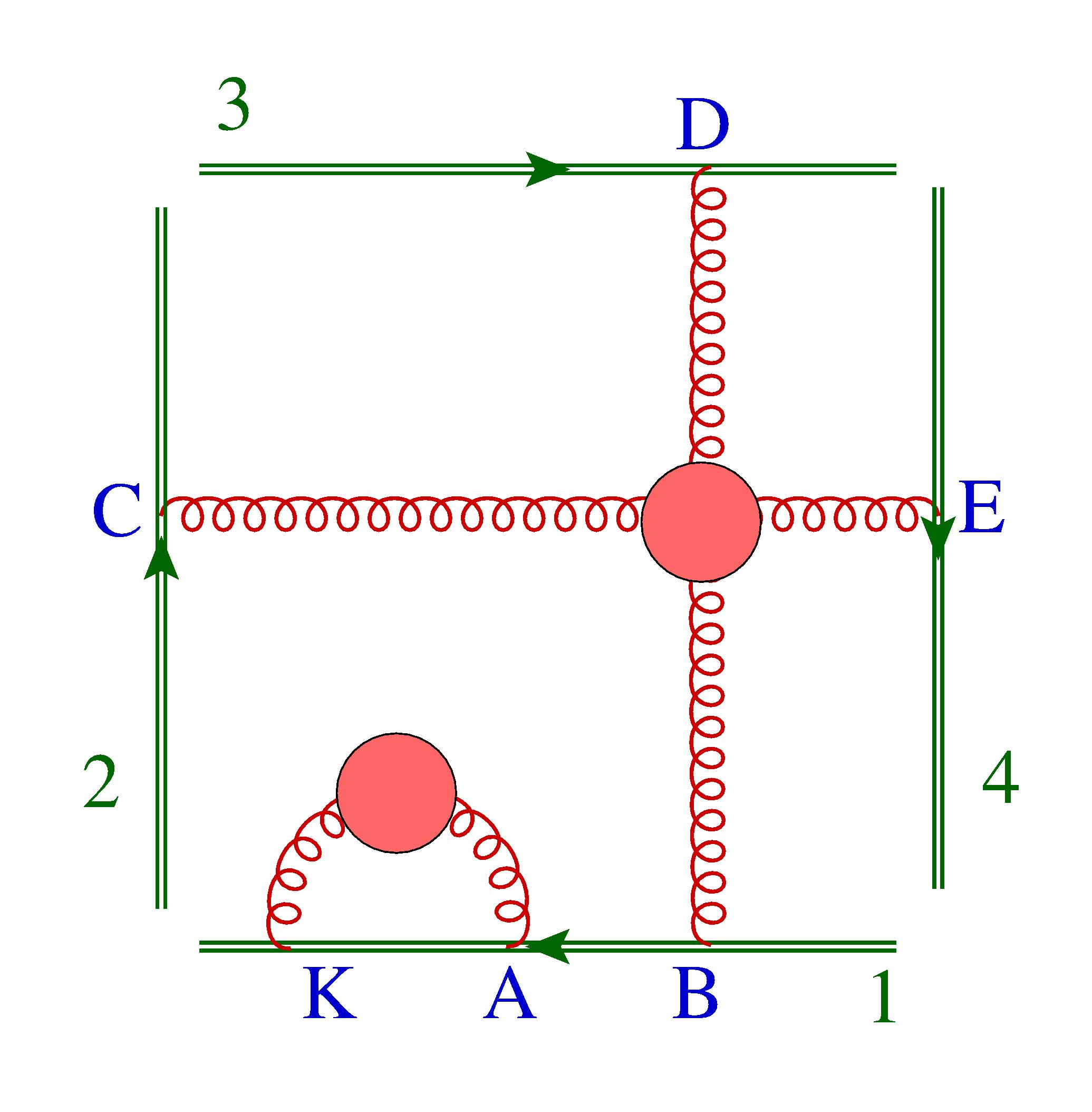} }
	\caption{Diagrams of Boomerang Cweb $\text{W}\,_{4}^{(1,0,1)}\,(3,1,1,1)$ }
	\label{fig:ExampleFour-Web}
\end{figure}

Boomerang Cweb $\text{W}\,_{4}^{(1,0,1)}\,(3,1,1,1)$, shown in fig~\eqref{fig:ExampleFour-Web}, connects four Wilson lines with one multi-pronged-boomerang correlator (m=2) and a four point gluon correlator. This Cweb has three possible shuffles on Wilson line $ 1 $. 
The shuffle of gluon attachments from the two correlators generates three diagrams as shown in fig.~\eqref{fig:ExampleFour-Web}; thus the order of mixing matrix for this Cweb is three. To proceed we choose the order of the diagrams as in the fig.~(\ref{fig:ExampleFour-Web}). This order is labeled using the order of attachments on line 1: $ C_1\,=\,\{ABK\} $, $ C_2\,=\,\{AKB\} $ and $ C_3\,=\,\{BAK\} $\,.
The mixing matrix $ R $ and the diagonalizing matrix $ Y $ has been calculated using  \texttt{CwebGen \,2.0}  and we get,  

\begin{align}\label{eq:4linew}
	R\,=\,\dfrac{1}{2}\left(
	\begin{array}{ccc}
		2 & -1 & -1 \\
		0 & \;\;\, 1 & -1 \\
		0 & -1 &\;\;\, 1 \\
	\end{array}
	\right) \,, \qquad\qquad\qquad
	Y\,=\,\left(
	\begin{array}{ccc}
		1 & -1 & 0 \\
		0 & -1 & 1 \\
		0 &\;\;\, 1 & 1 \\
	\end{array}
	\right)\,.
\end{align}
This matches with the result obtained in~\cite{Agarwal:2022xec}.

The column weight vector $ S $ for this Cweb can be calculated by determining the $ s $-factors~\cite{Gardi:2011yz} for each diagram. The $ s $-factor for $ C_1 $ is zero as one can not shrink any of the two correlators to the origin of Wilson lines. In the diagram $ C_2 $ one can shrink the multi-pronged-boomerang correlator first then the four point correlator, thus there is only one way to shrink all the correlators to origin, making $ s=1 $. The $ s $-factor for $ C_3 $ is also one except the fact that here the multi-pronged-boomerang correlator has to be  shrunk after the four point correlator. Thus the column weight vector becomes $S=\{0,1,1\}\,.$

The matrix $ R $ satisfies all the known properties of mixing matrix: it is idempotent, $ R^2=R $, satisfies the zero row sum rule as in each row of matrix sum of entries is zero, and, the conjectured column sum rule. The rank of this mixing matrix is two as there are only two independent rows in $ R $. We obtain the independent exponentiated colour factors given as, 
\begin{eqnarray}
	(YC)_1 \,=\,  0\,,\qquad
	(YC)_2 \,=\,  -i\, f^{abn} f^{bch} f^{deh} \, \ta 1 \tn 1 \tc 2 \td 3 \te 4\,.
\end{eqnarray}
Let us define 
\begin{equation}
	{N}_{ecf} = \textrm{number of non-vanishing independent ECFs}.
\end{equation}
For the above Cweb ${N}_{ecf} =1$, whereas the rank of the mixing matrix $ r(R) =2$. The vanishing of one of the independent ECFs  can be understood in terms of $\widetilde{C}$.
Applying $ R $ on the $C$ we get,
\begin{align}
	\widetilde{C}_1 =  \dfrac{1}{2}\left( 2C_1-C_2 - C_3\right),\qquad 
	\widetilde{C}_2 =  \dfrac{1}{2}\left( C_2 - C_3\right),\qquad
	\widetilde{C}_3 =  \dfrac{1}{2}\left( C_3 - C_2\right)\,.
\end{align} 
\noindent The colour factors of diagrams $ C_2 $ and $ C_3 $ are identical; the only difference is due to the placement of the Boomerang in each of the diagrams, however, it contributes the same factor $C_{A}$. Hence the exponentiated colour factors
$\widetilde{C}_2 $ and $\widetilde{C}_3$ in the above equations vanish.  This makes  ECFs for certain diagrams vanish for Boomerang Cwebs and it holds true to all orders in the perturbation theory~\cite{Gardi:2021gzz}. 
Recall that when a Cweb is Normal Ordered, the block $D$ gives mixing between the reducible diagrams. 
All these  diagrams contain Boomerang that give $C_{A}$ factors, and thus their ECFs vanish.
That is, 
\begin{align}
	{N}_{ecf} &= \text{rank}(R)  \hspace{0.5cm} \text{Cwebs without 
	\report{multi-pronged-boomerang correlators}}, \nonumber \\
	{N}_{ecf} &< \text{rank}(R)  \hspace{0.5cm} \text{Cwebs} \;
\text{\report{ containing multi-pronged-boomerang correlators}.}
\end{align}
The results presented in the appendix also exhibit this property. The mixing matrices and exponentiated colour factors for remaining seven Boomerang Cwebs that connect four Wilson lines are given in appendix~\ref{app:four-lines}.  

\subsubsection*{$\text{W}\,_{3,\text{I}}^{(2,1)}(3,3,1)$: a three-line Boomerang Cweb}

\begin{figure}
	\centering
	\vspace{-2mm}
	\includegraphics[scale=0.05]{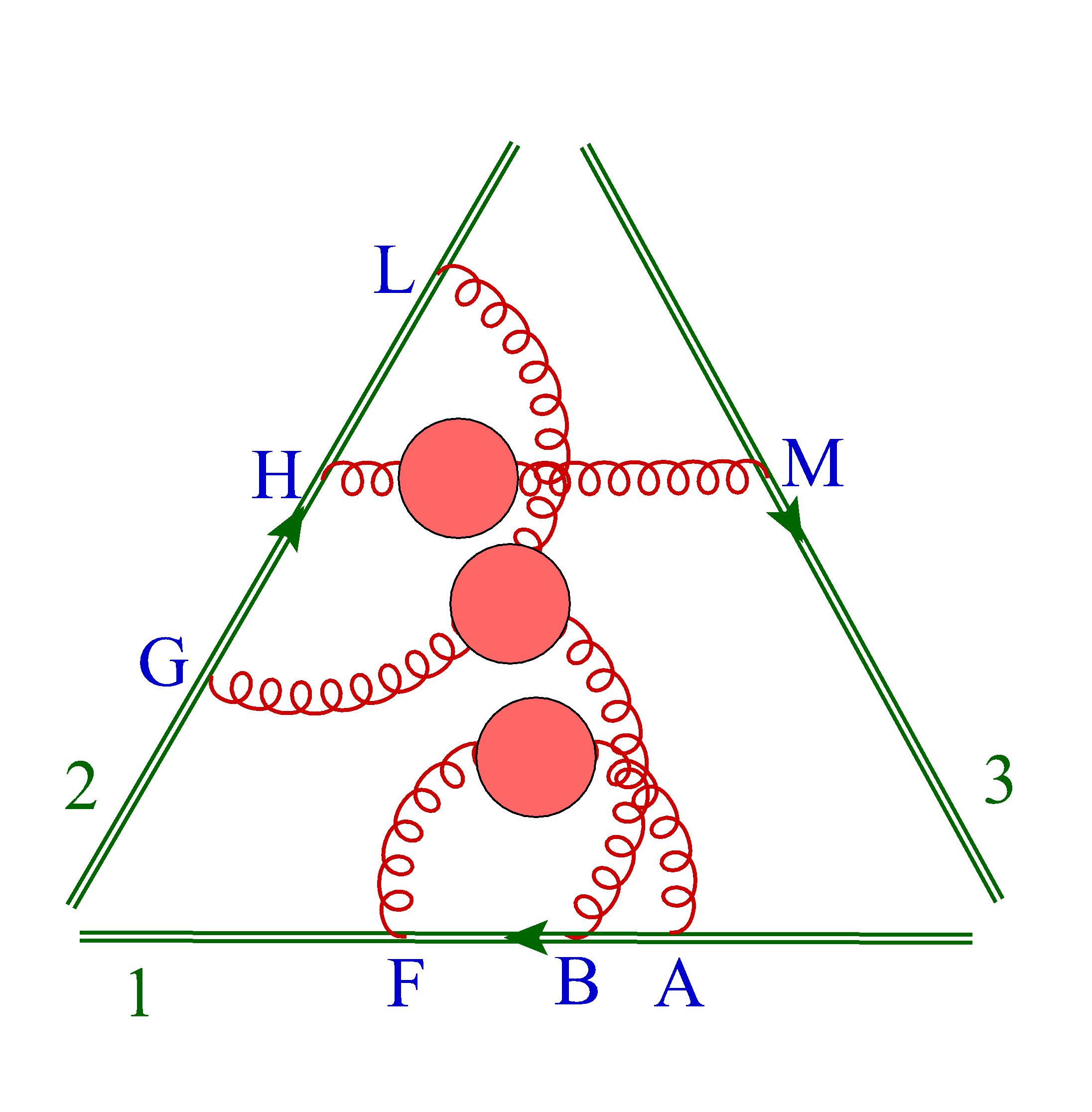} 
	\caption{Boomerang Cweb $\text{W}\,_{3,\text{I}}^{(2,1)}(3,3,1)$}\label{fig:example-2-three-line}
\end{figure}

In  $\text{W}\,_{3,\text{I}}^{(2,1)}(3,3,1)$, shown in fig.~\eqref{fig:example-2-three-line}, one multi-pronged-boomerang correlator (m=2), a two-point and a three-point gluon correlator connect the three lines. The subscript $ \text{``I''} $ indicates that there are more than one Cwebs with the same attachment and correlator content. The shuffle of attachments  generates nine diagrams; the order of diagrams  is given in table~\ref{tab:example-2-three-line} along with their $ s $-factors.

\begin{table}[H]
	\begin{minipage}[c]{0.5\textwidth}%
		\resizebox{0.95\textwidth}{!}{
			\begin{tabular}{|c|c|c|}
				\hline 
				\textbf{Diagrams}  & \textbf{Sequences}  & \textbf{s-factors}  \\ 
				\hline
				$C_{1}$&$\{\{ABF\},\{ GHL\}\}$&0\\\hline
				
				$C_{2}$&$\{\{AFB\},\{ GHL\}\}$&0\\\hline
				
				$C_{3}$&$\{\{ABF\},\{ GLH\}\}$&0\\\hline
				
				$C_{4}$&$\{\{ABF\},\{ HGL\}\}$&0\\\hline
				
				$C_{5}$&$\{\{BAF\},\{ GHL\}\}$&0\\\hline
		\end{tabular}	}
	\end{minipage}\hspace{0.7cm} %
	\begin{minipage}[c]{0.46\textwidth}
		\vspace{-0.51cm}
		\resizebox{0.95\textwidth}{!}{
			\begin{tabular}{|c|c|c|}
				\hline 
				\textbf{Diagrams}  & \textbf{Sequences}  & \textbf{s-factors}  \\ 
				\hline
				$C_{6}$&$\{\{AFB\},\{ GLH\}\}$&1\\\hline
				
				$C_{7}$&$\{\{BAF\},\{ HGL\}\}$&1\\\hline
				
				$C_{8}$&$\{\{AFB\},\{ HGL\}\}$&2\\\hline
				
				$C_{9}$&$\{\{BAF\},\{ GLH\}\}$&2\\\hline
		\end{tabular}}
	\end{minipage}
	\caption{Order of diagrams of Cweb $\text{W}\,_{3,\text{I}}^{(2,1)}(3,3,1)$ and their $ s $-factors}\label{tab:example-2-three-line} %
\end{table}

\noindent Using these $s$-factors we obtain the column weight vector $S$:
\begin{align}\label{eq:example-2-S}
	S\;=\;\{0,0,0,0,0,1,1,2,2\}\,.
\end{align}
Using {\it CwebGen 2.0} we obtain 
\begin{align}
	R=\frac{1}{6} \left(
	\begin{array}{ccccccccc}
		6 & -3 & -3 & -3 & -3 & 2 & 2 & 1 & 1 \\
		0 & 3 & 0 & 0 & -3 & -1 & 2 & -2 & 1 \\
		0 & 0 & 3 & -3 & 0 & -1 & 2 & 1 & -2 \\
		0 & 0 & -3 & 3 & 0 & 2 & -1 & -2 & 1 \\
		0 & -3 & 0 & 0 & 3 & 2 & -1 & 1 & -2 \\
		0 & 0 & 0 & 0 & 0 & 2 & 2 & -2 & -2 \\
		0 & 0 & 0 & 0 & 0 & 2 & 2 & -2 & -2 \\
		0 & 0 & 0 & 0 & 0 & -1 & -1 & 1 & 1 \\
		0 & 0 & 0 & 0 & 0 & -1 & -1 & 1 & 1 \\
	\end{array}
	\right)\,.
\end{align}
This satisfies all the known properties listed in section~\ref{sec:CwebProper}.
This Cweb has four independent exponentiated colour factors, two of which vanish due to the presence of a Boomerang.
The remaining two ECFs are, 
\begin{align}
	{(YC)}_3 &=
	i\,f^{edk} f^{abn} f^{bcd}\,\textbf{T}_1^{a} \textbf{T}_1^{n} \textbf{T}_2^{c} \textbf{T}_2^{k} \textbf{T}_3^{e} 
	+ i\,f^{eck}\,f^{abn}\,f^{bcd}\,\textbf{T}_1^{a}\,\textbf{T}_1^{n}\,\textbf{T}_2^{k}\,\textbf{T}_2^{d}\,\textbf{T}_3^{e}\,\nonumber\\
	& \nonumber\\
	{(YC)}_4 &=
	i\,f^{edk}\,f^{abn}\,f^{bcd}\,\textbf{T}_1^{a}\,\textbf{T}_1^{n}\,\textbf{T}_2^{c}\,\textbf{T}_2^{k}\,\textbf{T}_3^{e}\,.\nonumber	
\end{align}

\noindent The results for all the twenty Boomerang Cwebs that connect three Wilson lines at four  loops are given in appendix~\ref{app:three-lines}.

\section{Conclusions}
\label{sec:summary}
In this article we study Boomerang Cwebs that connect three and four Wilson lines at four loops
which are important ingredients in the studies of  scattering  involving massive particles. 

To enumerate Cwebs  at a given perturbative order, we have introduced an improved version of the algorithm that was developed in~\cite{Agarwal:2020nyc}. 
We presented some details of the current version of the in-house Mathematica code \texttt{CwebGen \,2.0}, that incorporates this new algorithm, and the replica trick, and it is a significantly faster  version of the codes that were used in 
\cite{Agarwal:2020nyc,Agarwal:2021him}.

\begin{figure}[t]
	\centering
	\includegraphics[scale=0.1]{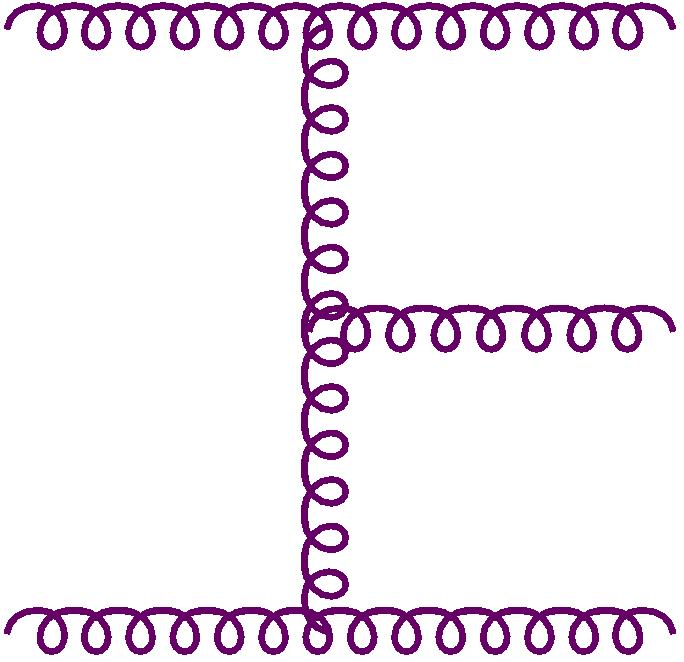}
	\caption{Exponentiated colour factor obtained from all the Boomerang Cwebs}
	\label{fig:ecf}
\end{figure}

We have computed the mixing matrices, the diagonalizing matrices and the exponentiated colour factors for all these Cwebs. 
We have verified that our results match with the predictions of the diagonal blocks presented in~\cite{Agarwal:2022xec}.  
We found that the general structure of exponentiated colour factor of all the twenty eight Boomerang Cwebs, shown in fig.~(\ref{fig:ecf}), is same as the general structure of \report{ECFs for the Cwebs without the multi-pronged-boomerang correlators at four  loops~\cite{Agarwal:2020nyc,Agarwal:2021him}. This is an artifact of the Cwebs with multi-pronged-boomerang correlator}, for which  quadratic Casimir $ C_A $ (the colour of a Boomerang) is absent from all the exponentiated colour factors. This is in agreement with the calculations of Boomerang Cwebs at three loops~\cite{Gardi:2021gzz}. 
We found that the ECFs corresponding to  the $ D $ block of mixing matrices vanish which is consistent with ~\cite{Gardi:2021gzz}.
The exponentiated colour factors have long expressions for Cwebs connecting three Wilson lines and thus we refrain from presenting it in this article. However, they can be obtained from the authors upon request as FORM files.

The interplay of colour and kinematics in the study of soft anomalous dimension has been an interesting object of study and it often puts constraints on the general structure of the anomalous dimension. It  will be an interesting study to determine the constraints on the massive soft anomalous dimension based on the exponentiated colour factors presented in this article.  

\section*{Acknowledgments} 

\noindent 
NA, SP, AT would like to thank Prof.~Lorenzo Magnea for collaboration on the related earlier projects. AS would like to thank CSIR, Govt. of India, for a SRF Fellowship (09/1001(0075)/2020-EMR-I). A part of the computing was performed in the TDP resources at the Physical Research Laboratory (PRL).

\section{Appendix}
\appendix

\section{Replica trick and mixing matrices}\label{sec:replAppen}
In this appendix we briefly discuss the replica trick algorithm, which will be used in the calculation of the mixing matrices for Boomerang Cwebs at four loops.  
To start with, we consider the path integral of the Wilson line correlators as, 
\begin{align}
	\mathcal{S}_n(\gamma_i)=\,\int \mathcal{D}A_\mu^a\,\exp(iS(A_\mu ^a)) \prod _{k=1}^n\phi_k(\gamma_k)=\exp[\mathcal{W}_n(\gamma_i)]\,,
\end{align}
where $ S(A_\mu ^a) $ is the classical action of the gauge fields. In order to proceed with the replica trick algorithm, one introduces $ N_r $ non-interacting identical copies of each gluon field $ A_\mu $, which means, we replace each $ A_\mu $ by $ A_\mu ^i $, where,  $ i=1,\ldots, N_r $. Now, for each replica, we associate a copy of each Wilson line, thereby, replacing each Wilson line by a product of $ N_r $ Wilson lines. Thus, in the replicated theory, the path integral of the Wilson line correlator can then be written as,
\begin{align}
	{\cal S}_n^{\, {\rm repl.}} \left( \gamma_i \right) \, = \,   \Big[ 
	{\cal S}_n \left( \gamma_i \right) \Big]^{N_r} \, = \, \exp \Big[ N_r \,
	{\cal W}_n (\gamma_i) \Big] \, =  \, {\bf 1} + N_r \, {\cal W}_n (\gamma_i) 
	+ {\cal O} (N_r^2) \, .
	\label{exprepl}
\end{align}
Now, using this equation, one can calculate $ \mathcal{W}_n $ by calculating $ \mathcal{O}(N_r) $ terms of the Wilson line correlator in the replicated theory. The method of replicas involves five steps, which are summarized below. 
\begin{enumerate}
	\item [-] Associate a replica number to each connected gluon correlator in a Cweb. 
	\item[-]  Define a replica ordering operator $ \textbf{R} $, which acts on the colour generators on each Wilson line and order them according to their replica numbers. Thus, if $ \textbf{T}_i $ denotes a colour generator for a correlator belonging to replica number $ i $, then action of $ \textbf{R} $ on $ \textbf{T}_i \textbf{T}_j$ preserves the order for $ i\leq j $, and reverses the order for $ i>j $. Thus, replica ordered colour factor for a diagram in a Cweb will always be a diagram of the same Cweb.
	\item [-] The next step in order to calculate the exponentiated colour factors, one needs to find the hierarchies between the replica numbers present in a Cweb. If a Cweb has $ m $ connected pieces, we call hierarchies $ h(m) $. $ h(m) $ are known as Bell number or Fubini number \cite{IntSeq} in the number theory and combinatorics. The first few Fubini numbers are given by $ h(m)=\{1,1,3,13,75,541\} $ for $ m= 0,1,2,3,4,5$. At four loops, the highest number of correlator in a Cweb is $ m_{\text{max}}=4 $, which corresponds to $ h_{\text{max}}=75 $.   
	\item [-]  The next object is to calculate $ M_{N_r}(h) $, which counts the number of appearances of a particular hierarchy in the presence of $ N_r $ replicas.  For a given hierarchy $ h $, which contains $ n_r(h) $ distinct replicas, the multiplicity $ M_{N_r}(h) $ is given by, 
	\begin{align}
		M_{N_r}(h) \, = \, \frac{N_r!}{\big( N_r - n_r(h) \big)! \,\, n_r(h)!}  \,.
	\end{align}  
	\item [-] The exponentiated colour factor for a diagram $ D $ is then given by, 
	\begin{align}
		C_{N_r}^{\, {\rm repl.}}  (D) \, = \, \sum_h M_{N_r} (h) \, \textbf{R} \big[ C(D) \big| h 
		\big]  \, ,
		\label{expocolf}
	\end{align}
	where $ \textbf{R} \big[ C(D) \big| h $ is the replica ordered colour factor of diagram $ D $, for hierarchy $ h $. Finally, the exponentiated colour factor for diagram $ D $ is computed by extracting the coefficient of $ \mathcal{O}(N_r) $ terms of the above equation.  	
\end{enumerate} 
\section{Calculation of $ s $-factor in different representations} \label{sec:shrinking}
\begin{figure}[h]
	\centering
	\vspace{-3mm}
	\hspace{-0.2cm}
	\subfloat[]{\hspace{0.8cm}\includegraphics[scale=0.6]{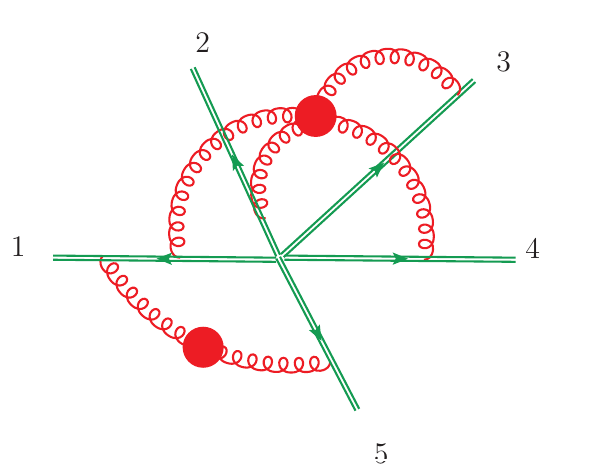} } 
	$ \; $
	\subfloat[]{\hspace{1.8cm}\includegraphics[scale=0.6]{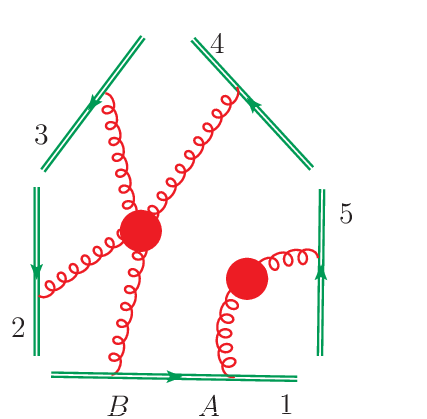} } 
	\caption{Two different representations of Cweb $W^{(1,0,1)}_5 (2,1, 1, 1, 1)$ }
	\label{fig:sfactorEx}
\end{figure}

In this section, we demonstrate the computation of $ s $-factors by shrinking correlators in two different representations. Let us consider the Cweb $W^{(1,0,1)}_5 (2, 1, 1, 1, 1)$. Here, in fig.~\ref{fig:sfactorEx}\textcolor{blue}{a} the hard interaction vertex is located at the point where tails of all the Wilson lines meet. The same diagram is drawn in fig.~\ref{fig:sfactorEx}\textcolor{blue}{b}  where the arrows on the Wilson lines are used to indicate the hard vertex: the tail ends of each of the Wilson lines meet at the hard vertex. 

In fig.~(\ref{fig:sfactorEx}\textcolor{blue}{a}) we first shrink the four gluon correlator to the hard interaction vertex, followed by the two gluon correlator this gives $ s=1$ for this diagram.
Now we describe the procedure for finding the $s$-factors when the diagrams are drawn as in 
fig.~(\ref{fig:sfactorEx}\textcolor{blue}{b}).
Pull each of the gluon correlators to the tail ends of the associated Wilson lines without dragging the attachments from other correlators. 
For example in fig.~(\ref{fig:sfactorEx}\textcolor{blue}{b}), we can pull the four gluon correlator to the tails of attached Wilson lines (1, 2, 3 and 4) followed by the pulling of two gluon correlator to the tails of Wilson lines 1 and 5. This is the only way we can sequentially pull these gluon correlators without dragging other correlators in the process, giving $s=1$. Following the same steps one can compute the $ s $-factors for any diagram in any of the representations.

\section{Boomerang Cwebs at Four loops}\label{sec:result}


\subsection{Boomerang Cwebs connecting four Wilson lines}
\label{app:four-lines}
\begin{itemize}
	\item[\textbf{1}.] $\textbf{W}\,_{4}^{(4)}(3,1,3,1)$

	This Cweb is made up with four two-gluon correlator. It contains eighteen diagrams, one of which is displayed below. The table below
	shows the chosen order of the eighteen shuffles of the gluon attachments,
	 their corresponding $s$-factors. \\%
	\begin{minipage}[c]{0.5\textwidth}%
		\begin{figure}[H]
			\vspace{-2mm}
			\includegraphics[width=4cm,height=4cm]{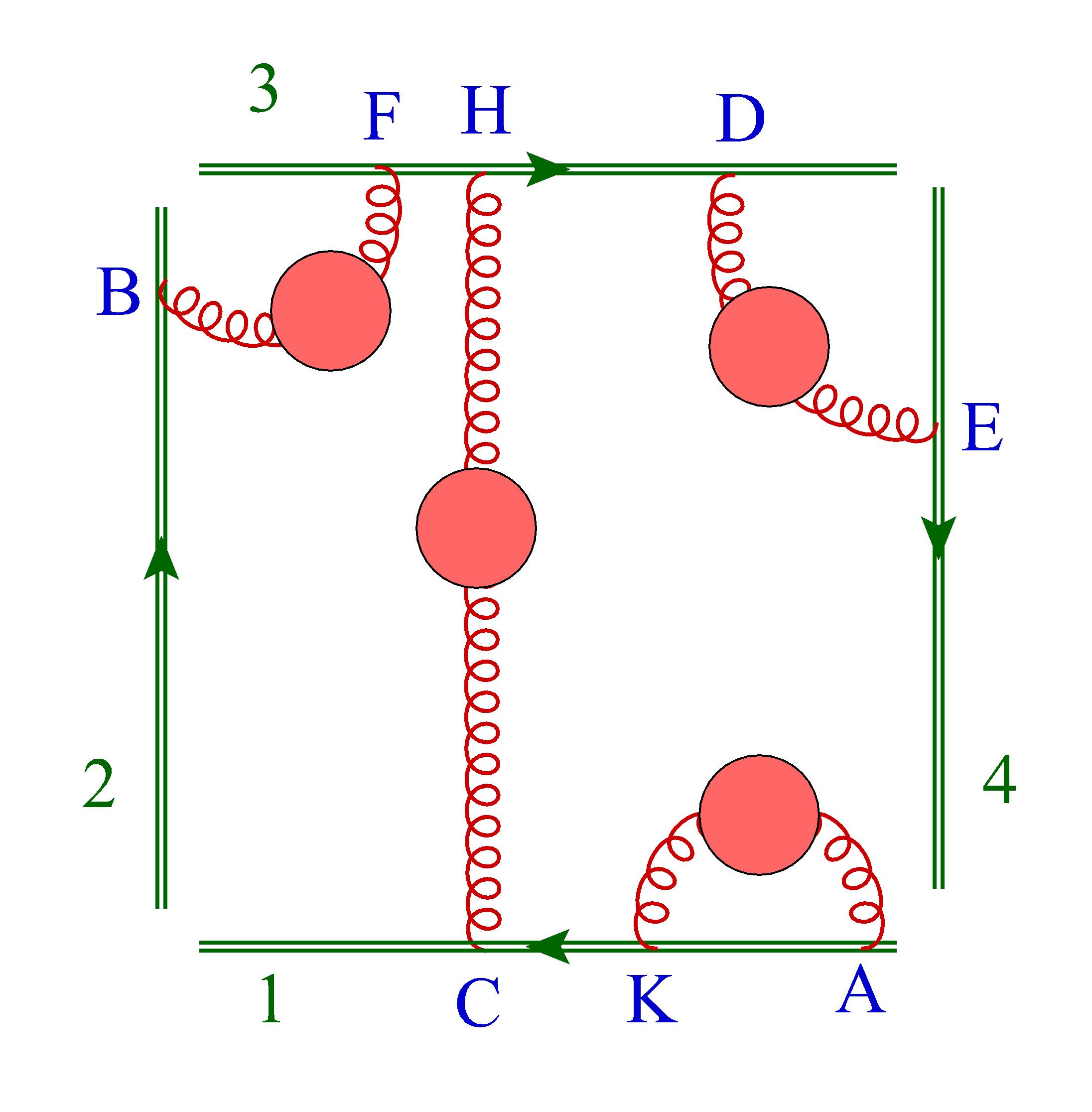} 
		\end{figure}
	\end{minipage}\hspace{-2cm} %
	\begin{minipage}[c]{0.46\textwidth}%
		\vspace{2cm}
		\begin{tabular}{|c|c|c|}
			\hline 
			\textbf{Diagrams}  & \textbf{Sequences}  & \textbf{s-factors}  \\ 
			\hline

			$C_{1}$  & $\lbrace\lbrace ACK\rbrace, \lbrace FHD\rbrace\rbrace$  & 0 \\ \hline
			$C_{2}$  & $\lbrace\lbrace ACK\rbrace, \lbrace FDH \rbrace\rbrace$  & 0 \\ \hline 
			$C_{3}$  & $\lbrace\lbrace ACK\rbrace, \lbrace HFD\rbrace\rbrace$  & 0 \\ \hline
			$C_{4}$  & $\lbrace\lbrace ACK\rbrace, \lbrace HDF\rbrace\rbrace$  & 0 \\ \hline
			$C_{5}$  & $\lbrace\lbrace ACK\rbrace, \lbrace DFH\rbrace\rbrace$  & 0 \\ \hline
			$C_{6}$  & $\lbrace\lbrace ACK\rbrace, \lbrace DHF\rbrace\rbrace$  & 0 \\ \hline 
			$C_{7}$  & $\lbrace\lbrace AKC\rbrace, \lbrace HFD\rbrace\rbrace$  &  1\\ \hline
			$C_{8}$  & $\lbrace\lbrace AKC\rbrace, \lbrace HDF\rbrace\rbrace$  &  1 \\ \hline
			$C_{9}$  & $\lbrace\lbrace CAK\rbrace, \lbrace FDH\rbrace\rbrace$  &  1\\ \hline
			$C_{10}$  & $\lbrace\lbrace CAK\rbrace, \lbrace DFH\rbrace\rbrace$  & 1 \\ \hline
			$C_{11}$  & $\lbrace\lbrace AKC\rbrace, \lbrace FHD\rbrace\rbrace$  & 2 \\ \hline
			$C_{12}$  & $\lbrace\lbrace AKC\rbrace, \lbrace FDH\rbrace\rbrace$  & 2 \\ \hline
			$C_{13}$  & $\lbrace\lbrace AKC\rbrace, \lbrace DFH\rbrace\rbrace$  & 2 \\ \hline 
			$C_{14}$  & $\lbrace\lbrace AKC\rbrace, \lbrace DHF\rbrace\rbrace$  & 2 \\ \hline 
			$C_{15}$  & $\lbrace\lbrace CAK\rbrace, \lbrace FHD\rbrace\rbrace$  & 2 \\ \hline 
			$C_{16}$  & $\lbrace\lbrace CAK\rbrace, \lbrace HFD\rbrace\rbrace$  & 2 \\ \hline 
			$C_{17}$  & $\lbrace\lbrace CAK\rbrace, \lbrace HDF\rbrace\rbrace$  & 2 \\ \hline 
			$C_{18}$  & $\lbrace\lbrace CAK\rbrace, \lbrace DHF\rbrace\rbrace$  & 2 \\ \hline  			
		\end{tabular}\label{tab:abcd1} 
	\end{minipage} \\ \\

	The mixing matrix, and the diagonal matrix are given by,
	\begin{align}
		\begin{split}
			\hspace{-0.5cm}
			R=\frac{1}{12} \scalemath{0.9}{\left(
				\begin{array}{cccccccccccccccccc}
					4 & -2 & -2 & -2 & 4 & -2 & 1 & -3 & -1 & 1 & 2 & -3 & 1 & 2 & 0 & 1 & -1 & 0 \\
					-2 & 4 & -2 & 4 & -2 & -2 & -3 & 1 & 1 & -1 & 2 & 1 & -3 & 2 & 0 & -1 & 1 & 0 \\
					-2 & -2 & 4 & -2 & -2 & 4 & 1 & 1 & 1 & 1 & -2 & 1 & 1 & -2 & -2 & 1 & 1 & -2 \\
					-2 & 4 & -2 & 4 & -2 & -2 & -1 & 1 & 1 & -3 & 0 & 1 & -1 & 0 & 2 & -3 & 1 & 2 \\
					4 & -2 & -2 & -2 & 4 & -2 & 1 & -1 & -3 & 1 & 0 & -1 & 1 & 0 & 2 & 1 & -3 & 2 \\
					-2 & -2 & 4 & -2 & -2 & 4 & 1 & 1 & 1 & 1 & -2 & 1 & 1 & -2 & -2 & 1 & 1 & -2 \\
					0 & 0 & 0 & 0 & 0 & 0 & 3 & -1 & 1 & -3 & -2 & -1 & 3 & -2 & 2 & -3 & 1 & 2 \\
					0 & 0 & 0 & 0 & 0 & 0 & -1 & 3 & -3 & 1 & -2 & 3 & -1 & -2 & 2 & 1 & -3 & 2 \\
					0 & 0 & 0 & 0 & 0 & 0 & 1 & -3 & 3 & -1 & 2 & -3 & 1 & 2 & -2 & -1 & 3 & -2 \\
					0 & 0 & 0 & 0 & 0 & 0 & -3 & 1 & -1 & 3 & 2 & 1 & -3 & 2 & -2 & 3 & -1 & -2 \\
					0 & 0 & 0 & 0 & 0 & 0 & -1 & -1 & 1 & 1 & 2 & -1 & -1 & 2 & -2 & 1 & 1 & -2 \\
					0 & 0 & 0 & 0 & 0 & 0 & -1 & 1 & -1 & 1 & 0 & 1 & -1 & 0 & 0 & 1 & -1 & 0 \\
					0 & 0 & 0 & 0 & 0 & 0 & 1 & -1 & 1 & -1 & 0 & -1 & 1 & 0 & 0 & -1 & 1 & 0 \\
					0 & 0 & 0 & 0 & 0 & 0 & -1 & -1 & 1 & 1 & 2 & -1 & -1 & 2 & -2 & 1 & 1 & -2 \\
					0 & 0 & 0 & 0 & 0 & 0 & 1 & 1 & -1 & -1 & -2 & 1 & 1 & -2 & 2 & -1 & -1 & 2 \\
					0 & 0 & 0 & 0 & 0 & 0 & -1 & 1 & -1 & 1 & 0 & 1 & -1 & 0 & 0 & 1 & -1 & 0 \\
					0 & 0 & 0 & 0 & 0 & 0 & 1 & -1 & 1 & -1 & 0 & -1 & 1 & 0 & 0 & -1 & 1 & 0 \\
					0 & 0 & 0 & 0 & 0 & 0 & 1 & 1 & -1 & -1 & -2 & 1 & 1 & -2 & 2 & -1 & -1 & 2 \\
				\end{array}
				\right)}\,,\nonumber 
		\end{split} \nonumber \\ \nonumber \\ \nonumber \\
		\mathcal{D}\,=\,&\D{4}\,.
	\end{align}
	
	The exponentiated colour factors are given by, 
	\begin{eqnarray}
		(YC)_1 &=& 0\,, \nonumber \\&& \nonumber \\
		(YC)_2 &=& 0\,, \nonumber  \\&&  \nonumber \\
		(YC)_3 &=& i f^{acn} f^{bkg} f^{cdk}  \ta 1 \tn 1 \tb 2 \tg 3 \td 4 \,,\nonumber \\&&  \nonumber \\
		(YC)_4 &=& i f^{acn} f^{bdr} f^{crm}  \ta 1 \tn 1 \tb 2 \tm 3 \td 4\,.  
	\end{eqnarray}
	\item[\textbf{2}.] $\textbf{W}\,_{4}^{(4)}(4,1,2,1)$

	This is a Boomerang Cweb which is made up with four two-gluon correlators. It has twenty-four diagrams, and one of them is shown below. The table below shows chosen order of shuffles and their corresponding $s$-factors. 
	
	\begin{minipage}[c]{0.5\textwidth}%
		\begin{figure}[H]
			\vspace{-2mm}
			\includegraphics[width=4cm,height=4cm]{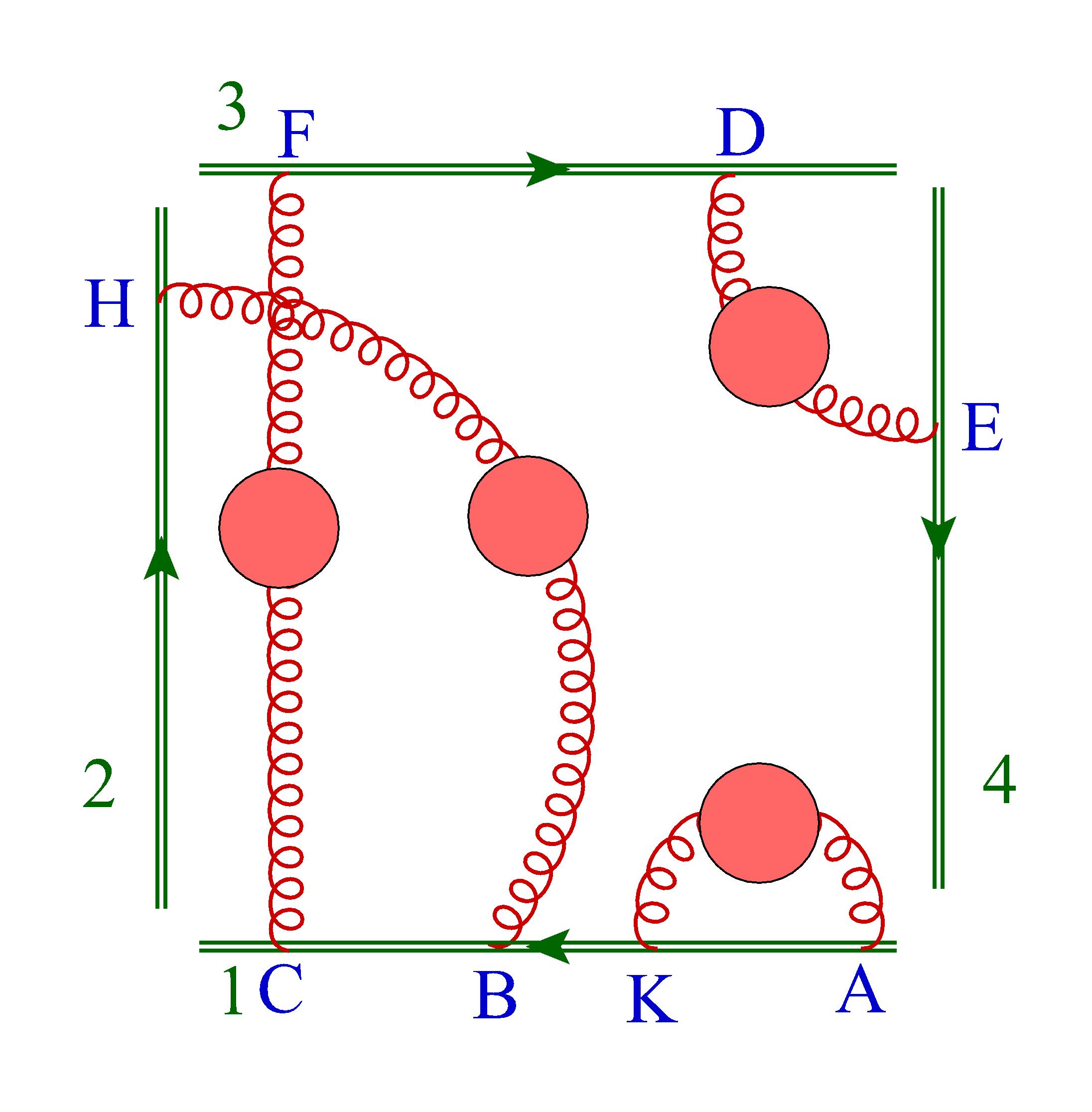} 
		\end{figure}
	\end{minipage}\hspace{-2cm} %
	\begin{minipage}[c]{0.46\textwidth}%
		\vspace{2cm}
		\begin{tabular}{|c|c|c|}
			\hline 
			\textbf{Diagrams}  & \textbf{Sequences}  & \textbf{s-factors}  \\ 
			\hline
			$C_{1}$&$\{\{ABKC\},\{ FD\}\}$&0\\\hline
			
			$C_{2}$&$\{\{ABKC\},\{ DF\}\}$&0\\\hline
			
			$C_{3}$&$\{\{ABCK\},\{ FD\}\}$&0\\\hline
			
			$C_{4}$&$\{\{ABCK\},\{ DF\}\}$&0\\\hline
			
			$C_{5}$&$\{\{ACKB\},\{ FD\}\}$&0\\\hline
			
			$C_{6}$&$\{\{ACKB\},\{ DF\}\}$&0\\\hline
			
			$C_{7}$&$\{\{ACBK\},\{ FD\}\}$&0\\\hline
			
			$C_{8}$&$\{\{ACBK\},\{ DF\}\}$&0\\\hline
			
			$C_{9}$&$\{\{BACK\},\{ FD\}\}$&0\\\hline
			
			$C_{10}$&$\{\{BACK\},\{ DF\}\}$&0\\\hline
			
			$C_{11}$&$\{\{CABK\},\{ FD\}\}$&0\\\hline
			
			$C_{12}$&$\{\{CABK\},\{ DF\}\}$&0\\\hline
			
			$C_{13}$&$\{\{AKBC\},\{ FD\}\}$&1\\\hline
			
			$C_{14}$&$\{\{BAKC\},\{ FD\}\}$&1\\\hline
			
			$C_{15}$&$\{\{CAKB\},\{ DF\}\}$&1\\\hline
			
			$C_{16}$&$\{\{CBAK\},\{ DF\}\}$&1\\\hline
			
			$C_{17}$&$\{\{AKBC\},\{ DF\}\}$&2\\\hline
			
			$C_{18}$&$\{\{AKCB\},\{ FD\}\}$&2\\\hline
			
			$C_{19}$&$\{\{AKCB\},\{ DF\}\}$&2\\\hline
			
			$C_{20}$&$\{\{BAKC\},\{ DF\}\}$&2\\\hline
			
			$C_{21}$&$\{\{BCAK\},\{ FD\}\}$&2\\\hline
			
			$C_{22}$&$\{\{BCAK\},\{ DF\}\}$&2\\\hline
			
			$C_{23}$&$\{\{CAKB\},\{ FD\}\}$&2\\\hline
			
			$C_{24}$&$\{\{CBAK\},\{ FD\}\}$&2\\\hline		
		\end{tabular}\label{tab:4legWeb2} %
	\end{minipage} \\
	
	The mixing matrix and the diagonal matrix for this Cweb are given by, 
	\begin{align}
		\begin{split}
			\hspace{-1cm}R= \frac{1}{12} \scalemath{0.8}{\left(
				\begin{array}{cccccccccccccccccccccccc}
					4 & -4 & 0 & 0 & 0 & 0 & 0 & 0 & 0 & 0 & -4 & 4 & -1 & -1 & -3 & -3 & 1 & -2 & 2 & 1 & -2 & 2 & 3 & 3 \\
					-2 & 2 & 0 & 0 & 0 & 0 & 0 & 0 & 0 & 0 & 2 & -2 & 1 & 1 & 1 & 1 & -1 & 0 & 0 & -1 & 0 & 0 & -1 & -1 \\
					-2 & 2 & 6 & -6 & -4 & 4 & 0 & 0 & -2 & 2 & -4 & 4 & -1 & 1 & -3 & -3 & 1 & 2 & -2 & -1 & -2 & 2 & 3 & 3 \\
					4 & -4 & -6 & 6 & 2 & -2 & 0 & 0 & 4 & -4 & 2 & -2 & 1 & -3 & 1 & 1 & -1 & -2 & 2 & 3 & 0 & 0 & -1 & -1 \\
					0 & 0 & 0 & 0 & 2 & -2 & 0 & 0 & -2 & 2 & 0 & 0 & -1 & 1 & 1 & 1 & 1 & 0 & 0 & -1 & 2 & -2 & -1 & -1 \\
					0 & 0 & 0 & 0 & -4 & 4 & 0 & 0 & 4 & -4 & 0 & 0 & 1 & -3 & -1 & -1 & -1 & 2 & -2 & 3 & -2 & 2 & 1 & 1 \\
					-2 & 2 & 0 & 0 & -4 & 4 & 6 & -6 & -2 & 2 & -4 & 4 & 1 & 1 & -3 & 1 & -1 & 0 & 0 & -1 & 2 & -2 & 3 & -1 \\
					4 & -4 & 0 & 0 & 2 & -2 & -6 & 6 & 4 & -4 & 2 & -2 & -3 & -3 & 1 & -1 & 3 & 2 & -2 & 3 & -2 & 2 & -1 & 1 \\
					0 & 0 & 0 & 0 & -4 & 4 & 0 & 0 & 4 & -4 & 0 & 0 & -1 & -1 & -3 & 1 & 1 & 2 & -2 & 1 & -2 & 2 & 3 & -1 \\
					0 & 0 & 0 & 0 & 2 & -2 & 0 & 0 & -2 & 2 & 0 & 0 & 1 & 1 & 1 & -1 & -1 & -2 & 2 & -1 & 0 & 0 & -1 & 1 \\
					-2 & 2 & 0 & 0 & 0 & 0 & 0 & 0 & 0 & 0 & 2 & -2 & 1 & 1 & 1 & 1 & -1 & 0 & 0 & -1 & 0 & 0 & -1 & -1 \\
					4 & -4 & 0 & 0 & 0 & 0 & 0 & 0 & 0 & 0 & -4 & 4 & -3 & -3 & -1 & -1 & 3 & 2 & -2 & 3 & 2 & -2 & 1 & 1 \\
					0 & 0 & 0 & 0 & 0 & 0 & 0 & 0 & 0 & 0 & 0 & 0 & 3 & -1 & 1 & -3 & -3 & -2 & 2 & 1 & -2 & 2 & -1 & 3 \\
					0 & 0 & 0 & 0 & 0 & 0 & 0 & 0 & 0 & 0 & 0 & 0 & -1 & 3 & -3 & 1 & 1 & -2 & 2 & -3 & -2 & 2 & 3 & -1 \\
					0 & 0 & 0 & 0 & 0 & 0 & 0 & 0 & 0 & 0 & 0 & 0 & 1 & -3 & 3 & -1 & -1 & 2 & -2 & 3 & 2 & -2 & -3 & 1 \\
					0 & 0 & 0 & 0 & 0 & 0 & 0 & 0 & 0 & 0 & 0 & 0 & -3 & 1 & -1 & 3 & 3 & 2 & -2 & -1 & 2 & -2 & 1 & -3 \\
					0 & 0 & 0 & 0 & 0 & 0 & 0 & 0 & 0 & 0 & 0 & 0 & -1 & 1 & -1 & 1 & 1 & 0 & 0 & -1 & 0 & 0 & 1 & -1 \\
					0 & 0 & 0 & 0 & 0 & 0 & 0 & 0 & 0 & 0 & 0 & 0 & -1 & -1 & 1 & 1 & 1 & 2 & -2 & 1 & 2 & -2 & -1 & -1 \\
					0 & 0 & 0 & 0 & 0 & 0 & 0 & 0 & 0 & 0 & 0 & 0 & 1 & 1 & -1 & -1 & -1 & -2 & 2 & -1 & -2 & 2 & 1 & 1 \\
					0 & 0 & 0 & 0 & 0 & 0 & 0 & 0 & 0 & 0 & 0 & 0 & 1 & -1 & 1 & -1 & -1 & 0 & 0 & 1 & 0 & 0 & -1 & 1 \\
					0 & 0 & 0 & 0 & 0 & 0 & 0 & 0 & 0 & 0 & 0 & 0 & -1 & -1 & 1 & 1 & 1 & 2 & -2 & 1 & 2 & -2 & -1 & -1 \\
					0 & 0 & 0 & 0 & 0 & 0 & 0 & 0 & 0 & 0 & 0 & 0 & 1 & 1 & -1 & -1 & -1 & -2 & 2 & -1 & -2 & 2 & 1 & 1 \\
					0 & 0 & 0 & 0 & 0 & 0 & 0 & 0 & 0 & 0 & 0 & 0 & -1 & 1 & -1 & 1 & 1 & 0 & 0 & -1 & 0 & 0 & 1 & -1 \\
					0 & 0 & 0 & 0 & 0 & 0 & 0 & 0 & 0 & 0 & 0 & 0 & 1 & -1 & 1 & -1 & -1 & 0 & 0 & 1 & 0 & 0 & -1 & 1 \\
				\end{array}
				\right)}
		\end{split} 
	\end{align} 
	\begin{align}
		\mathcal{D}\,=\,\D{6}
	\end{align}
	Thus, six independent exponentiated colour factors are given by,
	\begin{eqnarray}
		(YC)_1 &=&  0\,,
		\nonumber \\&& \nonumber \\
		(YC)_2 &=& i f^{abg} f^{cdk} f^{cng}  \ta 1 \tn 1 \tb 2 \tk 3 \td 4 \,,
		\nonumber  
		\\&& +\;  i f^{abg} f^{acn} f^{cdk}  \tn 1 \tg 1 \tb 2 \tk 3 \td 4 \,,
		; \nonumber \\
		(YC)_3 &=&  0\,,\nonumber \\&& \nonumber \\
		(YC)_4 &=& i\,f^{abn} f^{ack} f^{cdm}\, \tn 1 \tk 1 \tb 2 \tm 3 \td 4 \,,\nonumber \\&& \nonumber \\
		(YC)_5 &=&  i\,f^{ang} f^{bcn} f^{cdk}\, \ta 1 \tg 1 \tb 2 \tk 3 \td 4 \,,\nonumber \\&& \nonumber \\
		(YC)_6 &=& i\,f^{ack} f^{bkg} f^{cdr} \, \ta 1 \tg 1 \tb 2 \tr 3 \td 4 \,.
	\end{eqnarray}
	\item[\textbf{3}.] $\textbf{W}\,_{4}^{(4)}\,(3,1,2,2)$
	
	This Cweb has twelve diagrams, one of them is shown below. The table shows the chosen order of shuffle and the corresponding $s$-factors. \\ 
	\begin{minipage}[c]{0.5\textwidth}%
		\begin{figure}[H]
			\vspace{-2mm}
			\includegraphics[width=4cm,height=4cm]{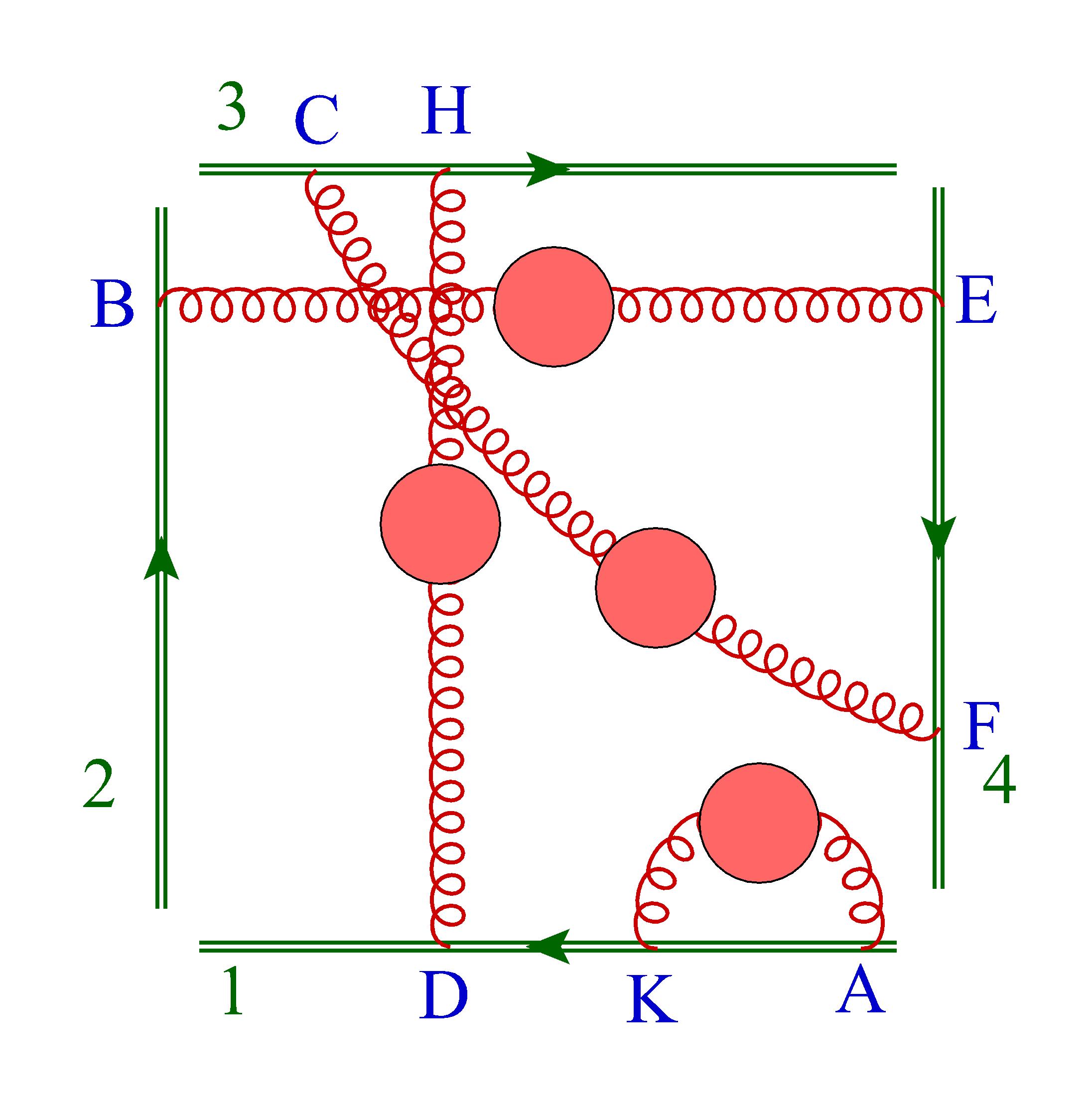} 
		\end{figure}
	\end{minipage}\hspace{-2cm} %
	\begin{minipage}[c]{0.46\textwidth}%
		\vspace{2cm}
		\begin{tabular}{|c|c|c|}
			\hline 
			\textbf{Diagrams}  & \textbf{Sequences}  & \textbf{s-factors}  \\ 
			\hline
			$C_{1}$&$\{\{ADK\},\{ CH\},\{EF\}\}$&0\\\hline
			
			$C_{2}$&$\{\{ADK\},\{ CH\},\{FE\}\}$&0\\\hline
			
			$C_{3}$&$\{\{ADK\},\{ HC\},\{EF\}\}$&0\\\hline
			
			$C_{4}$&$\{\{ADK\},\{ HC\},\{FE\}\}$&0\\\hline
			
			$C_{5}$&$\{\{AKD\},\{ HC\},\{FE\}\}$&1\\\hline
			
			$C_{6}$&$\{\{DAK\},\{ CH\},\{EF\}\}$&1\\\hline
			
			$C_{7}$&$\{\{AKD\},\{ CH\},\{EF\}\}$&2\\\hline
			
			$C_{8}$&$\{\{AKD\},\{ HC\},\{EF\}\}$&2\\\hline
			
			$C_{9}$&$\{\{DAK\},\{ CH\},\{FE\}\}$&2\\\hline
			
			$C_{10}$&$\{\{DAK\},\{ HC\},\{FE\}\}$&2\\\hline
			
			$C_{11}$&$\{\{AKD\},\{ CH\},\{FE\}\}$&4\\\hline
			
			$C_{12}$&$\{\{DAK\},\{ HC\},\{EF\}\}$&4\\\hline
		\end{tabular}\label{tab:4legWeb3} 
	\end{minipage}\\ \\ \\
	The mixing matrix, and the diagonal matrix are given by, 
	\begin{align}
		\begin{split}
			R=\,\frac{1}{12} \left(
			\begin{array}{cccccccccccc}
				4 & -4 & -4 & 4 & -3 & -1 & -3 & 3 & 1 & -1 & 3 & 1 \\
				-2 & 2 & 2 & -2 & 1 & 1 & 1 & -1 & -1 & 1 & -1 & -1 \\
				-2 & 2 & 2 & -2 & 1 & 1 & 1 & -1 & -1 & 1 & -1 & -1 \\
				4 & -4 & -4 & 4 & -1 & -3 & -1 & 1 & 3 & -3 & 1 & 3 \\
				0 & 0 & 0 & 0 & 3 & -3 & 3 & -3 & 3 & -3 & -3 & 3 \\
				0 & 0 & 0 & 0 & -3 & 3 & -3 & 3 & -3 & 3 & 3 & -3 \\
				0 & 0 & 0 & 0 & 1 & -1 & 1 & -1 & 1 & -1 & -1 & 1 \\
				0 & 0 & 0 & 0 & -1 & 1 & -1 & 1 & -1 & 1 & 1 & -1 \\
				0 & 0 & 0 & 0 & 1 & -1 & 1 & -1 & 1 & -1 & -1 & 1 \\
				0 & 0 & 0 & 0 & -1 & 1 & -1 & 1 & -1 & 1 & 1 & -1 \\
				0 & 0 & 0 & 0 & -1 & 1 & -1 & 1 & -1 & 1 & 1 & -1 \\
				0 & 0 & 0 & 0 & 1 & -1 & 1 & -1 & 1 & -1 & -1 & 1 \\
			\end{array}
			\right),\, \nonumber
		\end{split}
		\mathcal{D}\,=\,\D{2}\,.
	\end{align}

	The rank of the mixing matrix is 2, which corresponds to the following exponentiated colour factors, 
	\begin{eqnarray}
		(YC)_1 &=& 0 \,,\nonumber \\&& \nonumber \\
		(YC)_2 &=&  i\, f^{adg} f^{bcn} f^{cdk} \,\ta 1 \tg 1 \tb 2 \tk 3 \tn 4 \,.
	\end{eqnarray}
	\item[\textbf{4}.] $\textbf{W}\,_{4}^{(2,1)}\,(3,1,2,1)$
	
	This is first of two Cwebs which has the same correlator and attachment content. It has six diagrams, one of which is shown below. The table below shows the chosen order of shuffles, and their corresponding $s$-factors. \\
	\begin{minipage}[c]{0.5\textwidth}%
		\begin{figure}[H]
			\vspace{-2mm}
			\includegraphics[width=4cm,height=4cm]{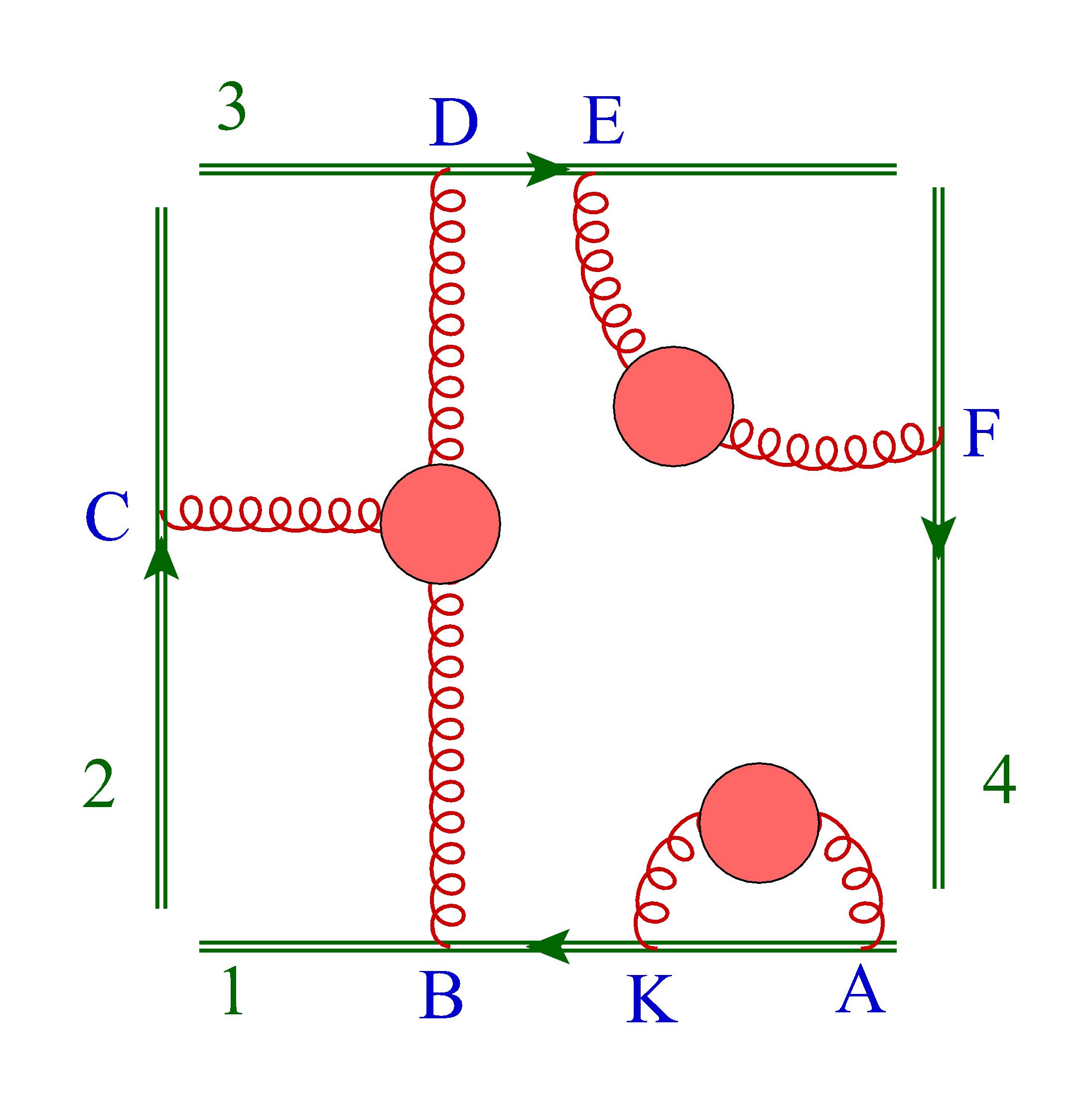} 
		\end{figure}
	\end{minipage}\hspace{-2cm} %
	\begin{minipage}[c]{0.46\textwidth}%
		\vspace{1cm}
		\begin{tabular}{|c|c|c|}
			\hline 
			\textbf{Diagrams}  & \textbf{Sequences}  & \textbf{s-factors}  \\ 
			\hline
			$C_{1}$  & $\lbrace\lbrace ABK\rbrace, \lbrace DE\rbrace\rbrace$  & 0 \\ \hline
			$C_{2}$  & $\lbrace\lbrace ABK\rbrace, \lbrace ED\rbrace\rbrace$  & 0 \\ \hline
			$C_{3}$  & $\lbrace\lbrace AKB\rbrace, \lbrace DE\rbrace\rbrace$  & 1 \\ \hline 
			$C_{4}$  & $\lbrace\lbrace BAK\rbrace, \lbrace DE\rbrace\rbrace$  & 1 \\ \hline
			$C_{5}$  & $\lbrace\lbrace BAK\rbrace, \lbrace ED\rbrace\rbrace$  & 2 \\ \hline
			$C_{6}$  & $\lbrace\lbrace AKB\rbrace, \lbrace ED\rbrace\rbrace$  & 2 \\ \hline
		\end{tabular}\label{tab:4legWeb4} %
	\end{minipage} \\ \\
	The mixing matrix, and the diagonal matrix are given by, 
	
	\begin{align}
		\begin{split}
			R=&\,\frac{1}{6} \left(
			\begin{array}{cccccc}
				3 & -3 & -1 & 2 & 1 & -2 \\
				-3 & 3 & 2 & -1 & -2 & 1 \\
				0 & 0 & 2 & 2 & -2 & -2 \\
				0 & 0 & 2 & 2 & -2 & -2 \\
				0 & 0 & -1 & -1 & 1 & 1 \\
				0 & 0 & -1 & -1 & 1 & 1 \\
			\end{array}
			\right)\,,\mathcal{D}\,=\,\D{2}\,.
		\end{split} 
	\end{align}
	Finally, the exponentiated colour factors are given by,  
	\begin{eqnarray}
		(YC)_1 &=& 0 \,,\nonumber \\&& \nonumber \\
		(YC)_2 &=& i\,f^{abn} f^{bcd} f^{dme}\, \ta 1 \tn 1 \tc 2 \tm 3 \te 4   \,.
	\end{eqnarray}

	\item[\textbf{5}.] $\textbf{W}\,_{4}^{(2,1)}(3,2,1,1)$
	
	This is the second Cweb with same correlator and attachment content. It has six diagrams, one of them is shown below. The table shows the chosen order of shuffle and their corresponding $s$-factors.  \\
	\begin{minipage}[c]{0.5\textwidth}%
		\begin{figure}[H]
			\vspace{-2mm}
			\includegraphics[width=4cm,height=4cm]{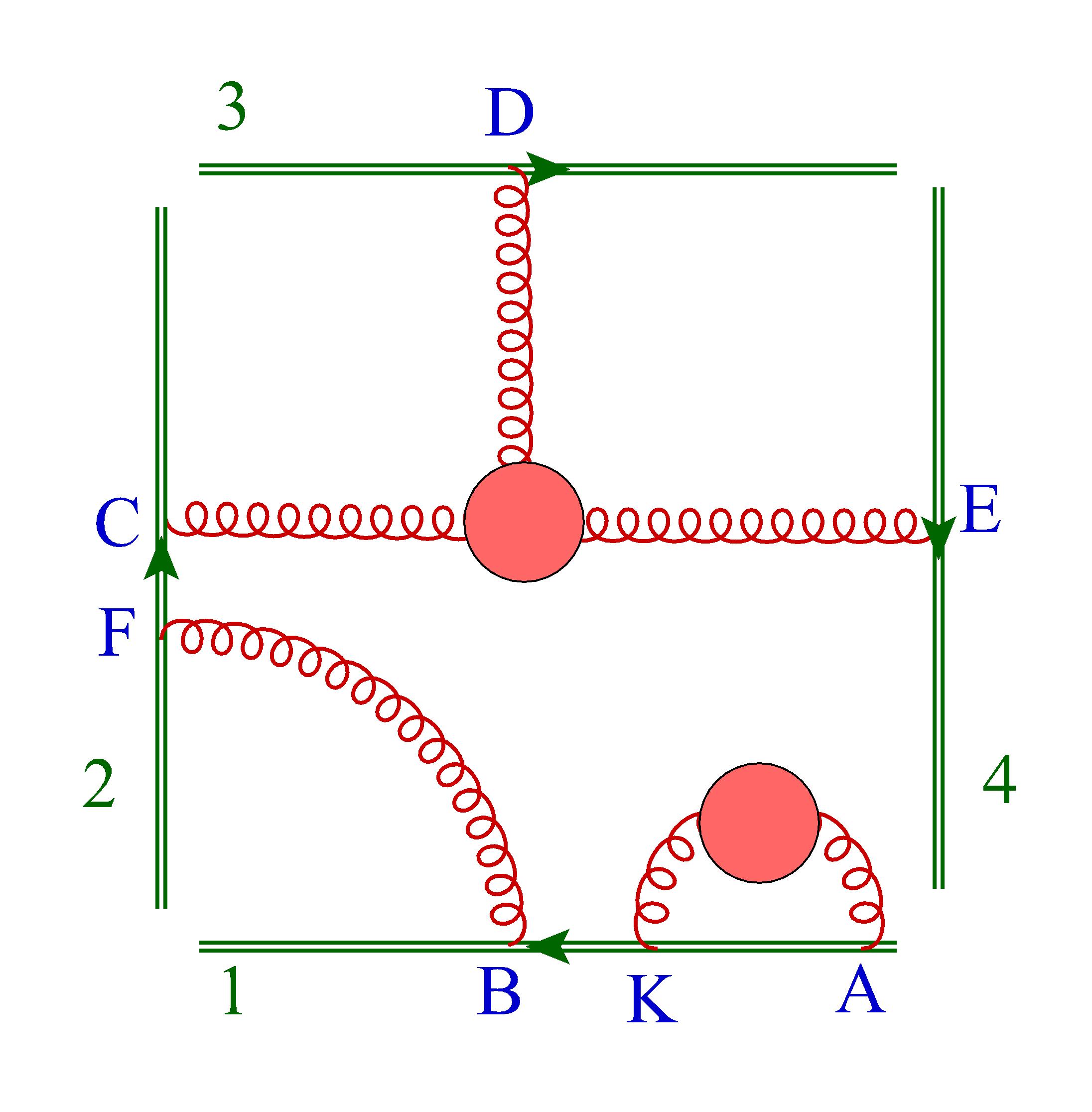} 
		\end{figure}
	\end{minipage}\hspace{-2cm} %
	\begin{minipage}[c]{0.46\textwidth}%
		\vspace{0.7cm}
		\begin{tabular}{|c|c|c|}
			\hline 
			\textbf{Diagrams}  & \textbf{Sequences}  & \textbf{s-factors}  \\ 
			\hline
			$C_{1}$  & $\lbrace\lbrace ABK\rbrace \lbrace CF \rbrace \rbrace$  & 0 \\ \hline
			$C_{2}$  & $\lbrace\lbrace ABK\rbrace \lbrace FC \rbrace \rbrace$  & 0 \\ \hline
			$C_{3}$  & $\lbrace\lbrace AKB\rbrace \lbrace FC \rbrace \rbrace$  & 1 \\ \hline
			$C_{4}$  & $\lbrace\lbrace BAK\rbrace \lbrace FC \rbrace \rbrace$  & 1 \\ \hline
			$C_{5}$  & $\lbrace\lbrace BAK\rbrace \lbrace CF \rbrace \rbrace$  & 2 \\ \hline
			$C_{6}$  & $\lbrace\lbrace AKB\rbrace \lbrace CF \rbrace \rbrace$  & 2 \\ \hline
		\end{tabular}\label{tab:4legWeb8} %
	\end{minipage} \\ \\ \\
	The mixing matrix and the diagonal matrix for this Cweb are given by, 
	\begin{align}
		\begin{split}
			R=\frac{1}{6} \left(
			\begin{array}{cccccc}
				3 & -3 & 2 & -1 & -2 & 1 \\
				-3 & 3 & -1 & 2 & 1 & -2 \\
				0 & 0 & 2 & 2 & -2 & -2 \\
				0 & 0 & 2 & 2 & -2 & -2 \\
				0 & 0 & -1 & -1 & 1 & 1 \\
				0 & 0 & -1 & -1 & 1 & 1 \\
			\end{array}
			\right),\,
			\mathcal{D}\,=\,\D{2}\,.
		\end{split}
	\end{align}
	Finally, the exponentiated colour factors are given by,  
	\begin{eqnarray}
		(YC)_1 &=& 0 \,,\nonumber \\&& \nonumber \\
		(YC)_2 &=& i\,f^{abk} f^{abg} f^{cde}\, \ta 1 \tg 1 \tk 2 \td 3 \te 4   \,.
	\end{eqnarray}

	\item[\textbf{6}.] $\textbf{W}\,_{4}^{(2,1)}\,(4,1,1,1)$

	This Cweb is made up with one three-gluon correlator, and two two-gluon correlators. This has 12 diagrams, one of them is shown below. The table shows the chosen order shuffle on Wilson line $ 1 $, and their corresponding $s$-factors. \\
	\begin{minipage}[c]{0.5\textwidth}%
		\begin{figure}[H]
			\vspace{-2mm}
			\includegraphics[width=4cm,height=4cm]{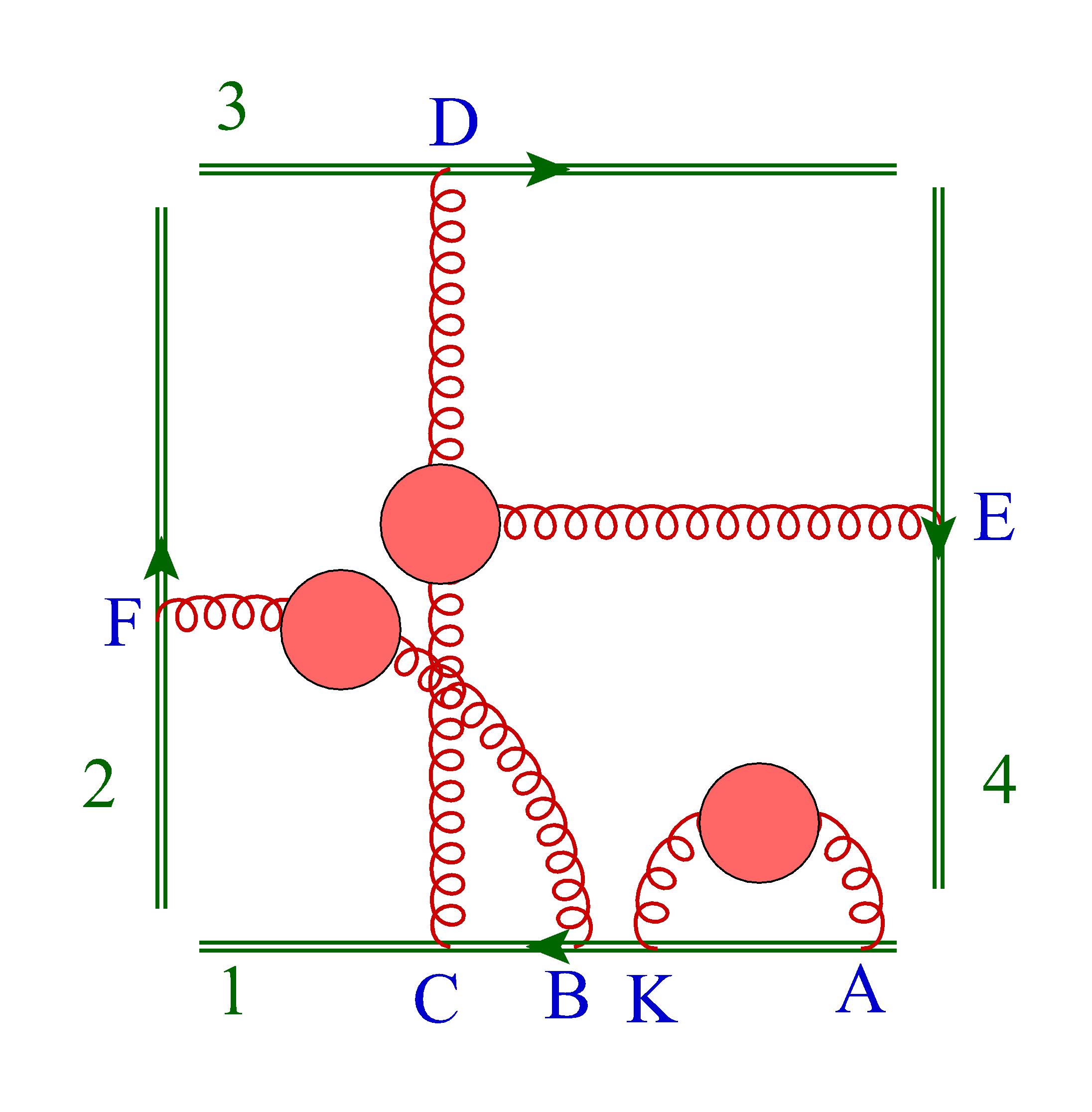} 
		\end{figure}
	\end{minipage}\hspace{-2cm} %
	\begin{minipage}[c]{0.46\textwidth}%
		\vspace{1.5cm}
		\begin{tabular}{|c|c|c|}
			\hline 
			\textbf{Diagrams}  & \textbf{Sequences}  & \textbf{s-factors}  \\ 
			\hline
			$C_{1}$  & $\lbrace\lbrace ABCK\rbrace\rbrace$  & 0 \\ \hline 
			$C_{2}$  & $\lbrace\lbrace ACBK\rbrace\rbrace$  & 0 \\ \hline
			$C_{3}$  & $\lbrace\lbrace ABKC\rbrace\rbrace$  & 0 \\ \hline 
			$C_{4}$  & $\lbrace\lbrace ACKB\rbrace\rbrace$  & 0 \\ \hline 
			$C_{5}$  & $\lbrace\lbrace BACK\rbrace\rbrace$  & 0 \\ \hline 
			$C_{6}$  & $\lbrace\lbrace CABK\rbrace\rbrace$  & 0 \\ \hline
			$C_{7}$  & $\lbrace\lbrace AKBC\rbrace\rbrace$  & 1\\ \hline
			$C_{8}$  & $\lbrace\lbrace AKCB\rbrace\rbrace$  & 1 \\ \hline 
			$C_{9}$  & $\lbrace\lbrace BAKC\rbrace\rbrace$  & 1 \\ \hline
			$C_{10}$  & $\lbrace\lbrace BCAK\rbrace\rbrace$  & 1 \\ \hline
			$C_{11}$  & $\lbrace\lbrace CAKB\rbrace\rbrace$  & 1 \\ \hline 
			$C_{12}$  & $\lbrace\lbrace CBAK\rbrace\rbrace$  & 1 \\ \hline     
		\end{tabular}\label{tab:4legWeb5} %
	\end{minipage} \\ \\
	The $ R $, and $ D $ matrices are given by,
	\begin{align}
		\begin{split}
			R=\frac{1}{6} \left(
			\begin{array}{cccccccccccc}
				6 & 0 & -3 & -3 & -3 & -3 & -1 & 2 & 2 & -1 & 2 & 2 \\
				0 & 6 & -3 & -3 & -3 & -3 & 2 & -1 & 2 & 2 & 2 & -1 \\
				0 & 0 & 3 & 0 & 0 & -3 & -1 & -1 & -1 & -1 & 2 & 2 \\
				0 & 0 & 0 & 3 & -3 & 0 & -1 & -1 & 2 & 2 & -1 & -1 \\
				0 & 0 & 0 & -3 & 3 & 0 & -1 & 2 & -1 & -1 & 2 & -1 \\
				0 & 0 & -3 & 0 & 0 & 3 & 2 & -1 & 2 & -1 & -1 & -1 \\
				0 & 0 & 0 & 0 & 0 & 0 & 2 & -1 & -1 & -1 & -1 & 2 \\
				0 & 0 & 0 & 0 & 0 & 0 & -1 & 2 & -1 & 2 & -1 & -1 \\
				0 & 0 & 0 & 0 & 0 & 0 & -1 & -1 & 2 & -1 & 2 & -1 \\
				0 & 0 & 0 & 0 & 0 & 0 & -1 & 2 & -1 & 2 & -1 & -1 \\
				0 & 0 & 0 & 0 & 0 & 0 & -1 & -1 & 2 & -1 & 2 & -1 \\
				0 & 0 & 0 & 0 & 0 & 0 & 2 & -1 & -1 & -1 & -1 & 2 \\
			\end{array}
			\right),\,
		\end{split}
		\mathcal{D}\,=\,\D{6} \,.
	\end{align}
	
	Finally, the exponentiated colour factors are given by, 
	\begin{eqnarray}
		(YC)_1 &=& 0\,,\nonumber \\&& \nonumber \\
		(YC)_2 &=&  -i\,f^{abg} f^{cde} f^{ckg} \, \ta 1 \tk 1 \tb 2 \td 3 \te 4\,\nonumber \\&& -i \,f^{abg} f^{ack} f^{cde}\, \tk 1 \tg 1 \tb 2 \td 3 \te 4\nonumber, \\
		(YC)_3 &=&  0\,,\nonumber \\&& \nonumber \\
		(YC)_4 &=&  -i\,f^{abg} f^{ack} f^{cde}\, \tg 1 \tk 1 \tb 2 \td 3 \te 4  \,,\nonumber \\&& \nonumber \\
		(YC)_5 &=&   i\,f^{akg} f^{bcg} f^{cde}\, \ta 1 \tk 1 \tb 2 \td 3 \te 4   \,,\nonumber \\&& \nonumber \\
		(YC)_6 &=&  -i\,f^{ack} f^{bkg} f^{cde} \ta 1 \tg 1 \tb 2 \td 3 \te 4   \,. 
	\end{eqnarray}
	\item[\textbf{7}.] $\textbf{W}\,_{4}^{(1,0,1)}\,(3,1,1,1)$

	This Cweb has one four-gluon correlator and one two-gluon correlator.  It has three diagrams, one of them is shown below. The figure below shows one of the three diagrams for this Cweb, and the table gives chosen order of shuffles and their corresponding $s$-factors.  \\
	\begin{minipage}[c]{0.5\textwidth}%
		\begin{figure}[H]
			\vspace{-2mm}
			\includegraphics[width=4cm,height=4cm]{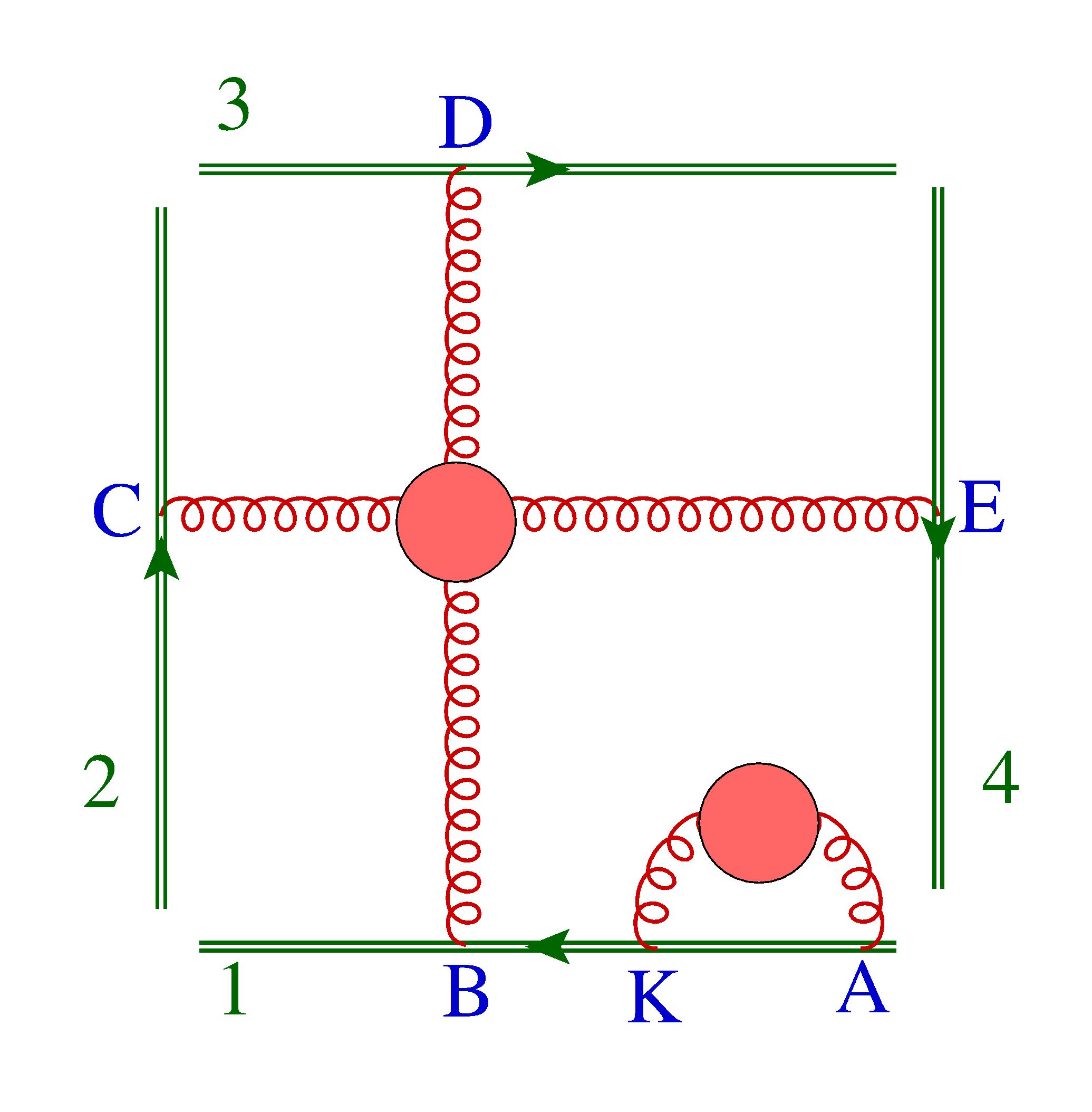} 
		\end{figure}
	\end{minipage}\hspace{-2cm} %
	\begin{minipage}[c]{0.46\textwidth}%
		\vspace{2cm}
		\begin{tabular}{|c|c|c|}
			\hline 
			\textbf{Diagrams}  & \textbf{Sequences}  & \textbf{s-factors}  \\ 
			\hline
			$C_{1}$  & $\lbrace\lbrace ABK\rbrace \rbrace$  & 0 \\ \hline
			$C_{2}$  & $\lbrace\lbrace AKB\rbrace \rbrace$  & 1\\ \hline 
			$C_{3}$  & $\lbrace\lbrace BAK\rbrace \rbrace$  & 1 \\ \hline 
		\end{tabular}\label{tab:4legWeb6} %
	\end{minipage} \\ \\
	The $ R $, and $ D $ matrices are given by, 
	\begin{align}
		\begin{split}
			R=\frac{1}{2} \left(
			\begin{array}{ccc}
				2 & -1 & -1 \\
				0 & 1 & -1 \\
				0 & -1 & 1 \\
			\end{array}
			\right),\,
			\mathcal{D}\,=\,&\D{2}\,.
		\end{split}
	\end{align}
	This mixing matrix agrees with the universal form of $ 3\times 3 $ mixing matrices computed in ~\cite{Agarwal:2021him}. 
	Finally, the exponentiated colour factors are given by, 
	\begin{eqnarray}
		(YC)_1 &=&  0\,,\nonumber \\&& \nonumber \\
		(YC)_2 &=&  -i\, f^{abn} f^{bch} f^{deh} \, \ta 1 \tn 1 \tc 2 \td 3 \te 4\,.
	\end{eqnarray}
	\item[\textbf{8}.] $\textbf{W}\,_{4}^{(4)}(5,1,1,1)$
	
	This is the largest Boomerang Cweb that can connect four Wilson lines. It has sixty diagrams, one of them is shown below.  The following table shows the chosen order of shuffle and their corresponding. We refrain to present the mixing matrix fo this Cweb here due to its larger dimension, however it can be found in the ancillary file \textit{Boomerang.nb}.    
	\begin{figure}[H]
		\begin{center}
			\vspace{-2mm}
			\includegraphics[width=3cm,height=3cm]{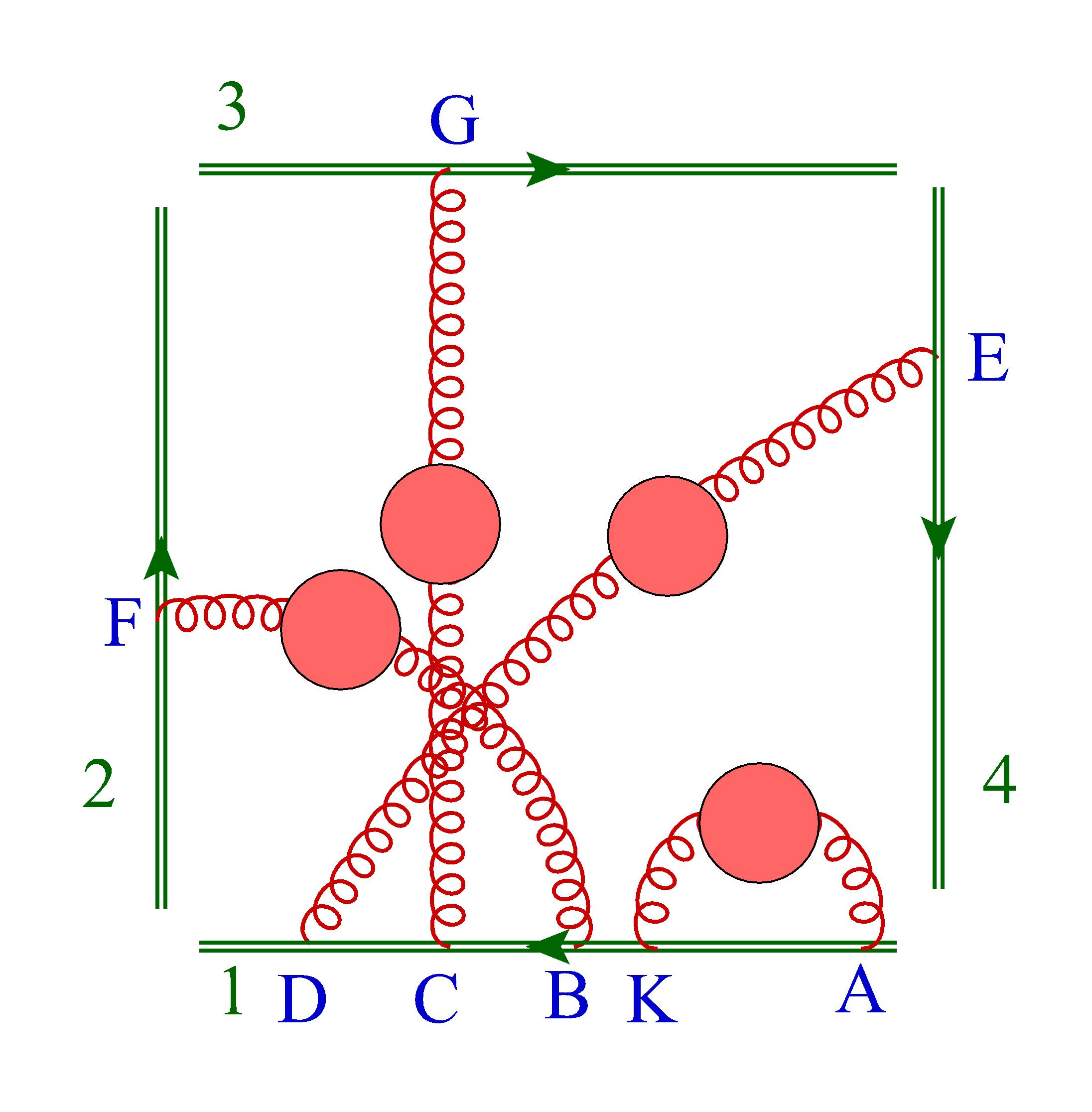}
		\end{center} 
	\end{figure}
	\begin{minipage}[c]{0.42\textwidth}%
		%
		\begin{tabular}{|c|c|c|}
			\hline 
			\textbf{Diagrams}  & \textbf{Sequences}  & \textbf{s-factors}  \\ 
			\hline
			$C_{1}$  & $\lbrace\lbrace ABCDK\rbrace \rbrace$  & 0\\ \hline 
			$C_{2}$  & $\lbrace\lbrace ABCDK\rbrace \rbrace$  & 0 \\ \hline
			$C_{3}$  & $\lbrace\lbrace ACBDK\rbrace \rbrace$  & 0 \\ \hline
			$C_{4}$  & $\lbrace\lbrace ACDBK\rbrace \rbrace$  & 0 \\ \hline
			$C_{5}$  & $\lbrace\lbrace ADBCK\rbrace \rbrace$  & 0 \\ \hline
			$C_{6}$  & $\lbrace\lbrace ADCBK\rbrace \rbrace$  & 0 \\ \hline
			$C_{7}$  & $\lbrace\lbrace ABKCD\rbrace \rbrace$  & 0 \\ \hline
			$C_{8}$  & $\lbrace\lbrace ABCDC\rbrace \rbrace$  & 0 \\ \hline
			$C_{9}$  & $\lbrace\lbrace ABCKD\rbrace \rbrace$  & 0 \\ \hline
			$C_{10}$  & $\lbrace\lbrace ABDKC\rbrace \rbrace$  & 0 \\ \hline
			$C_{11}$  & $\lbrace\lbrace ACKBD\rbrace \rbrace$  & 0 \\ \hline
			$C_{12}$  & $\lbrace\lbrace ACKDB\rbrace \rbrace$  & 0 \\ \hline
			$C_{13}$  & $\lbrace\lbrace ACBKD\rbrace \rbrace$ & 0  \\ \hline 
			$C_{14}$  & $\lbrace\lbrace ACDKB\rbrace \rbrace$  & 0 \\ \hline
			$C_{15}$  & $\lbrace\lbrace ADKBC\rbrace \rbrace$  & 0 \\ \hline
			$C_{16}$  & $\lbrace\lbrace ADKCB\rbrace \rbrace$  & 0 \\ \hline
			$C_{17}$  & $\lbrace\lbrace ADBKC\rbrace \rbrace$  & 0 \\ \hline
			$C_{18}$  & $\lbrace\lbrace ADCKB\rbrace \rbrace$  & 0 \\ \hline

			$C_{19}$  & $\lbrace\lbrace BACKD\rbrace \rbrace$  & 0 \\ \hline
			$C_{20}$  & $\lbrace\lbrace BACDK\rbrace \rbrace$  & 0 \\ \hline
			$C_{21}$  & $\lbrace\lbrace BADKC\rbrace \rbrace$  & 0 \\ \hline
			$C_{22}$  & $\lbrace\lbrace BADCK\rbrace \rbrace$  & 0 \\ \hline
			$C_{23}$  & $\lbrace\lbrace BCADK\rbrace \rbrace$  & 0 \\ \hline
			$C_{24}$  & $\lbrace\lbrace BDACK\rbrace \rbrace$  & 0 \\ \hline
			$C_{25}$  & $\lbrace\lbrace CABKD\rbrace \rbrace$  & 0 \\ \hline
			$C_{26}$  & $\lbrace\lbrace CABDK\rbrace \rbrace$  & 0 \\ \hline
			$C_{27}$  & $\lbrace\lbrace CADKB\rbrace \rbrace$  & 0 \\ \hline
			$C_{28}$  & $\lbrace\lbrace CADBK\rbrace \rbrace$  & 0 \\ \hline
			$C_{29}$  & $\lbrace\lbrace CBADK\rbrace \rbrace$  & 0 \\ \hline
			$C_{30}$  & $\lbrace\lbrace CDABK\rbrace \rbrace$  & 0 \\ \hline
				\end{tabular}\label{tab:4legWeb7} %
	\end{minipage}
\hspace{0.5cm}
	\begin{minipage}[c]{0.42\textwidth}%
		\begin{tabular}{|c|c|c|}
			\hline 
			\textbf{Diagrams}  & \textbf{Sequences}  & \textbf{s-factors}  \\ 
			\hline
		     $C_{31}$  & $\lbrace\lbrace DABKC\rbrace \rbrace$  & 0 \\ \hline
		     $C_{32}$  & $\lbrace\lbrace DABCK\rbrace \rbrace$  & 0 \\ \hline
		     $C_{33}$  & $\lbrace\lbrace DACKB\rbrace \rbrace$  & 0 \\ \hline
		     $C_{34}$  & $\lbrace\lbrace DACBK\rbrace \rbrace$  & 0 \\ \hline
		     $C_{35}$  & $\lbrace\lbrace DBACK\rbrace \rbrace$  & 0 \\ \hline	
		     $C_{36}$  & $\lbrace\lbrace DCABK\rbrace \rbrace$  & 0 \\ \hline
			$C_{37}$  & $\lbrace\lbrace AKBCD\rbrace \rbrace$  & 1 \\ \hline
			$C_{38}$  & $\lbrace\lbrace AKBDC\rbrace \rbrace$  & 1 \\ \hline
			$C_{39}$  & $\lbrace\lbrace AKCBD\rbrace \rbrace$  & 1 \\ \hline
			$C_{40}$  & $\lbrace\lbrace AKCDB\rbrace \rbrace$  & 1 \\ \hline
			$C_{41}$  & $\lbrace\lbrace AKDBC\rbrace \rbrace$  & 1 \\ \hline
			$C_{42}$  & $\lbrace\lbrace AKDCB\rbrace \rbrace$  & 1 \\ \hline
			$C_{43}$  & $\lbrace\lbrace BAKCD\rbrace \rbrace$  & 1 \\ \hline
			$C_{44}$  & $\lbrace\lbrace BAKDC\rbrace \rbrace$  & 1 \\ \hline
			$C_{45}$  & $\lbrace\lbrace BCAKD\rbrace \rbrace$  & 1 \\ \hline
			$C_{46}$  & $\lbrace\lbrace BCDAK\rbrace \rbrace$  & 1 \\ \hline
			$C_{47}$  & $\lbrace\lbrace BDAKC\rbrace \rbrace$  & 1 \\ \hline
			$C_{48}$  & $\lbrace\lbrace BDCAK\rbrace \rbrace$  & 1 \\ \hline
			$C_{49}$  & $\lbrace\lbrace CAKBD\rbrace \rbrace$  & 1 \\ \hline
			$C_{50}$  & $\lbrace\lbrace CAKDB\rbrace \rbrace$  & 1 \\ \hline
			$C_{51}$  & $\lbrace\lbrace CBAKD\rbrace \rbrace$  & 1 \\ \hline
			$C_{52}$  & $\lbrace\lbrace CBDAK\rbrace \rbrace$  & 1 \\ \hline
			$C_{53}$  & $\lbrace\lbrace CDAKB\rbrace \rbrace$  & 1 \\ \hline
			$C_{54}$  & $\lbrace\lbrace CDBAK\rbrace \rbrace$  & 1 \\ \hline
			$C_{55}$  & $\lbrace\lbrace DAKBC\rbrace \rbrace$  & 1 \\ \hline
			$C_{56}$  & $\lbrace\lbrace DAKCB\rbrace \rbrace$  & 1 \\ \hline
			$C_{57}$  & $\lbrace\lbrace DBAKC\rbrace \rbrace$  & 1 \\ \hline
			$C_{58}$  & $\lbrace\lbrace DBCAK\rbrace \rbrace$  & 1 \\ \hline
			$C_{59}$  & $\lbrace\lbrace DCAKB\rbrace \rbrace$  & 1 \\ \hline
			$C_{60}$  & $\lbrace\lbrace DCBAK\rbrace \rbrace$  & 1 \\ \hline
		\end{tabular}
	\end{minipage}

\end{itemize}

\noindent The diagonal matrix for this Cweb is given by, 		

\begin{align}
	\mathcal{D}\,=\,\D{24} 
\end{align}

There are twenty-four independent exponentiated colour factors for this Cweb given as
\begin{eqnarray}
	(YC)_{1} &=& 0\,,\nonumber  \\	
	(YC)_{2} &=&
	- i\, f^{abg} f^{cdk} f^{nkg}\,\ta 1 \tn 1 \tb 2 \tc 3 \td 4 
	- i\,f^{abg} f^{ank} f^{cdk}\,\tn 1 \tg 1 \tb 2 \tc 3 \td 4 
	\,,\nonumber \\&& \nonumber \\
	(YC)_{3} &=& 0\,,\nonumber  \\
	(YC)_{4} &=& 0\,,\nonumber  \\
	(YC)_{5} &=&
	i\, f^{acg} f^{bdk} f^{nkg}\, \tn 1 \ta 1 \tb 2 \tc 3 \td 4 
	+i\, f^{acg} f^{ank} f^{bdk}\, \tg 1 \tn 1 \tb 2 \tc 3 \td 4  
	\,,\nonumber \\&& \nonumber \\
	(YC)_{6} &=& 0\,,\nonumber  \\
	(YC)_{7} &=&
	i\,f^{ack} f^{bng} f^{dnk}\, \ta 1 \tg 1 \tb 2 \tc 3 \td 4 
	-i\, f^{ack} f^{adg} f^{bng}\, \tn 1 \tkk 1 \tb 2 \tc 3 \td 4 \nonumber\\
	&\quad& - i\,f^{abg} f^{ack} f^{dnk}\,\tg 1 \tn 1 \tb 2 \tc 3 \td 4 
	-i\,f^{ack} f^{adg} f^{bnk}\, \tg 1 \tn 1 \tb 2 \tc 3 \td 4 
	\,,\nonumber \\&& \nonumber \\
	(YC)_{8} &=&
	i\,f^{abg} f^{cnk} f^{dng}\, \ta 1 \tkk 1 \tb 2 \tc 3 \td 4 
	-i \,f^{abg} f^{adn} f^{adz}\, \tn 1 \tkk 1 \tb 2 \tc 3 \td 4 \nonumber \\
	&\quad&-i\, f^{abg} f^{ack} f^{dng}\, \tkk 1 \tn 1 \tb 2 \tc 3 \td 4 
	+i\,f^{abg} f^{adn} f^{cnk}\, \tkk 1 \tg 1 \tb 2 \tc 3 \td 4 
	\,,\nonumber \\&& \nonumber \\
	(YC)_{9} &=& 0\,,\nonumber  \\
	(YC)_{10} &=& 0\,,\nonumber  \\
	(YC)_{11} &=&
	-i \,f^{adn} f^{akg} f^{bcg}\, \tkk 1 \tn 1 \tb 2 \tc 3 \td 4 
	\,,\nonumber \\&& \nonumber \\
	(YC)_{12} &=&
	- i\,f^{abg} f^{adk} f^{cng}\,\tn 1 \tkk 1 \tb 2 \tc 3 \td 4 
	- i\,f^{adk} f^{ang} f^{bcg}\,\tn 1 \tkk 1 \tb 2 \tc 3 \td 4  
	\,,\nonumber \\&& \nonumber \\
	(YC)_{13} &=&
	i\,f^{ang} f^{bkg} f^{cdk}\,\tn 1 \ta 1 \tb 2 \tc 3 \td 4 	\,,\nonumber \\
	(YC)_{14} &=&
	-i \,f^{akg} f^{bng} f^{cdk}\, \ta 1 \tn 1 \tb 2 \tc 3 \td 4  
	\,, \nonumber \\
	(YC)_{15} &=&
	- i\,f^{abk} f^{bdg} f^{cng}\,\ta 1 \tkk 1 \tb 2 \tc 3 \td 4  
	\,, \nonumber \\
	(YC)_{16} &=&
	i \,f^{akg} f^{bdg} f^{cnk}\, \ta 1 \tn 1 \tb 2 \tc 3 \td 4  
	\,, \nonumber \\
	(YC)_{17} &=&
	i\,f^{adn} f^{bkg} f^{cnk} \,\ta 1 \tg 1 \tb 2 \tc 3 \td 4 
	\,, \nonumber \\
	(YC)_{18} &=&
	i\,f^{adn} f^{bnk} f^{adz}\,\ta 1 \tg 1 \tb 2 \tc 3 \td 4  
	\,,\nonumber \\
	(YC)_{19} &=&
	-i\,f^{acg} f^{adk} f^{bng}\, \tn 1 \tkk 1 \tb 2 \tc 3 \td 4  
	-i\,f^{acg} f^{ank} f^{bdk}\, \tg 1 \tn 1 \tb 2 \tc 3 \td 4  
	\,,\nonumber \\&& \nonumber \\
	(YC)_{20} &=&
	-i\,f^{acg} f^{adk} f^{bng}\, \tn 1 \tkk 1 \tb 2 \tc 3 \td 4 
	-i\,f^{acg} f^{adk} f^{bnk}\, \tg 1 \tn 1 \tb 2 \tc 3 \td 4 
	\,,\nonumber \\&& \nonumber \\
	(YC)_{21} &=&
	i\,f^{adz} f^{ank} f^{bck}\, \tn 1 \tz 1 \tb 2 \tc 3 \td 4 
	-i\,f^{ank} f^{bck} f^{dny}\, \ty 1 \ta 1 \tb 2 \tc 3 \td 4 
	\,,\nonumber \\&& \nonumber \\
	(YC)_{22} &=&
	- i\,f^{abg} f^{cdk} f^{nkg}\,\ta 1 \tn 1 \tb 2 \tc 3 \td 4 
	- i\,f^{abg} f^{ack} f^{dnk}\,\tn 1 \tg 1 \tb 2 \tc 3 \td 4 \nonumber \\
	&\quad&	- i\,f^{abg} f^{ank} f^{cdk}\,\tn 1 \tg 1 \tb 2 \tc 3 \td 4 
	-i\,f^{abg} f^{ack} f^{dng}\, \tkk 1 \tn 1 \tb 2 \tc 3 \td 4 
	\,,\nonumber \\&& \nonumber \\
	(YC)_{23} &=&
	-i\,f^{abn} f^{adk} f^{ckg}\, \tn 1 \tg 1 \tb 2 \tc 3 \td 4 
	+ i\,f^{abn} f^{akg} f^{cdk}\,\tn 1 \tg 1 \tb 2 \tc 3 \td 4 
	\,,\nonumber \\&& \nonumber \\
	(YC)_{24} &=&
	-i\, f^{abg} f^{adk} f^{cng}\, \tn 1 \tkk 1 \tb 2 \tc 3 \td 4  
	-i\,f^{adk} f^{ang} f^{bcg}\, \tn 1 \tkk 1 \tb 2 \tc 3 \td 4 \nonumber\\
	&\quad&	-i\,f^{abg} f^{adk} f^{cnk}\, \tg 1 \tn 1 \tb 2 \tc 3 \td 4 
	\,.
\end{eqnarray}


\subsection{Boomerang Cwebs connecting three Wilson lines}
\label{app:three-lines}
\begin{itemize}
	\item[\textbf{1}.] $\textbf{W}\,_{3}^{(4)}(4,1,3)$ \\
	This is the first of the two Cwebs with same correlator and attachment content. It has thirty-six diagrams, one of them is displayed below. The table shows the chosen order of shuffles and their corresponding $ s $-factors. We do not to present the mixing matrix fo this Cweb here due to its larger dimension, however it can be found in the ancillary file \textit{Boomerang.nb}.\\
	
	\begin{center}
			\includegraphics[width=5cm,height=5cm]{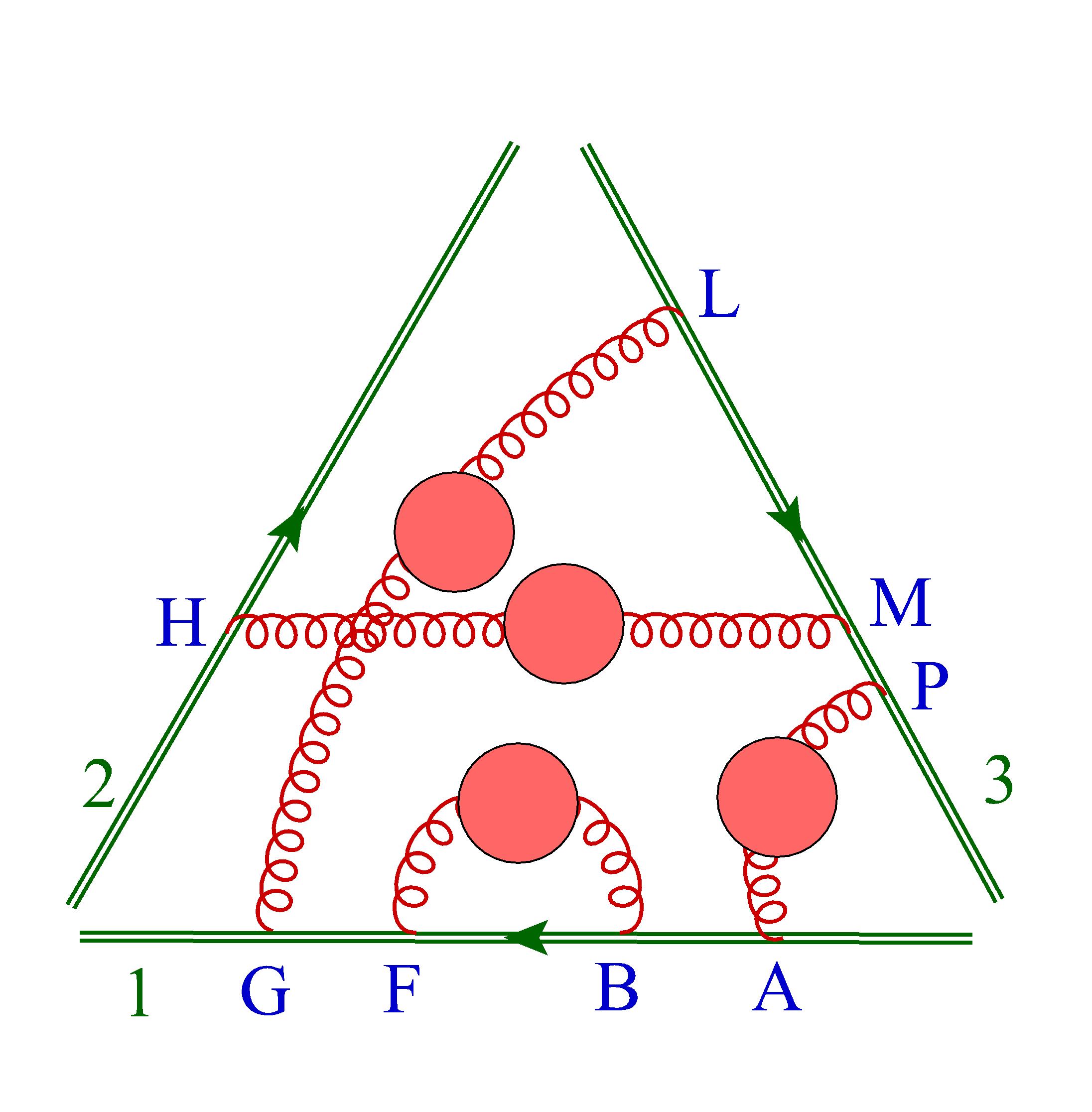} 
	\end{center}

			\begin{minipage}[c]{0.42\textwidth}%
		\begin{tabular}{|c|c|c|}
			\hline 
			\textbf{Diagrams}  & \textbf{Sequences}  & \textbf{s}  \\ 
			\hline
			$C_{1}$&$\{\{ABFG\},\{ LMP\}\}$&0\\\hline
			
			$C_{2}$&$\{\{ABGF\},\{ LMP\}\}$&0\\\hline
			
			$C_{3}$&$\{\{BAFG\},\{ LMP\}\}$&0\\\hline
			
			$C_{4}$&$\{\{BAGF\},\{ PML\}\}$&0\\\hline
			
			$C_{5}$&$\{\{BAGF\},\{ LMP\}\}$&0\\\hline
			
			$C_{6}$&$\{\{ABFG\},\{ LPM\}\}$&0\\\hline
			
			$C_{7}$&$\{\{ABFG\},\{ MLP\}\}$&0\\\hline
			
			$C_{8}$&$\{\{ABGF\},\{ PLM\}\}$&0\\\hline
			
			$C_{9}$&$\{\{ABGF\},\{ PML\}\}$&0\\\hline
			
			$C_{10}$&$\{\{ABGF\},\{ LPM\}\}$&0\\\hline
			
			$C_{11}$&$\{\{ABGF\},\{ MPL\}\}$&0\\\hline
			
			$C_{12}$&$\{\{ABGF\},\{ MLP\}\}$&0\\\hline
			
			$C_{13}$&$\{\{AGBF\},\{ LPM\}\}$&0\\\hline
			
			$C_{14}$&$\{\{AGBF\},\{ LMP\}\}$&0\\\hline
			
			$C_{15}$&$\{\{AGBF\},\{ MLP\}\}$&0\\\hline
			
			$C_{16}$&$\{\{BAFG\},\{ PLM\}\}$&0\\\hline
			
			$C_{17}$&$\{\{BAFG\},\{ PML\}\}$&0\\\hline
			
			$C_{18}$&$\{\{BAFG\},\{ LPM\}\}$&0\\\hline
		   \end{tabular}
	   \end{minipage}
   \hspace{0.7cm}
			\begin{minipage}[c]{0.42\textwidth}%
				\begin{tabular}{|c|c|c|}
					\hline 
					\textbf{Diagrams}  & \textbf{Sequences}  & \textbf{s}  \\ 
					\hline
			
			$C_{19}$&$\{\{BAFG\},\{ MPL\}\}$&0\\\hline
			
			$C_{20}$&$\{\{BAFG\},\{ MLP\}\}$&0\\\hline
			
			$C_{21}$&$\{\{BAGF\},\{ PLM\}\}$&0\\\hline
			
			$C_{22}$&$\{\{BAGF\},\{ LPM\}\}$&0\\\hline
			
			$C_{23}$&$\{\{BAGF\},\{ MPL\}\}$&0\\\hline
			
			$C_{24}$&$\{\{BAGF\},\{ MLP\}\}$&0\\\hline
			
			$C_{25}$&$\{\{BFAG\},\{ LPM\}\}$&0\\\hline
			
			$C_{26}$&$\{\{BFAG\},\{ LMP\}\}$&0\\\hline
			
			$C_{27}$&$\{\{BFAG\},\{ MLP\}\}$&0\\\hline
			
			$C_{28}$&$\{\{ABFG\},\{ PLM\}\}$&1\\\hline
			
			$C_{29}$&$\{\{ABFG\},\{ MPL\}\}$&1\\\hline
			
			$C_{30}$&$\{\{AGBF\},\{ PML\}\}$&1\\\hline
			
			$C_{31}$&$\{\{AGBF\},\{ MPL\}\}$&1\\\hline
			
			$C_{32}$&$\{\{BFAG\},\{ PLM\}\}$&1\\\hline
			
			$C_{33}$&$\{\{BFAG\},\{ PML\}\}$&1\\\hline
			
			$C_{34}$&$\{\{ABFG\},\{ PML\}\}$&2\\\hline
			
			$C_{35}$&$\{\{AGBF\},\{ PLM\}\}$&2\\\hline
			
			$C_{36}$&$\{\{BFAG\},\{ MPL\}\}$&2\\\hline
		\end{tabular}\label{tab:3legWeb1} %
	\end{minipage}\\

	The diagonal matrix is given as
	\begin{align}
		\mathcal{D}\,=\,\D{16}
	\end{align}

	\item[\textbf{2}.] $\textbf{W}\,_{3}^{(4)}(3,4,1)$ \\
	This is the second Cweb with same correlator and attachment content. The shuffle of correlators on Wilson line 1 and 2 generates thirty-six diagrams, out of which one is shown below. The tables show the chosen order of shuffle and the corresponding $ s $-factors. We do not to present the mixing matrix fo this Cweb here due to its larger dimension, however it can be found in the ancillary file \textit{Boomerang.nb}.

		\begin{center}
			\vspace{-2mm}
			\includegraphics[width=4cm,height=4cm]{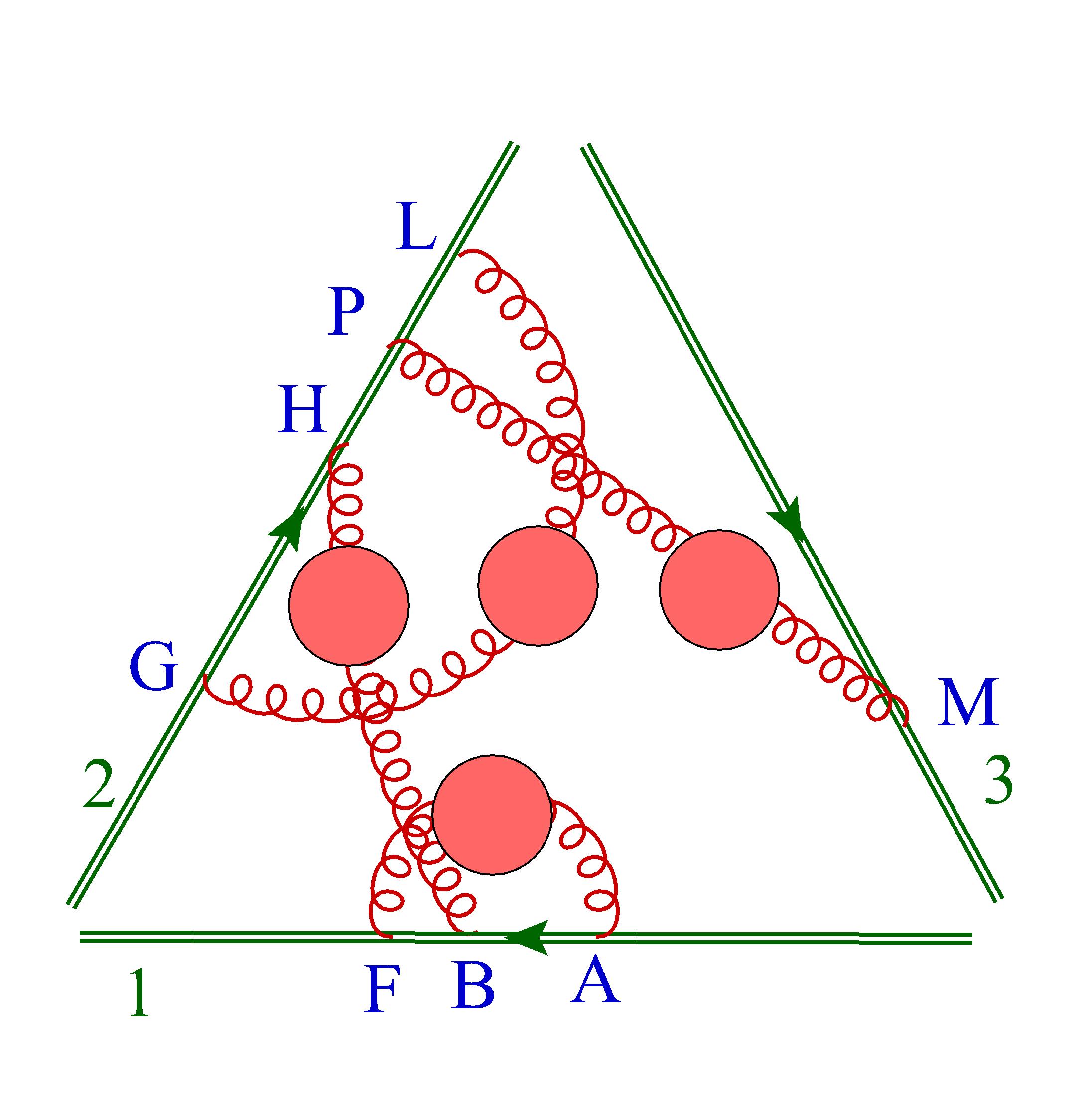}
		\end{center}

	\hspace{-1.5cm}\begin{minipage}[c]{0.46\textwidth}%
		%
		\begin{tabular}{|c|c|c|}
			\hline 
			\textbf{Diagrams}  & \textbf{Sequences}  & \textbf{s-factors}  \\ 
			\hline
			$C_{1}$&$\{\{ABF\},\{ GHPL\}\} $&0\\\hline
			
			$C_{2}$&$\{\{ABF\},\{ GPHL\}\} $&0\\\hline
			
			$C_{3}$&$\{\{AFB\},\{ HGPL\}\} $&0\\\hline
			
			$C_{4}$&$\{\{AFB\},\{ GHLP\}\} $&0\\\hline
			
			$C_{5}$&$\{\{AFB\},\{ GHPL\}\} $&0\\\hline
			
			$C_{6}$&$\{\{AFB\},\{ GPHL\}\} $&0\\\hline
			
			$C_{7}$&$\{\{AFB\},\{ GPLH\}\} $&0\\\hline
			
			$C_{8}$&$\{\{AFB\},\{ PGHL\}\} $&0\\\hline
			
			$C_{9}$&$\{\{ABF\},\{ HGLP\}\} $&0\\\hline
			
			$C_{10}$&$\{\{ABF\},\{ HGPL\}\} $&0\\\hline
			
			$C_{11}$&$\{\{ABF\},\{ HPGL\}\} $&0\\\hline
			
			$C_{12}$&$\{\{ABF\},\{ GHLP\}\} $&0\\\hline
			
			$C_{13}$&$\{\{ABF\},\{ GLHP\}\} $&0\\\hline
			
			$C_{14}$&$\{\{ABF\},\{ GLPH\}\} $&0\\\hline
			
			$C_{15}$&$\{\{ABF\},\{ GPLH\}\} $&0\\\hline
			
			$C_{16}$&$\{\{ABF\},\{ PHGL\}\} $&0\\\hline
			
			$C_{17}$&$\{\{ABF\},\{ PGHL\}\} $&0\\\hline
			
			$C_{18}$&$\{\{ABF\},\{ PGLH\}\} $&0\\\hline

		\end{tabular}\label{tab:3legWeb4} %
	\end{minipage} 
	\hspace{1.8cm}
	\begin{minipage}[c]{0.46\textwidth}%
		%
		\begin{tabular}{|c|c|c|}
			\hline 
			\textbf{Diagrams}  & \textbf{Sequences}  & \textbf{s-factors}  \\ 
			\hline
			$C_{19}$&$\{\{BAF\},\{ HGPL\}\} $&0\\\hline
			
			$C_{20}$&$\{\{BAF\},\{ GHLP\}\} $&0\\\hline
			
			$C_{21}$&$\{\{BAF\},\{ GHPL\}\} $&0\\\hline
			
			$C_{22}$&$\{\{BAF\},\{ GPHL\}\} $&0\\\hline
			
			$C_{23}$&$\{\{BAF\},\{ GPLH\}\} $&0\\\hline
			
			$C_{24}$&$\{\{BAF\},\{ PGHL\}\} $&0\\\hline
			
			$C_{25}$&$\{\{AFB\},\{ HGLP\}\} $&1\\\hline
			
			$C_{26}$&$\{\{AFB\},\{ HPGL\}\} $&1\\\hline
			
			$C_{27}$&$\{\{BAF\},\{ GLPH\}\} $&1\\\hline
			
			$C_{28}$&$\{\{BAF\},\{ PGLH\}\} $&1\\\hline
			
			$C_{29}$&$\{\{AFB\},\{ GLHP\}\} $&2\\\hline
			
			$C_{30}$&$\{\{AFB\},\{ GLPH\}\} $&2\\\hline
			
			$C_{31}$&$\{\{AFB\},\{ PHGL\}\} $&2\\\hline
			
			$C_{32}$&$\{\{AFB\},\{ PGLH\}\} $&2\\\hline
			
			$C_{33}$&$\{\{BAF\},\{ HGLP\}\} $&2\\\hline
			
			$C_{34}$&$\{\{BAF\},\{ HPGL\}\} $&2\\\hline
			
			$C_{35}$&$\{\{BAF\},\{ GLHP\}\} $&2\\\hline
			
			$C_{36}$&$\{\{BAF\},\{ PHGL\}\} $&2\\\hline
		\end{tabular}\label{tab:3legWeb4b} %
	\end{minipage}

	The diagonal matrix for this Cweb is given by, 

	\begin{align}
		\mathcal{D}\,=\,\D{12}
	\end{align}

	\item[\textbf{3}.] $\textbf{W}\,_{3}^{(4)}(4,2,2)$ \\
	This is a Cweb made out of four two-gluon correlators, and has 48 diagrams. We present one of the diagrams below. The table gives the chosen order of shuffle and their corresponding $ s $-factors. We do not to present the mixing matrix fo this Cweb here due to its larger dimension, however it can be found in the ancillary file \textit{Boomerang.nb}.
	
	\begin{center}
		\includegraphics[width=4cm,height=4cm]{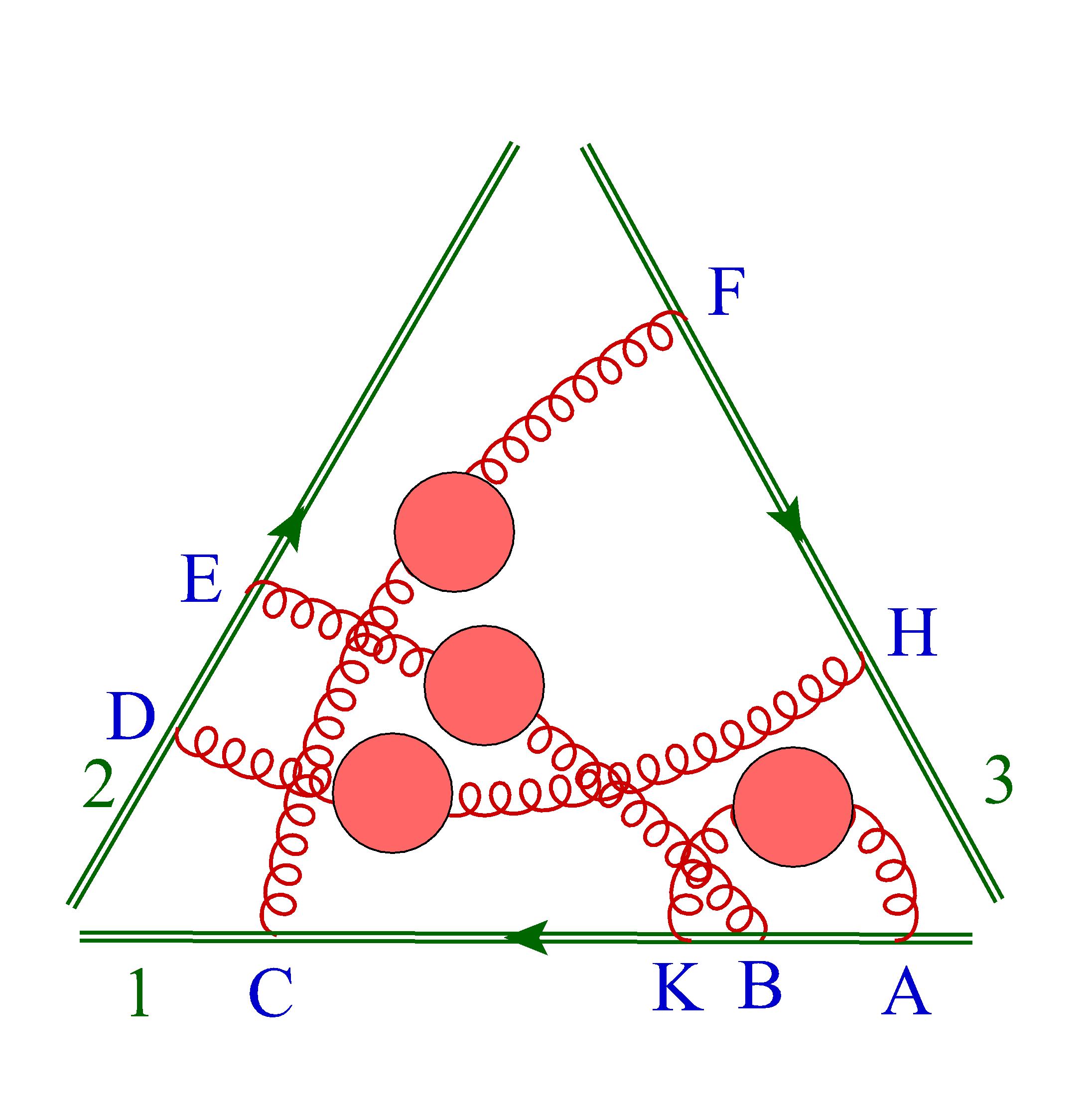}
	\end{center}
	
		\begin{minipage}[c]{0.42\textwidth}
			\resizebox{0.97\textwidth}{!}{
		\begin{tabular}{|c|c|c|}
			\hline 
			\textbf{Diagrams}  & \textbf{Sequences}  & \textbf{s-factors}  \\ 
			\hline
			$C_{1}$&$\{\{ABKC\},\{ DE\},\{FH\}\}$&0\\\hline
			
			$C_{2}$&$\{\{ABCK\},\{ ED\},\{HF\}\}$&0\\\hline
			
			$C_{3}$&$\{\{ABCK\},\{ DE\},\{FH\}\}$&0\\\hline
			
			$C_{4}$&$\{\{ACKB\},\{ ED\},\{HF\}\}$&0\\\hline
			
			$C_{5}$&$\{\{ACBK\},\{ ED\},\{HF\}\}$&0\\\hline
			
			$C_{6}$&$\{\{ACBK\},\{ DE\},\{FH\}\}$&0\\\hline
			
			$C_{7}$&$\{\{BAKC\},\{ DE\},\{FH\}\}$&0\\\hline
			
			$C_{8}$&$\{\{BACK\},\{ DE\},\{FH\}\}$&0\\\hline
			
			$C_{9}$&$\{\{CAKB\},\{ ED\},\{HF\}\}$&0\\\hline
			
			$C_{10}$&$\{\{CABK\},\{ ED\},\{HF\}\}$&0\\\hline
			
			$C_{11}$&$\{\{AKBC\},\{ DE\},\{FH\}\}$&0\\\hline
			
			$C_{12}$&$\{\{AKCB\},\{ ED\},\{HF\}\}$&0\\\hline
			
			$C_{13}$&$\{\{ABKC\},\{ ED\},\{FH\}\}$&0\\\hline
			
			$C_{14}$&$\{\{ABKC\},\{ ED\},\{HF\}\}$&0\\\hline
			
			$C_{15}$&$\{\{ABKC\},\{ DE\},\{HF\}\}$&0\\\hline
			
			$C_{16}$&$\{\{ABCK\},\{ ED\},\{FH\}\}$&0\\\hline
			
			$C_{17}$&$\{\{ABCK\},\{ DE\},\{HF\}\}$&0\\\hline
			
			$C_{18}$&$\{\{ACKB\},\{ ED\},\{FH\}\}$&0\\\hline
			
			$C_{19}$&$\{\{ACKB\},\{ DE\},\{FH\}\}$&0\\\hline
			
			$C_{20}$&$\{\{ACKB\},\{ DE\},\{HF\}\}$&0\\\hline
			
			$C_{21}$&$\{\{ACBK\},\{ ED\},\{FH\}\}$&0\\\hline
			
			$C_{22}$&$\{\{ACBK\},\{ DE\},\{HF\}\}$&0\\\hline
			
			$C_{23}$&$\{\{BACK\},\{ ED\},\{FH\}\}$&0\\\hline
			
			$C_{24}$&$\{\{BACK\},\{ ED\},\{HF\}\}$&0\\\hline
		\end{tabular}}\label{tab:3legWeb1} 
	\end{minipage} 
\hspace{0.6cm}
    \begin{minipage}[c]{0.42\textwidth}%
    	\resizebox{0.97\textwidth}{!}{
	\begin{tabular}{|c|c|c|}
		\hline 
		\textbf{Diagrams}  & \textbf{Sequences}  & \textbf{s-factors}  \\ 
		\hline
		$C_{25}$&$\{\{BACK\},\{ DE\},\{HF\}\}$&0\\\hline
		
		$C_{26}$&$\{\{BCAK\},\{ DE\},\{FH\}\}$&0\\\hline
		
		$C_{27}$&$\{\{CABK\},\{ ED\},\{FH\}\}$&0\\\hline
		
		$C_{28}$&$\{\{CABK\},\{ DE\},\{FH\}\}$&0\\\hline
		
		$C_{29}$&$\{\{CABK\},\{ DE\},\{HF\}\}$&0\\\hline
		
		$C_{30}$&$\{\{CBAK\},\{ ED\},\{HF\}\}$&0\\\hline
		
		$C_{31}$&$\{\{AKBC\},\{ ED\},\{FH\}\}$&1\\\hline
		
		$C_{32}$&$\{\{AKBC\},\{ ED\},\{HF\}\}$&1\\\hline
		
		$C_{33}$&$\{\{AKCB\},\{ ED\},\{FH\}\}$&1\\\hline
		
		$C_{34}$&$\{\{AKCB\},\{ DE\},\{FH\}\}$&1\\\hline
		
		$C_{35}$&$\{\{BAKC\},\{ ED\},\{FH\}\}$&1\\\hline
		
		$C_{36}$&$\{\{BAKC\},\{ DE\},\{HF\}\}$&1\\\hline
		
		$C_{37}$&$\{\{BCAK\},\{ ED\},\{HF\}\}$&1\\\hline
		
		$C_{38}$&$\{\{BCAK\},\{ DE\},\{HF\}\}$&1\\\hline
		
		$C_{39}$&$\{\{CAKB\},\{ ED\},\{FH\}\}$&1\\\hline
		
		$C_{40}$&$\{\{CAKB\},\{ DE\},\{HF\}\}$&1\\\hline
		
		$C_{41}$&$\{\{CBAK\},\{ DE\},\{FH\}\}$&1\\\hline
		
		$C_{42}$&$\{\{CBAK\},\{ DE\},\{HF\}\}$&1\\\hline
		
		$C_{43}$&$\{\{AKBC\},\{ DE\},\{HF\}\}$&2\\\hline
		
		$C_{44}$&$\{\{AKCB\},\{ DE\},\{HF\}\}$&2\\\hline
		
		$C_{45}$&$\{\{BAKC\},\{ ED\},\{HF\}\}$&2\\\hline
		
		$C_{46}$&$\{\{BCAK\},\{ ED\},\{FH\}\}$&2\\\hline
		
		$C_{47}$&$\{\{CAKB\},\{ DE\},\{FH\}\}$&2\\\hline
		
		$C_{48}$&$\{\{CBAK\},\{ ED\},\{FH\}\}$&2\\\hline
	\end{tabular}}
\end{minipage}

	\noindent The diagonal matrix for this Cweb is given by, 		

	\begin{align}
		\mathcal{D}\,=\,\D{22} 
	\end{align}	
	
	\item[\textbf{4}.] $\textbf{W}\,_{3,\text{I}}^{(4)}(3,2,3)$ \\
	
	This is a Cweb made out of four two-gluon correlators, and has 18 diagrams. We present one of the diagrams below. The table gives the chosen order of shuffle and their corresponding $ s $-factors. 
	
	\begin{center}
		\includegraphics[width=4cm,height=4cm]{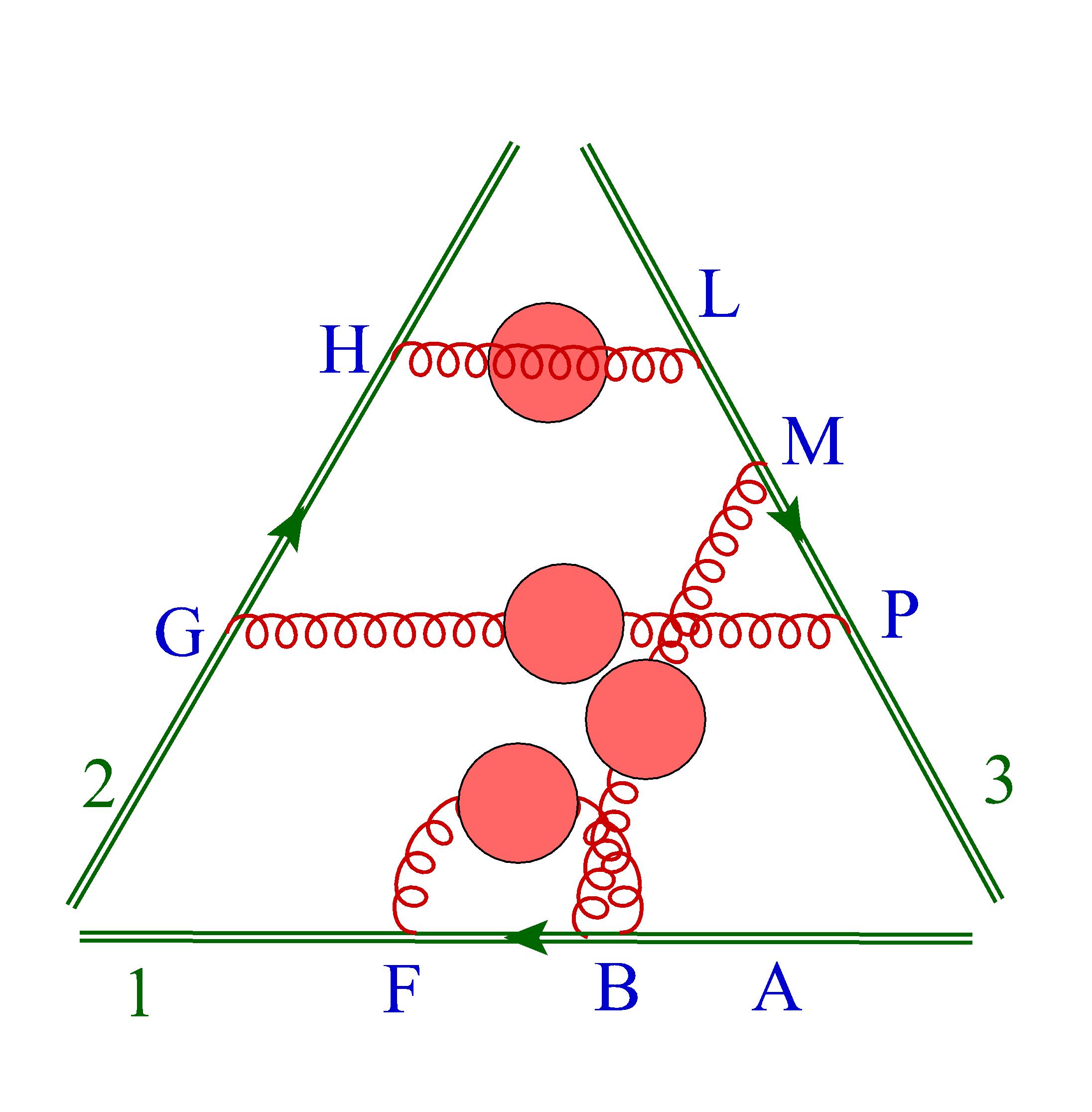}
	
		\resizebox{0.6\textwidth}{!}{
		\begin{tabular}{|c|c|c|}
			\hline 
			\textbf{Diagrams}  & \textbf{Sequences}  & \textbf{s-factors}  \\ 
			\hline
			$C_{1}$&$\{\{ABF\},\{ GH\},\{ LMP\}\}$&0\\\hline
			
			$C_{2}$&$\{\{AFB\},\{ GH\},\{ MLP\}\}$&0\\\hline
			
			$C_{3}$&$\{\{AFB\},\{ GH\},\{ LMP\}\}$&0\\\hline
			
			$C_{4}$&$\{\{AFB\},\{ GH\},\{ LPM\}\}$&0\\\hline
			
			$C_{5}$&$\{\{ABF\},\{ GH\},\{ MPL\}\}$&0\\\hline
			
			$C_{6}$&$\{\{ABF\},\{ GH\},\{ MLP\}\}$&0\\\hline
			
			$C_{7}$&$\{\{ABF\},\{ GH\},\{ PML\}\}$&0\\\hline
			
			$C_{8}$&$\{\{ABF\},\{ GH\},\{ PLM\}\}$&0\\\hline
			
			$C_{9}$&$\{\{ABF\},\{ GH\},\{ LPM\}\}$&0\\\hline
			
			$C_{10}$&$\{\{BAF\},\{ GH\},\{ MLP\}\}$&0\\\hline
			
			$C_{11}$&$\{\{BAF\},\{ GH\},\{ LMP\}\}$&0\\\hline
			
			$C_{12}$&$\{\{BAF\},\{ GH\},\{ LPM\}\}$&0\\\hline
			
			$C_{13}$&$\{\{AFB\},\{ GH\},\{ MPL\}\}$&1\\\hline
			
			$C_{14}$&$\{\{BAF\},\{ GH\},\{ PLM\}\}$&1\\\hline
			
			$C_{15}$&$\{\{AFB\},\{ GH\},\{ PML\}\}$&2\\\hline
			
			$C_{16}$&$\{\{AFB\},\{ GH\},\{ PLM\}\}$&2\\\hline
			
			$C_{17}$&$\{\{BAF\},\{ GH\},\{ MPL\}\}$&2\\\hline
			
			$C_{18}$&$\{\{BAF\},\{ GH\},\{ PML\}\}$&2\\\hline
		\end{tabular}}\label{tab:3legWeb3} %
\end{center}

The mixing and diagonalizing matrices are given as

	\begin{align}
		\begin{split}
			R=&\,\frac{1}{12} \left(
			\begin{array}{cccccccccccccccccc}
				12 & 4 & -6 & 2 & 2 & -6 & -4 & 2 & -6 & 2 & -6 & 4 & -2 & -2 & 2 & 0 & 0 & 2 \\
				0 & 4 & 0 & -4 & 0 & 0 & 0 & 0 & 0 & -4 & 0 & 4 & -2 & -6 & -4 & 6 & 2 & 4 \\
				0 & -2 & 6 & -4 & 0 & 0 & 0 & 0 & 0 & 2 & -6 & 4 & 0 & -2 & -2 & 2 & 0 & 2 \\
				0 & -2 & 0 & 2 & 0 & 0 & 0 & 0 & 0 & 2 & 0 & -2 & 2 & 2 & 0 & -2 & -2 & 0 \\
				0 & 0 & 0 & 0 & 2 & 0 & -4 & 2 & 0 & 0 & 0 & 0 & 0 & -2 & 0 & 0 & -2 & 4 \\
				0 & -2 & 0 & 2 & -4 & 6 & -4 & 8 & -6 & -4 & 0 & 4 & 2 & -6 & 0 & -2 & 2 & 4 \\
				0 & 0 & 0 & 0 & -4 & 0 & 8 & -4 & 0 & 0 & 0 & 0 & 2 & 2 & -4 & 2 & 2 & -4 \\
				0 & 0 & 0 & 0 & 2 & 0 & -4 & 2 & 0 & 0 & 0 & 0 & -2 & 0 & 4 & -2 & 0 & 0 \\
				0 & 4 & 0 & -4 & 8 & -6 & -4 & -4 & 6 & 2 & 0 & -2 & -6 & 2 & 4 & 2 & -2 & 0 \\
				0 & -2 & 0 & 2 & 0 & 0 & 0 & 0 & 0 & 2 & 0 & -2 & 2 & 2 & 0 & -2 & -2 & 0 \\
				0 & 4 & -6 & 2 & 0 & 0 & 0 & 0 & 0 & -4 & 6 & -2 & -2 & 0 & 2 & 0 & 2 & -2 \\
				0 & 4 & 0 & -4 & 0 & 0 & 0 & 0 & 0 & -4 & 0 & 4 & -6 & -2 & 4 & 2 & 6 & -4 \\
				0 & 0 & 0 & 0 & 0 & 0 & 0 & 0 & 0 & 0 & 0 & 0 & 2 & -2 & -4 & 2 & -2 & 4 \\
				0 & 0 & 0 & 0 & 0 & 0 & 0 & 0 & 0 & 0 & 0 & 0 & -2 & 2 & 4 & -2 & 2 & -4 \\
				0 & 0 & 0 & 0 & 0 & 0 & 0 & 0 & 0 & 0 & 0 & 0 & -2 & 2 & 4 & -2 & 2 & -4 \\
				0 & 0 & 0 & 0 & 0 & 0 & 0 & 0 & 0 & 0 & 0 & 0 & 0 & 0 & 0 & 0 & 0 & 0 \\
				0 & 0 & 0 & 0 & 0 & 0 & 0 & 0 & 0 & 0 & 0 & 0 & 0 & 0 & 0 & 0 & 0 & 0 \\
				0 & 0 & 0 & 0 & 0 & 0 & 0 & 0 & 0 & 0 & 0 & 0 & 2 & -2 & -4 & 2 & -2 & 4 \\
			\end{array}
			\right) \\
			\mathcal{D}\,=\,&\,\D{6}
		\end{split}
	\end{align}

	\item[\textbf{5}.] $\textbf{W}\,_{3,\text{II}}^{(4)}(3,2,3)$ 
	
	This is the second Cweb with same correlator and attachment content. It has eighteen diagrams, one of them is shown below. The table shows the chosen order of shuffle and their corresponding $ s $-factors. 

\begin{center}

\includegraphics[width=5cm,height=5cm]{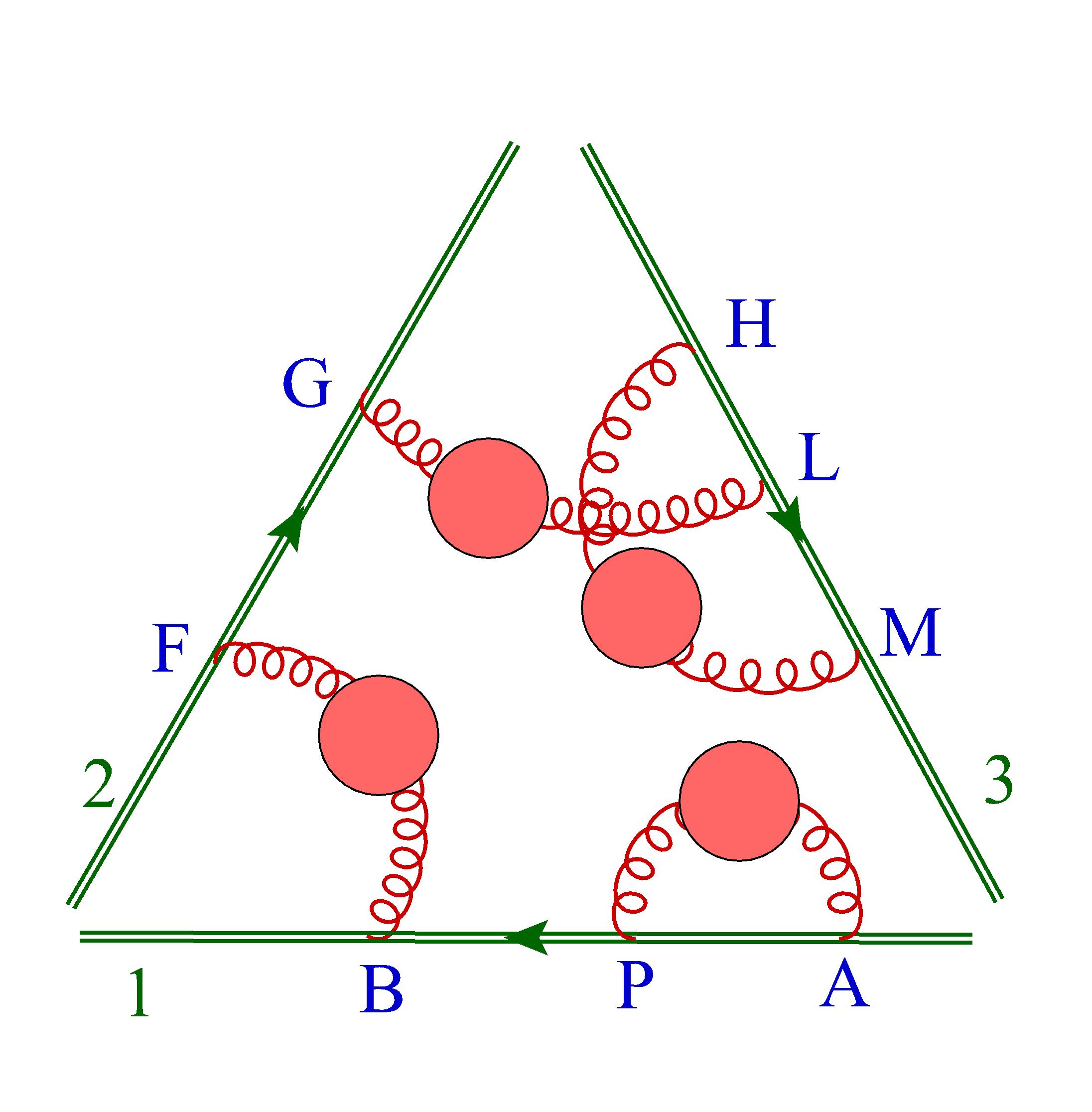} 

	\begin{minipage}[c]{0.46\textwidth}%
		%
		\begin{tabular}{|c|c|c|}
			\hline 
			\textbf{Diagrams}  & \textbf{Sequences}  & \textbf{s-factors}  \\ 
			\hline
			$C_{1}$&$\{\{APB\},\{ FG\},\{HLM\}\}$&0\\\hline
			
			$C_{2}$&$\{\{APB\},\{ GF\},\{HLM\}\}$&0\\\hline
			
			$C_{3}$&$\{\{ABP\},\{ FG\},\{LHM\}\}$&0\\\hline
			
			$C_{4}$&$\{\{ABP\},\{ FG\},\{HLM\}\}$&0\\\hline
			
			$C_{5}$&$\{\{ABP\},\{ FG\},\{HML\}\}$&0\\\hline
			
			$C_{6}$&$\{\{ABP\},\{ GF\},\{LHM\}\}$&0\\\hline
			
			$C_{7}$&$\{\{ABP\},\{ GF\},\{HLM\}\}$&0\\\hline
			
			$C_{8}$&$\{\{ABP\},\{ GF\},\{HML\}\}$&0\\\hline
			
			$C_{9}$&$\{\{BAP\},\{ FG\},\{HLM\}\}$&0\\\hline
			
			$C_{10}$&$\{\{BAP\},\{ GF\},\{HLM\}\}$&0\\\hline
			
			$C_{11}$&$\{\{APB\},\{ FG\},\{LHM\}\}$&1\\\hline
			
			$C_{12}$&$\{\{BAP\},\{ GF\},\{HML\}\}$&1\\\hline
			
			$C_{13}$&$\{\{APB\},\{ FG\},\{HML\}\}$&2\\\hline
			
			$C_{14}$&$\{\{APB\},\{ GF\},\{HML\}\}$&2\\\hline
			
			$C_{15}$&$\{\{BAP\},\{ FG\},\{LHM\}\}$&2\\\hline
			
			$C_{16}$&$\{\{BAP\},\{ GF\},\{LHM\}\}$&2\\\hline
			
			$C_{17}$&$\{\{APB\},\{ GF\},\{LHM\}\}$&4\\\hline
			
			$C_{18}$&$\{\{BAP\},\{ FG\},\{HML\}\}$&4\\\hline
		\end{tabular}\label{tab:4legWeb5} %
	\end{minipage}
	
\end{center}

The mixing and diagonalizing matrices are given as

	\begin{align}
		\begin{split}
			R=\frac{1}{12} \left(
			\begin{array}{cccccccccccccccccc}
				4 & -4 & 0 & 0 & 0 & 0 & 0 & 0 & -4 & 4 & -1 & -3 & -3 & 3 & 1 & -1 & 1 & 3 \\
				-2 & 2 & 0 & 0 & 0 & 0 & 0 & 0 & 2 & -2 & 1 & 1 & 1 & -1 & -1 & 1 & -1 & -1 \\
				0 & 0 & 4 & 0 & -4 & -4 & 0 & 4 & 0 & 0 & -1 & -3 & 1 & -1 & -3 & 3 & 1 & 3 \\
				-2 & 2 & -2 & 6 & -4 & 2 & -6 & 4 & -4 & 4 & 1 & -3 & 1 & -1 & 1 & -1 & -1 & 3 \\
				0 & 0 & -2 & 0 & 2 & 2 & 0 & -2 & 0 & 0 & 1 & 1 & -1 & 1 & 1 & -1 & -1 & -1 \\
				0 & 0 & -2 & 0 & 2 & 2 & 0 & -2 & 0 & 0 & 1 & 1 & -1 & 1 & 1 & -1 & -1 & -1 \\
				4 & -4 & 4 & -6 & 2 & -4 & 6 & -2 & 2 & -2 & -3 & 1 & -1 & 1 & -1 & 1 & 3 & -1 \\
				0 & 0 & 4 & 0 & -4 & -4 & 0 & 4 & 0 & 0 & -3 & -1 & 3 & -3 & -1 & 1 & 3 & 1 \\
				-2 & 2 & 0 & 0 & 0 & 0 & 0 & 0 & 2 & -2 & 1 & 1 & 1 & -1 & -1 & 1 & -1 & -1 \\
				4 & -4 & 0 & 0 & 0 & 0 & 0 & 0 & -4 & 4 & -3 & -1 & -1 & 1 & 3 & -3 & 3 & 1 \\
				0 & 0 & 0 & 0 & 0 & 0 & 0 & 0 & 0 & 0 & 3 & -3 & -3 & 3 & -3 & 3 & -3 & 3 \\
				0 & 0 & 0 & 0 & 0 & 0 & 0 & 0 & 0 & 0 & -3 & 3 & 3 & -3 & 3 & -3 & 3 & -3 \\
				0 & 0 & 0 & 0 & 0 & 0 & 0 & 0 & 0 & 0 & -1 & 1 & 1 & -1 & 1 & -1 & 1 & -1 \\
				0 & 0 & 0 & 0 & 0 & 0 & 0 & 0 & 0 & 0 & 1 & -1 & -1 & 1 & -1 & 1 & -1 & 1 \\
				0 & 0 & 0 & 0 & 0 & 0 & 0 & 0 & 0 & 0 & -1 & 1 & 1 & -1 & 1 & -1 & 1 & -1 \\
				0 & 0 & 0 & 0 & 0 & 0 & 0 & 0 & 0 & 0 & 1 & -1 & -1 & 1 & -1 & 1 & -1 & 1 \\
				0 & 0 & 0 & 0 & 0 & 0 & 0 & 0 & 0 & 0 & -1 & 1 & 1 & -1 & 1 & -1 & 1 & -1 \\
				0 & 0 & 0 & 0 & 0 & 0 & 0 & 0 & 0 & 0 & 1 & -1 & -1 & 1 & -1 & 1 & -1 & 1 \\
			\end{array}
			\right) \nonumber
		\end{split}
	\end{align}
	\begin{align}
		\mathcal{D}\,=\,\D{4}
	\end{align}

	\item[\textbf{6}.] $\textbf{W}\,_{3,\text{I}}^{(4)}(5,1,2)$ 
	
		This Cweb has four two point correlator including two boomerang on line 1. It has thirty diagrams, one of them is shown below. The table shows the chosen order of shuffle and their corresponding $ s $-factors.  We do not to present the mixing matrix fo this Cweb here due to its larger dimension, however it can be found in the ancillary file \textit{Boomerang.nb}.

\begin{center}
	
			\includegraphics[width=5cm,height=5cm]{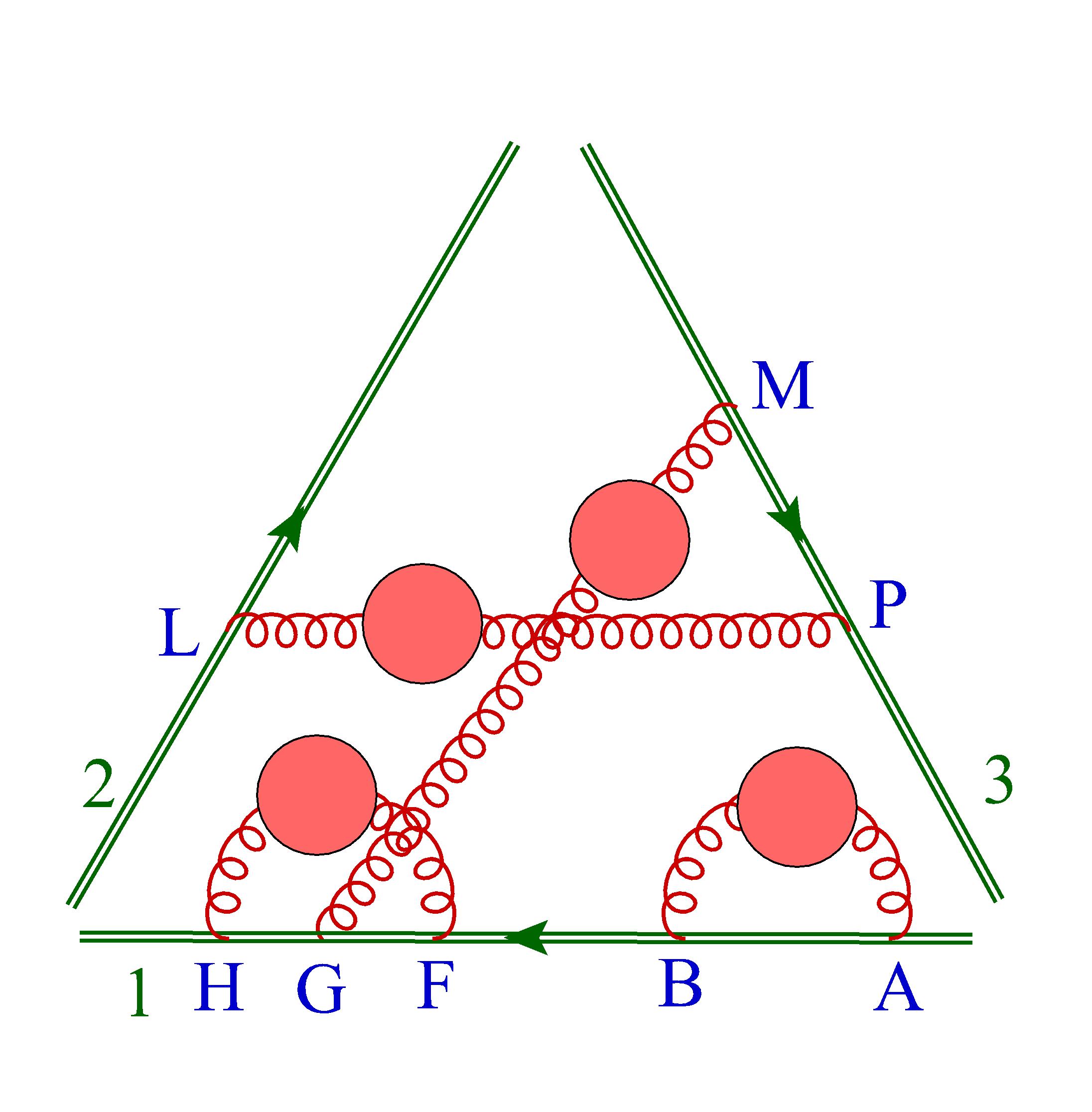} 

\begin{minipage}[c]{0.46\textwidth}%
	%
	\begin{tabular}{|c|c|c|}
		\hline 
		\textbf{Diagrams}  & \textbf{Sequences}  & \textbf{s}  \\ 
		\hline
		$C_{1}$&$\{\{ABFGH\},\{ MP\}\}$&0\\\hline
		
		$C_{2}$&$\{\{ABFGH\},\{ PM\}\}$&0\\\hline
		
		$C_{3}$&$\{\{AFBHG\},\{ MP\}\}$&0\\\hline
		
		$C_{4}$&$\{\{AFBHG\},\{ PM\}\}$&0\\\hline
		
		$C_{5}$&$\{\{AFBGH\},\{ MP\}\}$&0\\\hline
		
		$C_{6}$&$\{\{AFBGH\},\{ PM\}\}$&0\\\hline
		
		$C_{7}$&$\{\{AFHBG\},\{ MP\}\}$&0\\\hline
		
		$C_{8}$&$\{\{AFHBG\},\{ PM\}\}$&0\\\hline
		
		$C_{9}$&$\{\{AFHGB\},\{ MP\}\}$&0\\\hline
		
		$C_{10}$&$\{\{AFHGB\},\{ PM\}\}$&0\\\hline
		
		$C_{11}$&$\{\{AFGBH\},\{ MP\}\}$&0\\\hline
		
		$C_{12}$&$\{\{AFGBH\},\{ PM\}\}$&0\\\hline
		
		$C_{13}$&$\{\{AFGHB\},\{ MP\}\}$&0\\\hline
		
		$C_{14}$&$\{\{AFGHB\},\{ PM\}\}$&0\\\hline
		
		$C_{15}$&$\{\{AGBFH\},\{ MP\}\}$&0\\\hline
	\end{tabular}
\end{minipage}
	\begin{minipage}[c]{0.46\textwidth}%
		%
		\begin{tabular}{|c|c|c|}
			\hline 
			\textbf{Diagrams}  & \textbf{Sequences}  & \textbf{s}  \\ 
			\hline			
			$C_{16}$&$\{\{AGBFH\},\{ PM\}\}$&0\\\hline
			
			$C_{17}$&$\{\{AGFBH\},\{ MP\}\}$&0\\\hline
			
			$C_{18}$&$\{\{AGFBH\},\{ PM\}\}$&0\\\hline
			
			$C_{19}$&$\{\{AGFHB\},\{ MP\}\}$&0\\\hline
			
			$C_{20}$&$\{\{AGFHB\},\{ PM\}\}$&0\\\hline
			
			$C_{21}$&$\{\{GAFBH\},\{ MP\}\}$&0\\\hline
			
			$C_{22}$&$\{\{GAFBH\},\{ PM\}\}$&0\\\hline
			
			$C_{23}$&$\{\{GAFHB\},\{ MP\}\}$&0\\\hline
			
			$C_{24}$&$\{\{GAFHB\},\{ PM\}\}$&0\\\hline
			
			$C_{25}$&$\{\{ABFHG\},\{ MP\}\}$&1\\\hline
			
			$C_{26}$&$\{\{GABFH\},\{ PM\}\}$&1\\\hline
			
			$C_{27}$&$\{\{ABFHG\},\{ PM\}\}$&2\\\hline
			
			$C_{28}$&$\{\{ABGFH\},\{ MP\}\}$&2\\\hline
			
			$C_{29}$&$\{\{ABGFH\},\{ PM\}\}$&2\\\hline
			
			$C_{30}$&$\{\{GABFH\},\{ MP\}\}$&2\\\hline
		\end{tabular}\label{tab:3legWeb6} %
	\end{minipage}\\

\end{center}

The diagonalizing matrix is given as

	\begin{align}
		\mathcal{D}\,=\,\D{10}
	\end{align}

	\item[\textbf{7}.] $\textbf{W}\,_{3,\text{II}}^{(4)}(5,1,2)$ 
	
	This is second Cweb with same correlator content as previous one. It has sixty diagrams, one of them is shown below. The table shows chosen order of shuffle and their corresponding $ s $-factors.  We do not to present the mixing matrix fo this Cweb here due to its larger dimension, however it can be found in the ancillary file \textit{Boomerang.nb}.
	\begin{center}
			\includegraphics[width=5cm,height=5cm]{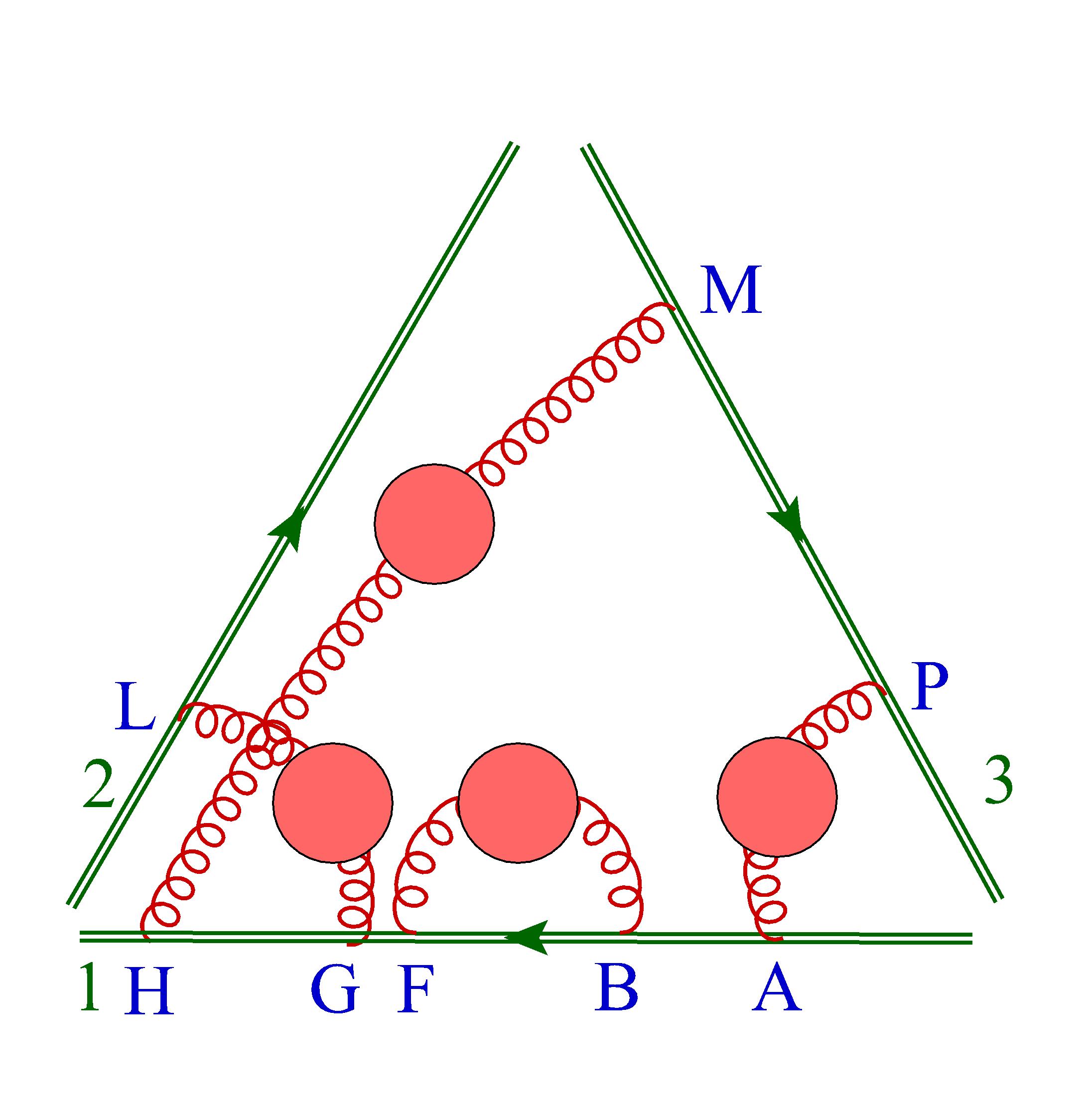} 

	\begin{minipage}[c]{0.42\textwidth}%
			\begin{tabular}{|c|c|c|}
			\hline 
			\textbf{Diagrams}  & \textbf{Sequences}  & \textbf{s}  \\ 
			\hline
			$C_{1}$&$\{\{ABFGH\},\{ MP\}\}$&0\\\hline
			
			$C_{2}$&$\{\{ABHGF\},\{ MP\}\}$&0\\\hline
			
			$C_{3}$&$\{\{ABGFH\},\{ MP\}\}$&0\\\hline
			
			$C_{4}$&$\{\{ABGHF\},\{ MP\}\}$&0\\\hline
			
			$C_{5}$&$\{\{AGBFH\},\{ MP\}\}$&0\\\hline
			
			$C_{6}$&$\{\{AGBHF\},\{ MP\}\}$&0\\\hline
			
			$C_{7}$&$\{\{BAFGH\},\{ MP\}\}$&0\\\hline
			
			$C_{8}$&$\{\{BAHGF\},\{ PM\}\}$&0\\\hline
			
			$C_{9}$&$\{\{BAHGF\},\{ MP\}\}$&0\\\hline
			
			$C_{10}$&$\{\{BAGFH\},\{ MP\}\}$&0\\\hline
			
			$C_{11}$&$\{\{BAGHF\},\{ PM\}\}$&0\\\hline
			
			$C_{12}$&$\{\{BAGHF\},\{ MP\}\}$&0\\\hline
			
			$C_{13}$&$\{\{BGAFH\},\{ MP\}\}$&0\\\hline
			
			$C_{14}$&$\{\{BGAHF\},\{ PM\}\}$&0\\\hline
			
			$C_{15}$&$\{\{BGAHF\},\{ MP\}\}$&0\\\hline
			
			$C_{16}$&$\{\{ABFHG\},\{ MP\}\}$&0\\\hline
			
			$C_{17}$&$\{\{ABHFG\},\{ PM\}\}$&0\\\hline
			
			$C_{18}$&$\{\{ABHFG\},\{ MP\}\}$&0\\\hline
			
			$C_{19}$&$\{\{ABHGF\},\{ PM\}\}$&0\\\hline
			
			$C_{20}$&$\{\{ABGFH\},\{ PM\}\}$&0\\\hline
			
			$C_{21}$&$\{\{ABGHF\},\{ PM\}\}$&0\\\hline
			
			$C_{22}$&$\{\{AHBFG\},\{ MP\}\}$&0\\\hline
			
			$C_{23}$&$\{\{AHBGF\},\{ PM\}\}$&0\\\hline
			
			$C_{24}$&$\{\{AHBGF\},\{ MP\}\}$&0\\\hline
			
			$C_{25}$&$\{\{AHGBF\},\{ MP\}\}$&0\\\hline
			
			$C_{26}$&$\{\{AGBHF\},\{ PM\}\}$&0\\\hline
			
			$C_{27}$&$\{\{AGHBF\},\{ MP\}\}$&0\\\hline
			
			$C_{28}$&$\{\{BAFHG\},\{ PM\}\}$&0\\\hline
			
			$C_{29}$&$\{\{BAFHG\},\{ MP\}\}$&0\\\hline
			
			$C_{30}$&$\{\{BAFGH\},\{ PM\}\}$&0\\\hline
		\end{tabular}\label{tab:4legWeb2} %
	\end{minipage}
\hspace{0.5cm}
		\begin{minipage}[c]{0.42\textwidth}%
		\begin{tabular}{|c|c|c|}
			\hline 
			\textbf{Diagrams}  & \textbf{Sequences}  & \textbf{s}  \\ 
			\hline
			$C_{31}$&$\{\{BAHFG\},\{ PM\}\}$&0\\\hline
			
			$C_{32}$&$\{\{BAHFG\},\{ MP\}\}$&0\\\hline
			
			$C_{33}$&$\{\{BAGFH\},\{ PM\}\}$&0\\\hline
			
			$C_{34}$&$\{\{BFAHG\},\{ MP\}\}$&0\\\hline
			
			$C_{35}$&$\{\{BFAGH\},\{ MP\}\}$&0\\\hline
			
			$C_{36}$&$\{\{BFGAH\},\{ MP\}\}$&0\\\hline
			
			$C_{37}$&$\{\{BGAFH\},\{ PM\}\}$&0\\\hline
			
			$C_{38}$&$\{\{BGFAH\},\{ PM\}\}$&0\\\hline
			
			$C_{39}$&$\{\{BGFAH\},\{ MP\}\}$&0\\\hline
			
			$C_{40}$&$\{\{GABFH\},\{ MP\}\}$&0\\\hline
			
			$C_{41}$&$\{\{GABHF\},\{ PM\}\}$&0\\\hline
			
			$C_{42}$&$\{\{GABHF\},\{ MP\}\}$&0\\\hline
			
			$C_{43}$&$\{\{GAHBF\},\{ MP\}\}$&0\\\hline
			
			$C_{44}$&$\{\{GBAFH\},\{ PM\}\}$&0\\\hline
			
			$C_{45}$&$\{\{GBAFH\},\{ MP\}\}$&0\\\hline
			
			$C_{46}$&$\{\{GBAHF\},\{ PM\}\}$&0\\\hline
			
			$C_{47}$&$\{\{GBAHF\},\{ MP\}\}$&0\\\hline
			
			$C_{48}$&$\{\{GBFAH\},\{ MP\}\}$&0\\\hline
			
			$C_{49}$&$\{\{ABFHG\},\{ PM\}\}$&1\\\hline
			
			$C_{50}$&$\{\{ABFGH\},\{ PM\}\}$&1\\\hline
			
			$C_{51}$&$\{\{AHBFG\},\{ PM\}\}$&1\\\hline
			
			$C_{52}$&$\{\{AHGBF\},\{ PM\}\}$&1\\\hline
			
			$C_{53}$&$\{\{AGBFH\},\{ PM\}\}$&1\\\hline
			
			$C_{54}$&$\{\{AGHBF\},\{ PM\}\}$&1\\\hline
			
			$C_{55}$&$\{\{BFAHG\},\{ PM\}\}$&1\\\hline
			
			$C_{56}$&$\{\{BFAGH\},\{ PM\}\}$&1\\\hline
			
			$C_{57}$&$\{\{BFGAH\},\{ PM\}\}$&1\\\hline
			
			$C_{58}$&$\{\{GABFH\},\{ PM\}\}$&1\\\hline
			
			$C_{59}$&$\{\{GAHBF\},\{ PM\}\}$&1\\\hline
			
			$C_{60}$&$\{\{GBFAH\},\{ PM\}\}$&1\\\hline
		\end{tabular}\label{tab:4legWeb2} %
	\end{minipage}\\

\end{center}
	
	Diagonalizing matrix is given as 
		\begin{align}
		\mathcal{D}\,=\,\D{32}
	\end{align}

	\item[\textbf{8}.] $\textbf{W}\,_{3,\text{I}}^{(2,1)}(4,2,1)$ \\
	This is the first Cweb that connects three Wilson lines and is made out of a three-gluon correlator and 2 two-gluon correlators. We present one representative diagram of this Cweb. The table shows the chosen order of shuffles and their corresponding $ s $-factors. \\   
	\begin{minipage}[c]{0.5\textwidth}%
		\begin{figure}[H]
			\vspace{-2mm}
			\includegraphics[width=5cm,height=5cm]{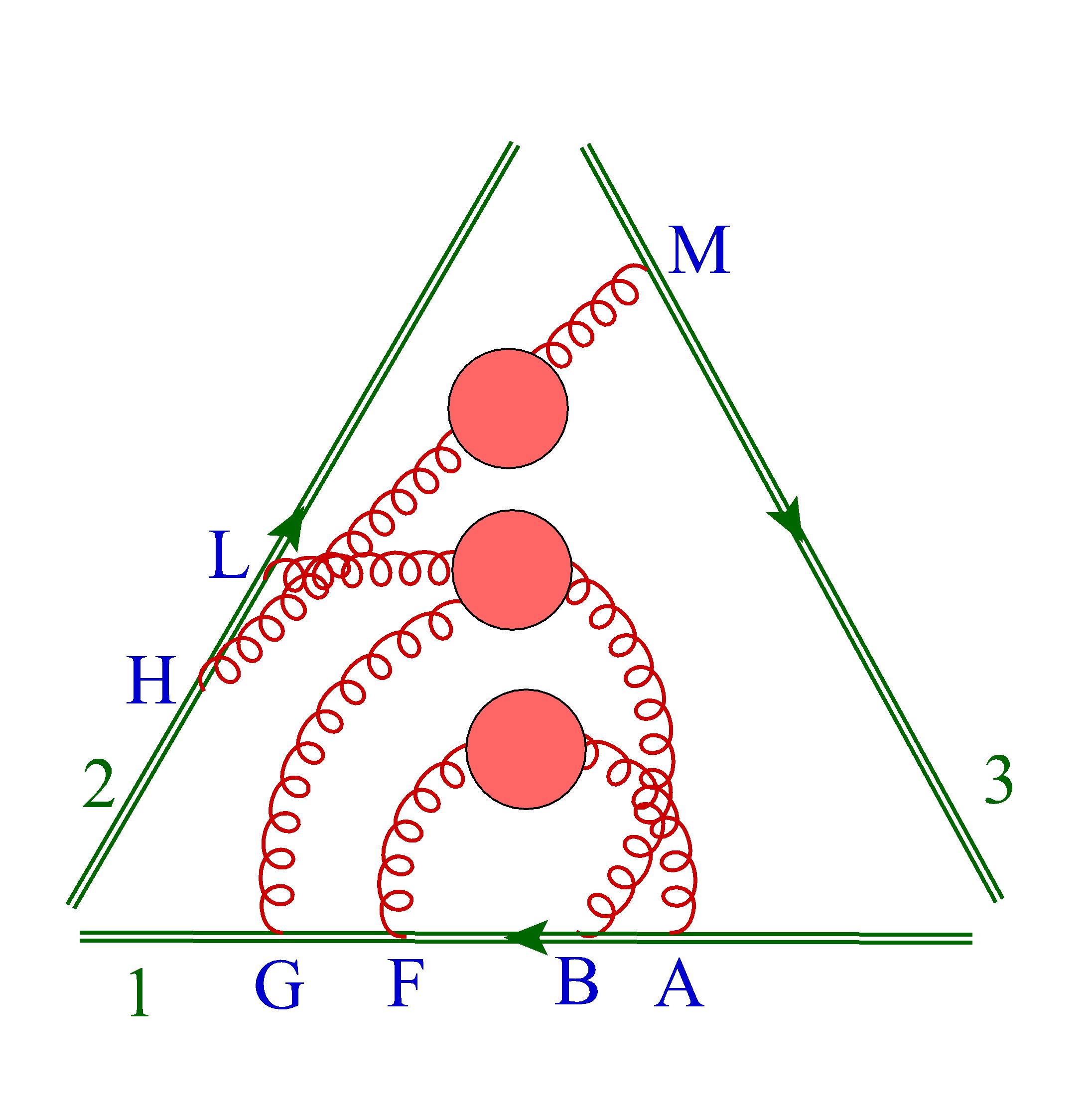} 
		\end{figure}
	\end{minipage}\hspace{-2cm} %
	\begin{minipage}[c]{0.46\textwidth}%
		\vspace{1cm}
		\begin{tabular}{|c|c|c|}
			\hline 
			\textbf{Diagrams}  & \textbf{Sequences}  & \textbf{s-factors}  \\ 
			\hline
			$C_{1}$&$\{\{ABFG\},\{ HL\}\}$&0\\\hline
			
			$C_{2}$&$\{\{ABFG\},\{ LH\}\}$&0\\\hline
			
			$C_{3}$&$\{\{ABGF\},\{ HL\}\}$&0\\\hline
			
			$C_{4}$&$\{\{ABGF\},\{ LH\}\}$&0\\\hline
			
			$C_{5}$&$\{\{BAFG\},\{ HL\}\}$&0\\\hline
			
			$C_{6}$&$\{\{BAFG\},\{ LH\}\}$&0\\\hline
			
			$C_{7}$&$\{\{BAGF\},\{ HL\}\}$&0\\\hline
			
			$C_{8}$&$\{\{BAGF\},\{ LH\}\}$&0\\\hline
			
			$C_{9}$&$\{\{AFBG\},\{ LH\}\}$&1\\\hline
			
			$C_{10}$&$\{\{BGAF\},\{ HL\}\}$&1\\\hline
			
			$C_{11}$&$\{\{AFBG\},\{ HL\}\}$&2\\\hline
			
			$C_{12}$&$\{\{BGAF\},\{ LH\}\}$&2\\\hline
		\end{tabular}\label{tab:3legWeb7} %
	\end{minipage} \\ \\
	The mixing matrix and the diagonal matrix for this Cweb are given by,  
	\begin{align}
		\begin{split}
			R=&\frac{1}{6} \left(
			\begin{array}{cccccccccccc}
				3 & -3 & 0 & 0 & 0 & 0 & 0 & 0 & 2 & -1 & -2 & 1 \\
				-3 & 3 & 0 & 0 & 0 & 0 & 0 & 0 & -1 & 2 & 1 & -2 \\
				0 & 0 & 3 & -3 & 0 & 0 & 0 & 0 & 2 & -1 & -2 & 1 \\
				0 & 0 & -3 & 3 & 0 & 0 & 0 & 0 & -1 & 2 & 1 & -2 \\
				0 & 0 & 0 & 0 & 3 & -3 & 0 & 0 & 2 & -1 & -2 & 1 \\
				0 & 0 & 0 & 0 & -3 & 3 & 0 & 0 & -1 & 2 & 1 & -2 \\
				0 & 0 & 0 & 0 & 0 & 0 & 3 & -3 & 2 & -1 & -2 & 1 \\
				0 & 0 & 0 & 0 & 0 & 0 & -3 & 3 & -1 & 2 & 1 & -2 \\
				0 & 0 & 0 & 0 & 0 & 0 & 0 & 0 & 2 & 2 & -2 & -2 \\
				0 & 0 & 0 & 0 & 0 & 0 & 0 & 0 & 2 & 2 & -2 & -2 \\
				0 & 0 & 0 & 0 & 0 & 0 & 0 & 0 & -1 & -1 & 1 & 1 \\
				0 & 0 & 0 & 0 & 0 & 0 & 0 & 0 & -1 & -1 & 1 & 1 \\
			\end{array}
			\right)
		\end{split},
		\mathcal{D}\,=\,\D{5}\,.
	\end{align}

	\item[\textbf{9}.] $\textbf{W}\,_{3,\text{II}}^{(2,1)}(4,2,1)$ \\
	This is the second Cweb with same correlator and attachment content. It has twenty four diagrams, one of them is shown below. The table lists all possible shuffles and their corresponding $ s $-factors. \\
	\begin{minipage}[c]{0.5\textwidth}%
		\begin{figure}[H]
			\vspace{-2mm}
			\includegraphics[width=5cm,height=5cm]{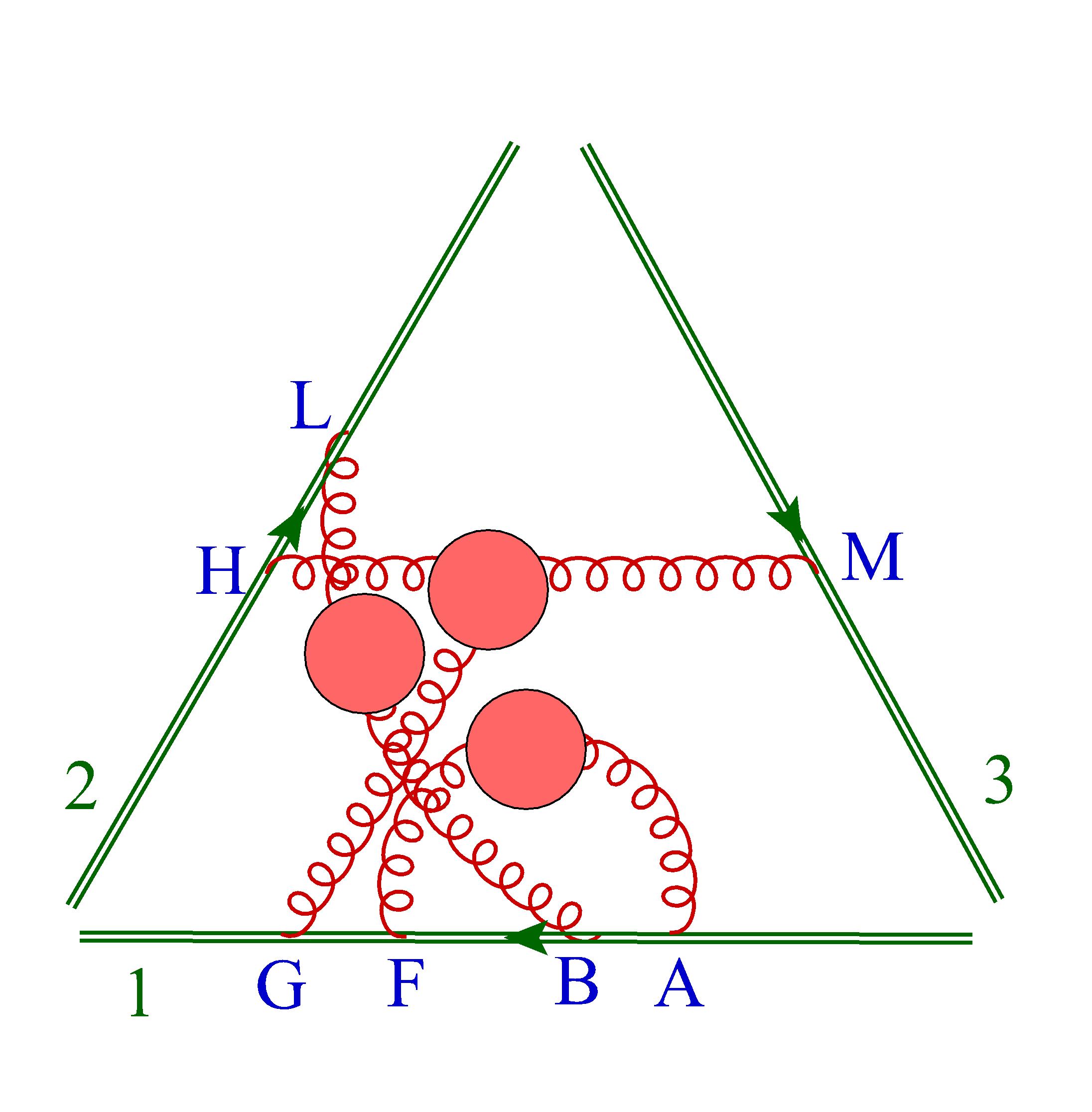} 
		\end{figure}
	\end{minipage}\hspace{-2cm} %
	\begin{minipage}[c]{0.46\textwidth}%
		\vspace{2cm}
		\begin{tabular}{|c|c|c|}
			\hline 
			\textbf{Diagrams}  & \textbf{Sequences}  & \textbf{s-factors}  \\ 
			\hline
			$C_{1}$&$\{\{ABFG\},\{ HL\}\}$&0\\\hline
			
			$C_{2}$&$\{\{ABGF\},\{ LH\}\}$&0\\\hline
			
			$C_{3}$&$\{\{ABGF\},\{ HL\}\}$&0\\\hline
			
			$C_{4}$&$\{\{AGFB\},\{ LH\}\}$&0\\\hline
			
			$C_{5}$&$\{\{AGBF\},\{ LH\}\}$&0\\\hline
			
			$C_{6}$&$\{\{AGBF\},\{ HL\}\}$&0\\\hline
			
			$C_{7}$&$\{\{BAFG\},\{ HL\}\}$&0\\\hline
			
			$C_{8}$&$\{\{BAGF\},\{ HL\}\}$&0\\\hline
			
			$C_{9}$&$\{\{GAFB\},\{ LH\}\}$&0\\\hline
			
			$C_{10}$&$\{\{GABF\},\{ LH\}\}$&0\\\hline
			
			$C_{11}$&$\{\{AFBG\},\{ HL\}\}$&0\\\hline
			
			$C_{12}$&$\{\{AFGB\},\{ LH\}\}$&0\\\hline
			
			$C_{13}$&$\{\{ABFG\},\{ LH\}\}$&0\\\hline
			
			$C_{14}$&$\{\{AGFB\},\{ HL\}\}$&0\\\hline
			
			$C_{15}$&$\{\{BAGF\},\{ LH\}\}$&0\\\hline
			
			$C_{16}$&$\{\{BGAF\},\{ HL\}\}$&0\\\hline
			
			$C_{17}$&$\{\{GABF\},\{ HL\}\}$&0\\\hline
			
			$C_{18}$&$\{\{GBAF\},\{ LH\}\}$&0\\\hline
			
			$C_{19}$&$\{\{AFBG\},\{ LH\}\}$&1\\\hline
			
			$C_{20}$&$\{\{AFGB\},\{ HL\}\}$&1\\\hline
			
			$C_{21}$&$\{\{BAFG\},\{ LH\}\}$&1\\\hline
			
			$C_{22}$&$\{\{BGAF\},\{ LH\}\}$&1\\\hline
			
			$C_{23}$&$\{\{GAFB\},\{ HL\}\}$&1\\\hline
			
			$C_{24}$&$\{\{GBAF\},\{ HL\}\}$&1\\\hline
		\end{tabular}\label{tab:3linesWeb12} %
	\end{minipage}\\
	The mixing matrix $ R $, and the diagonal matrix $ D $ are given by, 
	\begin{align}
		\begin{split}
			\hspace{-1cm}R=\scalemath{0.9}{\frac{1}{6} \left(
				\begin{array}{cccccccccccccccccccccccc}
					6 & 0 & 0 & 0 & 0 & 0 & 0 & 0 & 0 & 0 & -3 & 0 & -3 & 0 & 0 & -3 & -3 & 0 & 2 & -1 & -1 & 2 & 2 & 2 \\
					0 & 6 & 0 & 0 & 0 & 0 & 0 & 0 & 0 & 0 & 0 & 0 & -3 & -3 & -3 & 0 & -3 & 0 & -1 & 2 & 2 & -1 & 2 & 2 \\
					0 & 0 & 6 & 0 & 0 & 0 & 0 & 0 & 0 & 0 & -3 & 0 & -3 & -3 & -3 & -3 & -3 & 0 & 2 & 2 & 2 & 2 & 2 & 2 \\
					0 & 0 & 0 & 6 & 0 & 0 & 0 & 0 & 0 & 0 & 0 & -3 & 0 & -3 & -3 & 0 & 0 & -3 & -1 & 2 & 2 & 2 & -1 & 2 \\
					0 & 0 & 0 & 0 & 6 & 0 & 0 & 0 & 0 & 0 & 0 & -3 & -3 & -3 & -3 & 0 & -3 & -3 & 2 & 2 & 2 & 2 & 2 & 2 \\
					0 & 0 & 0 & 0 & 0 & 6 & 0 & 0 & 0 & 0 & 0 & 0 & -3 & -3 & -3 & 0 & -3 & 0 & 2 & -1 & 2 & 2 & 2 & -1 \\
					0 & 0 & 0 & 0 & 0 & 0 & 6 & 0 & 0 & 0 & -3 & 0 & 0 & 0 & 0 & -3 & 0 & 0 & 2 & -1 & -4 & 2 & 2 & -1 \\
					0 & 0 & 0 & 0 & 0 & 0 & 0 & 6 & 0 & 0 & -3 & 0 & 0 & -3 & -3 & -3 & 0 & 0 & 2 & 2 & -1 & 2 & 2 & -1 \\
					0 & 0 & 0 & 0 & 0 & 0 & 0 & 0 & 6 & 0 & 0 & -3 & 0 & 0 & 0 & 0 & 0 & -3 & -1 & 2 & 2 & -1 & -4 & 2 \\
					0 & 0 & 0 & 0 & 0 & 0 & 0 & 0 & 0 & 6 & 0 & -3 & -3 & 0 & 0 & 0 & -3 & -3 & 2 & 2 & 2 & -1 & -1 & 2 \\
					0 & 0 & 0 & 0 & 0 & 0 & 0 & 0 & 0 & 0 & 3 & 0 & 0 & 0 & 0 & -3 & 0 & 0 & -1 & -1 & -1 & 2 & -1 & 2 \\
					0 & 0 & 0 & 0 & 0 & 0 & 0 & 0 & 0 & 0 & 0 & 3 & 0 & 0 & 0 & 0 & 0 & -3 & -1 & -1 & -1 & 2 & -1 & 2 \\
					0 & 0 & 0 & 0 & 0 & 0 & 0 & 0 & 0 & 0 & 0 & 0 & 3 & 0 & 0 & 0 & -3 & 0 & -1 & -1 & -1 & -1 & 2 & 2 \\
					0 & 0 & 0 & 0 & 0 & 0 & 0 & 0 & 0 & 0 & 0 & 0 & 0 & 3 & -3 & 0 & 0 & 0 & -1 & -1 & 2 & 2 & -1 & -1 \\
					0 & 0 & 0 & 0 & 0 & 0 & 0 & 0 & 0 & 0 & 0 & 0 & 0 & -3 & 3 & 0 & 0 & 0 & -1 & 2 & -1 & -1 & 2 & -1 \\
					0 & 0 & 0 & 0 & 0 & 0 & 0 & 0 & 0 & 0 & -3 & 0 & 0 & 0 & 0 & 3 & 0 & 0 & 2 & 2 & -1 & -1 & -1 & -1 \\
					0 & 0 & 0 & 0 & 0 & 0 & 0 & 0 & 0 & 0 & 0 & 0 & -3 & 0 & 0 & 0 & 3 & 0 & 2 & -1 & 2 & -1 & -1 & -1 \\
					0 & 0 & 0 & 0 & 0 & 0 & 0 & 0 & 0 & 0 & 0 & -3 & 0 & 0 & 0 & 0 & 0 & 3 & 2 & 2 & -1 & -1 & -1 & -1 \\
					0 & 0 & 0 & 0 & 0 & 0 & 0 & 0 & 0 & 0 & 0 & 0 & 0 & 0 & 0 & 0 & 0 & 0 & 2 & -1 & -1 & -1 & -1 & 2 \\
					0 & 0 & 0 & 0 & 0 & 0 & 0 & 0 & 0 & 0 & 0 & 0 & 0 & 0 & 0 & 0 & 0 & 0 & -1 & 2 & -1 & 2 & -1 & -1 \\
					0 & 0 & 0 & 0 & 0 & 0 & 0 & 0 & 0 & 0 & 0 & 0 & 0 & 0 & 0 & 0 & 0 & 0 & -1 & -1 & 2 & -1 & 2 & -1 \\
					0 & 0 & 0 & 0 & 0 & 0 & 0 & 0 & 0 & 0 & 0 & 0 & 0 & 0 & 0 & 0 & 0 & 0 & -1 & 2 & -1 & 2 & -1 & -1 \\
					0 & 0 & 0 & 0 & 0 & 0 & 0 & 0 & 0 & 0 & 0 & 0 & 0 & 0 & 0 & 0 & 0 & 0 & -1 & -1 & 2 & -1 & 2 & -1 \\
					0 & 0 & 0 & 0 & 0 & 0 & 0 & 0 & 0 & 0 & 0 & 0 & 0 & 0 & 0 & 0 & 0 & 0 & 2 & -1 & -1 & -1 & -1 & 2 \\
				\end{array}
				\right)}\nonumber
		\end{split}
	\end{align}
	\begin{align}
		\mathcal{D}\,=\,\D{16}
	\end{align}

	\item[\textbf{10}.] $\textbf{W}\,_{3,\text{III}}^{(2,1)}(4,2,1)$ \\
	This is the fourth Cweb with same correlator and attachment content. It has twelve diagrams, one of them is displayed below. \\
	\begin{minipage}[c]{0.5\textwidth}%
		\begin{figure}[H]
			\vspace{-2mm}
			\includegraphics[width=5cm,height=5cm]{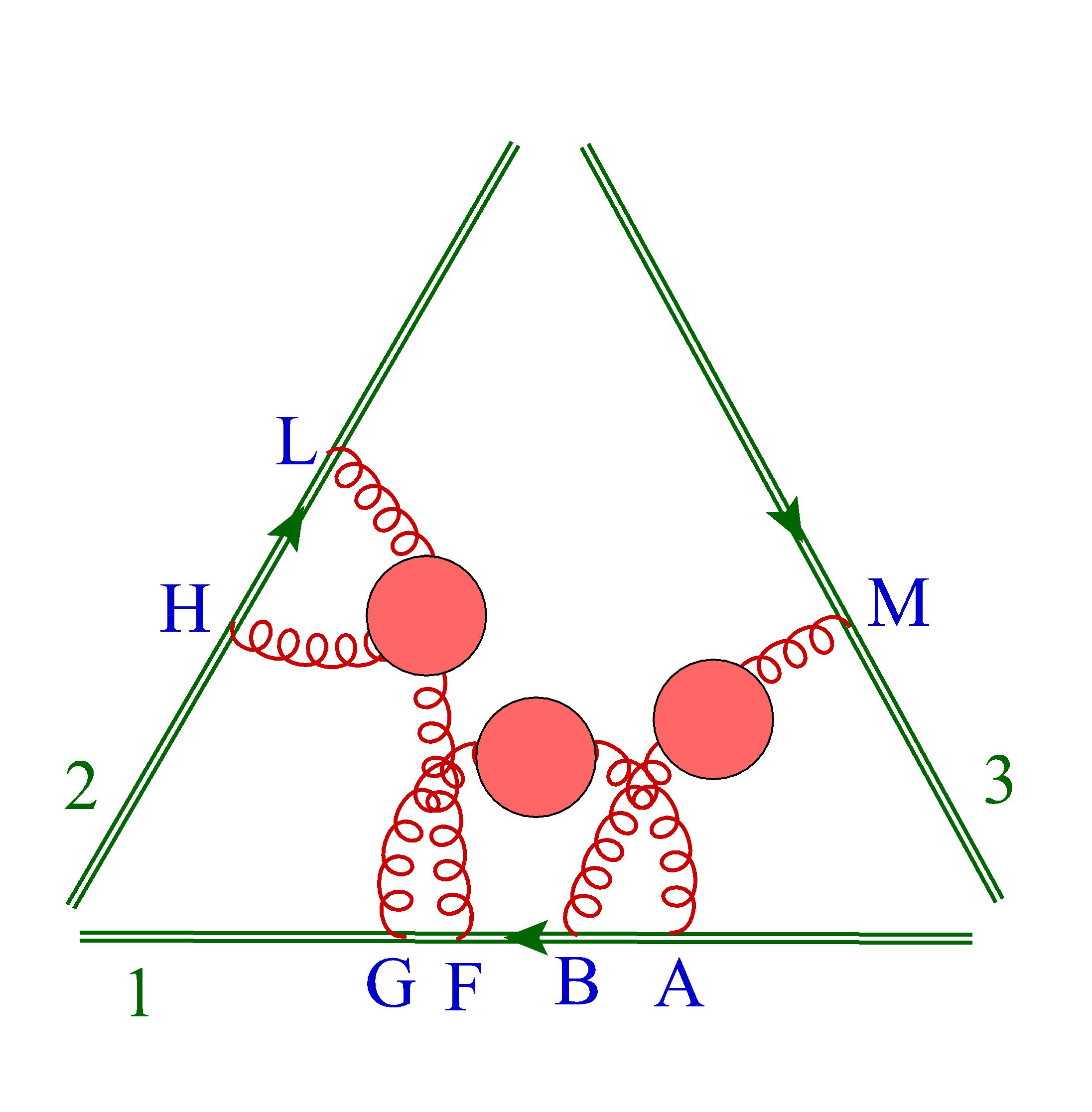} 
		\end{figure}
	\end{minipage}\hspace{-2cm} %
	\begin{minipage}[c]{0.46\textwidth}%
		\vspace{2cm}
		\begin{tabular}{|c|c|c|}
			\hline 
			\textbf{Diagrams}  & \textbf{Sequences}  & \textbf{s-factors}  \\ 
			\hline
			$C_{1}$&$\{\{ABFA\},\{ HL\}\}$&0\\\hline
			
			$C_{2}$&$\{\{AFBA\},\{ HL\}\}$&0\\\hline
			
			$C_{3}$&$\{\{ABAF\},\{ HL\}\}$&0\\\hline
			
			$C_{4}$&$\{\{AFAB\},\{ HL\}\}$&0\\\hline
			
			$C_{5}$&$\{\{BAFA\},\{ HL\}\}$&0\\\hline
			
			$C_{6}$&$\{\{FABA\},\{ HL\}\}$&0\\\hline
			
			$C_{7}$&$\{\{AABF\},\{ HL\}\}$&1\\\hline
			
			$C_{8}$&$\{\{AAFB\},\{ HL\}\}$&1\\\hline
			
			$C_{9}$&$\{\{BAAF\},\{ HL\}\}$&1\\\hline
			
			$C_{10}$&$\{\{BFAA\},\{ HL\}\}$&1\\\hline
			
			$C_{11}$&$\{\{FAAB\},\{ HL\}\}$&1\\\hline
			
			$C_{12}$&$\{\{FBAA\},\{ HL\}\}$&1\\\hline
		\end{tabular}\label{tab:3legWeb21} %
	\end{minipage}\\ \\
	The mixing matrix, and the diagonal matrix are given by, 
	\begin{align}
		\begin{split}
			R=\frac{1}{6} \left(
			\begin{array}{cccccccccccc}
				6 & 0 & -3 & -3 & -3 & -3 & -1 & 2 & 2 & -1 & 2 & 2 \\
				0 & 6 & -3 & -3 & -3 & -3 & 2 & -1 & 2 & 2 & 2 & -1 \\
				0 & 0 & 3 & 0 & 0 & -3 & -1 & -1 & -1 & -1 & 2 & 2 \\
				0 & 0 & 0 & 3 & -3 & 0 & -1 & -1 & 2 & 2 & -1 & -1 \\
				0 & 0 & 0 & -3 & 3 & 0 & -1 & 2 & -1 & -1 & 2 & -1 \\
				0 & 0 & -3 & 0 & 0 & 3 & 2 & -1 & 2 & -1 & -1 & -1 \\
				0 & 0 & 0 & 0 & 0 & 0 & 2 & -1 & -1 & -1 & -1 & 2 \\
				0 & 0 & 0 & 0 & 0 & 0 & -1 & 2 & -1 & 2 & -1 & -1 \\
				0 & 0 & 0 & 0 & 0 & 0 & -1 & -1 & 2 & -1 & 2 & -1 \\
				0 & 0 & 0 & 0 & 0 & 0 & -1 & 2 & -1 & 2 & -1 & -1 \\
				0 & 0 & 0 & 0 & 0 & 0 & -1 & -1 & 2 & -1 & 2 & -1 \\
				0 & 0 & 0 & 0 & 0 & 0 & 2 & -1 & -1 & -1 & -1 & 2 \\
			\end{array}
			\right),\,
			\mathcal{D}\,=\,\D{6}\,.
		\end{split}
	\end{align}

	\item[\textbf{11}.] $\textbf{W}\,_{3,\text{I}}^{(2,1)}(3,2,2)$ \\
	This Cweb also has twelve diagrams, and one of them is shown below. The table shows twelve shuffles and their corresponding $ s $-factors. \\
	\begin{minipage}[c]{0.5\textwidth}%
		\begin{figure}[H]
			\vspace{-2mm}
			\includegraphics[width=5cm,height=5cm]{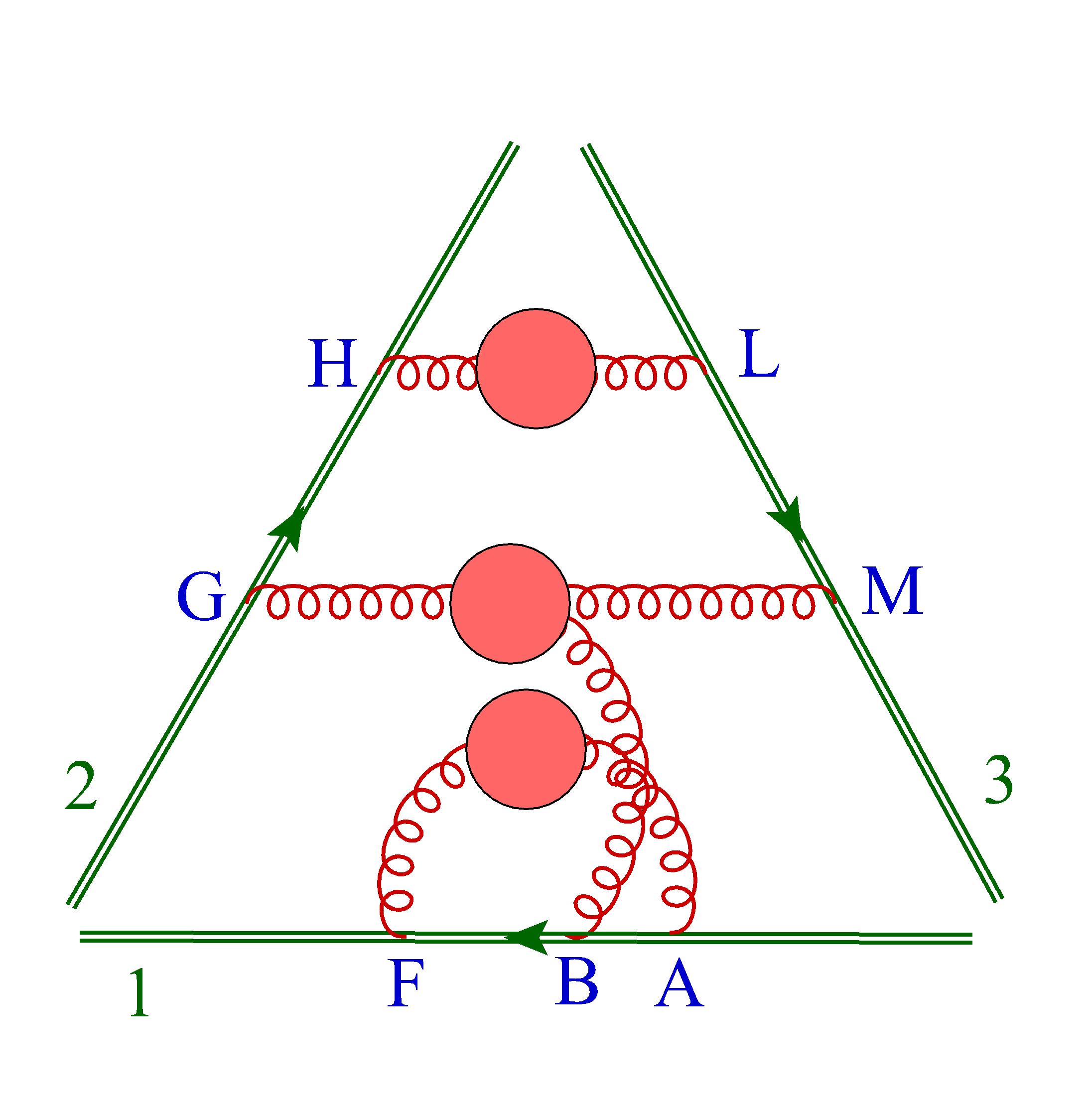} 
		\end{figure}
	\end{minipage}\hspace{-2cm} %
	\begin{minipage}[c]{0.46\textwidth}%
		\vspace{1cm}
		\begin{tabular}{|c|c|c|}
			\hline 
			\textbf{Diagrams}  & \textbf{Sequences}  & \textbf{s-factors}  \\ 
			\hline
			$C_{1}$&$\{\{ABF\},\{ GH\},\{LM\}\}$&0\\\hline
			
			$C_{2}$&$\{\{ABF\},\{ HG\},\{ML\}\}$&0\\\hline
			
			$C_{3}$&$\{\{AFB\},\{ GH\},\{LM\}\}$&0\\\hline
			
			$C_{4}$&$\{\{AFB\},\{ HG\},\{ML\}\}$&0\\\hline
			
			$C_{5}$&$\{\{ABF\},\{ GH\},\{ML\}\}$&0\\\hline
			
			$C_{6}$&$\{\{ABF\},\{ HG\},\{LM\}\}$&0\\\hline
			
			$C_{7}$&$\{\{BAF\},\{ GH\},\{LM\}\}$&0\\\hline
			
			$C_{8}$&$\{\{BAF\},\{ HG\},\{ML\}\}$&0\\\hline
			
			$C_{9}$&$\{\{AFB\},\{ GH\},\{ML\}\}$&1\\\hline
			
			$C_{10}$&$\{\{BAF\},\{ HG\},\{LM\}\}$&1\\\hline
			
			$C_{11}$&$\{\{AFB\},\{ HG\},\{LM\}\}$&2\\\hline
			
			$C_{12}$&$\{\{BAF\},\{ GH\},\{ML\}\}$&2\\\hline
		\end{tabular}\label{tab:3legWeb8} %
	\end{minipage} \\ \\
	The mixing matrix, and the diagonal matrix are given by, 
	\begin{align}
		\begin{split}
			R=&\frac{1}{6} \left(
			\begin{array}{cccccccccccc}
				6 & 0 & -3 & 0 & -3 & -3 & -3 & 0 & 2 & 2 & 1 & 1 \\
				0 & 6 & 0 & -3 & -3 & -3 & 0 & -3 & 2 & 2 & 1 & 1 \\
				0 & 0 & 3 & 0 & 0 & 0 & -3 & 0 & -1 & 2 & -2 & 1 \\
				0 & 0 & 0 & 3 & 0 & 0 & 0 & -3 & -1 & 2 & -2 & 1 \\
				0 & 0 & 0 & 0 & 3 & -3 & 0 & 0 & -1 & 2 & 1 & -2 \\
				0 & 0 & 0 & 0 & -3 & 3 & 0 & 0 & 2 & -1 & -2 & 1 \\
				0 & 0 & -3 & 0 & 0 & 0 & 3 & 0 & 2 & -1 & 1 & -2 \\
				0 & 0 & 0 & -3 & 0 & 0 & 0 & 3 & 2 & -1 & 1 & -2 \\
				0 & 0 & 0 & 0 & 0 & 0 & 0 & 0 & 2 & 2 & -2 & -2 \\
				0 & 0 & 0 & 0 & 0 & 0 & 0 & 0 & 2 & 2 & -2 & -2 \\
				0 & 0 & 0 & 0 & 0 & 0 & 0 & 0 & -1 & -1 & 1 & 1 \\
				0 & 0 & 0 & 0 & 0 & 0 & 0 & 0 & -1 & -1 & 1 & 1 \\
			\end{array}
			\right)
		\end{split}\nonumber\\ \nonumber\\
		\mathcal{D}\,=\,&\D{6}
	\end{align}

	\item[\textbf{12}.] $\textbf{W}\,_{3,\text{II}}^{(2,1)}(3,2,2)$ \\
	This Cweb has six diagrams, one of which is shown below. The table shows the chosen order of shuffles and their corresponding $ s $-factors. \\ 
	\begin{minipage}[c]{0.5\textwidth}%
		\begin{figure}[H]
			\vspace{-2mm}
			\includegraphics[width=5cm,height=5cm]{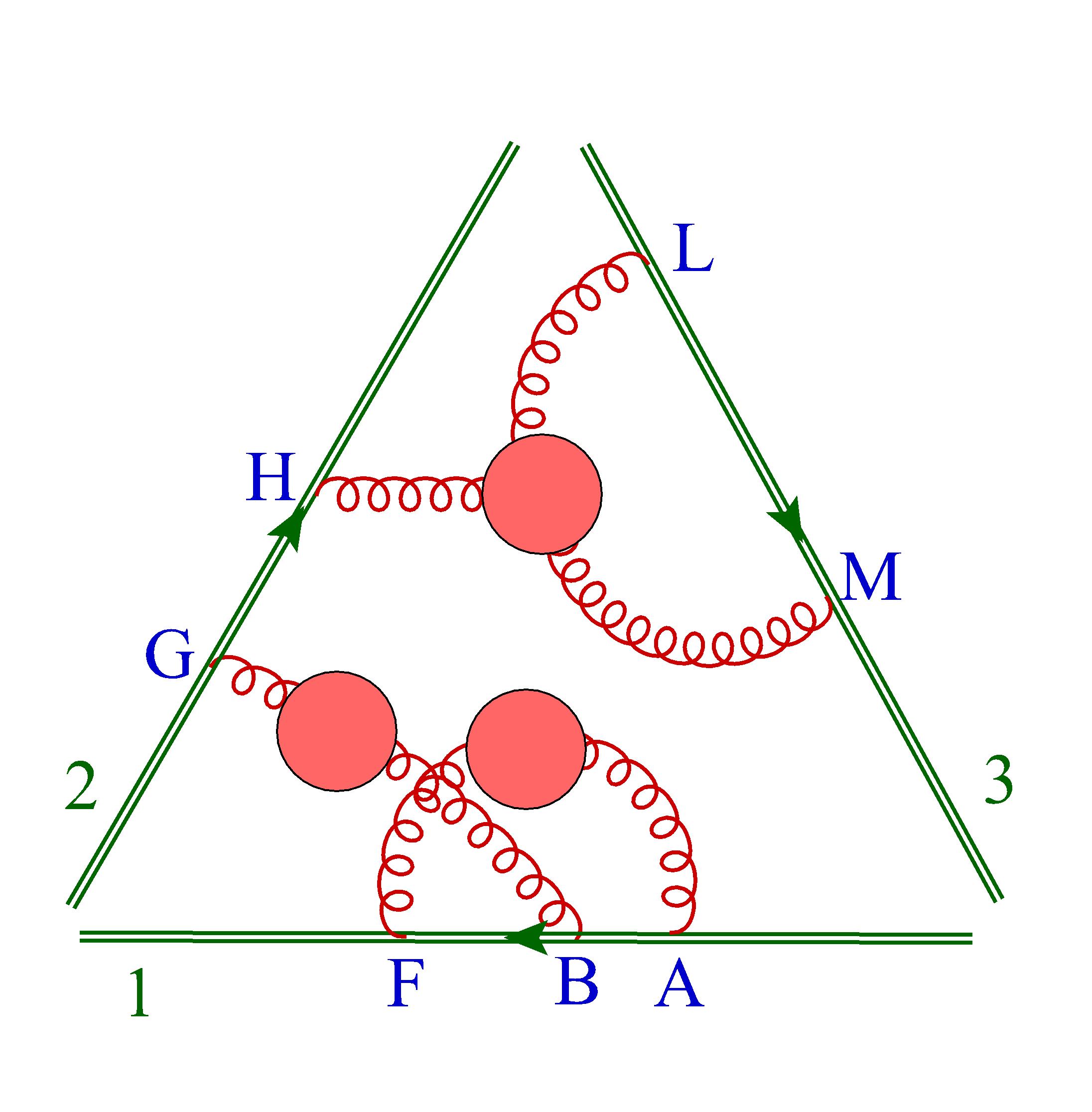} 
		\end{figure}
	\end{minipage}\hspace{-2cm} %
	\begin{minipage}[c]{0.46\textwidth}%
		\vspace{1cm}

		\begin{tabular}{|c|c|c|}
			\hline 
			\textbf{Diagrams}  & \textbf{Sequences}  & \textbf{s-factors}  \\ 
			\hline
			$C_{1}$&$\{\{ABF\},\{ GH\},\{LM\}\}$&0\\\hline
			
			$C_{2}$&$\{\{ABF\},\{ HG\},\{LM\}\}$&0\\\hline
			
			$C_{3}$&$\{\{AFB\},\{ GH\},\{LM\}\}$&1\\\hline
			
			$C_{4}$&$\{\{BAF\},\{ HG\},\{LM\}\}$&1\\\hline
			
			$C_{5}$&$\{\{AFB\},\{ HG\},\{LM\}\}$&2\\\hline
			
			$C_{6}$&$\{\{BAF\},\{ GH\},\{LM\}\}$&2\\\hline
		\end{tabular}\label{tab:4legWeb2} %
	\end{minipage} \\ \\
	The $ R $ and $ D $ matrices are given by, 
	\begin{align}
		\begin{split}
			R=&\frac{1}{6} \left(
			\begin{array}{cccccc}
				3 & -3 & -1 & 2 & 1 & -2 \\
				-3 & 3 & 2 & -1 & -2 & 1 \\
				0 & 0 & 2 & 2 & -2 & -2 \\
				0 & 0 & 2 & 2 & -2 & -2 \\
				0 & 0 & -1 & -1 & 1 & 1 \\
				0 & 0 & -1 & -1 & 1 & 1 \\
			\end{array}
			\right)\end{split} ,\mathcal{D}\,=\,\D{2}\,.\end{align}

	\item[\textbf{13}.] $\textbf{W}\,_{3,\text{I}}^{(2,1)}(3,3,1)$ \\
	This is the first Cweb with same correlator and attachment content. It  Cweb has nine diagrams, one of them is shown below. The table shows chosen order of shuffle and their corresponding $ s $-factors. \\
	\begin{minipage}[c]{0.5\textwidth}%
		\begin{figure}[H]
			\vspace{-2mm}
			\includegraphics[width=5cm,height=5cm]{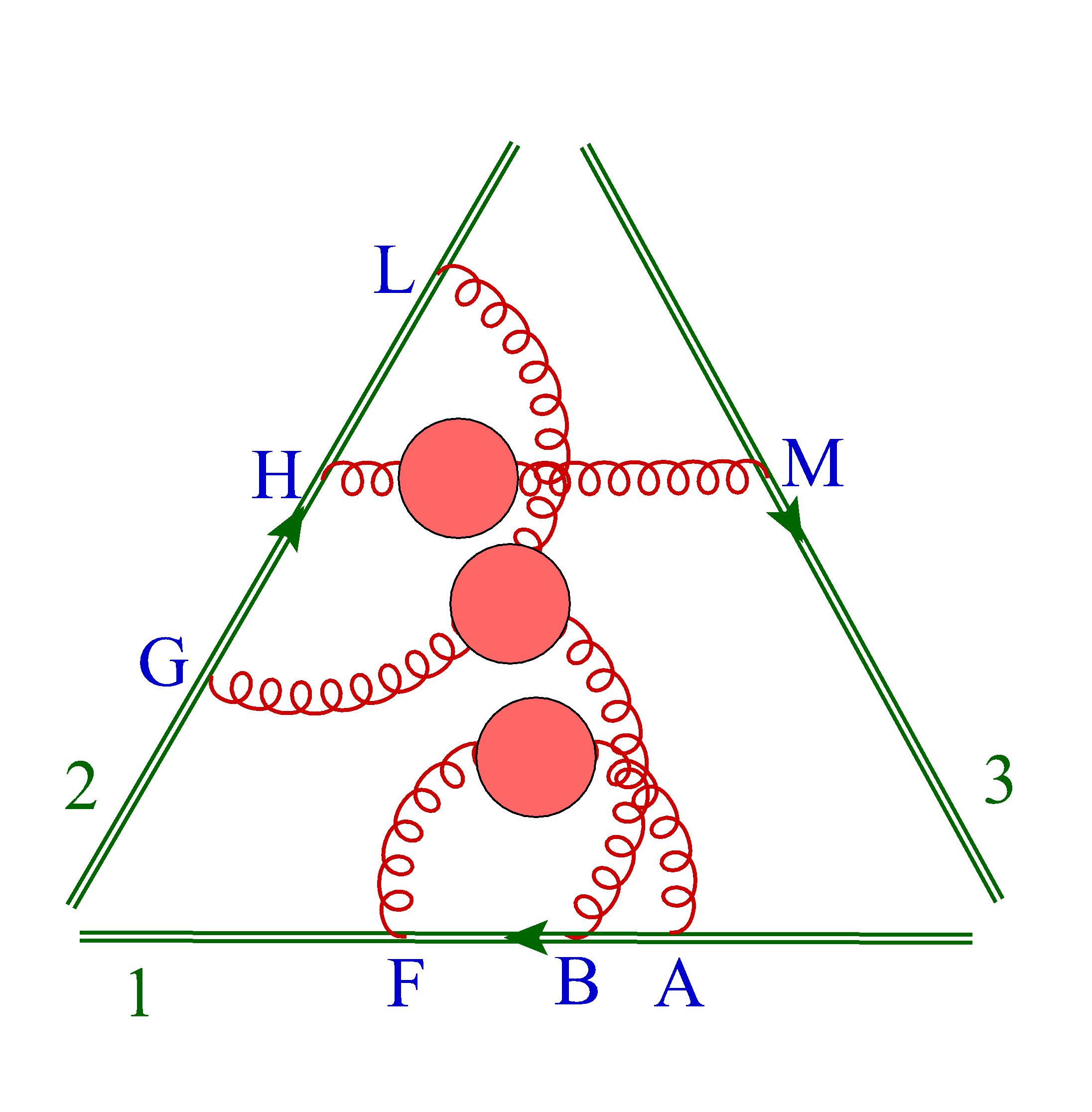} 
		\end{figure}
	\end{minipage}\hspace{-2cm} %
	\begin{minipage}[c]{0.46\textwidth}%
		\vspace{1cm}
		\begin{tabular}{|c|c|c|}
			\hline 
			\textbf{Diagrams}  & \textbf{Sequences}  & \textbf{s-factors}  \\ 
			\hline
			$C_{1}$&$\{\{ABF\},\{ GHL\}\}$&0\\\hline
			
			$C_{2}$&$\{\{AFB\},\{ GHL\}\}$&0\\\hline
			
			$C_{3}$&$\{\{ABF\},\{ GLH\}\}$&0\\\hline
			
			$C_{4}$&$\{\{ABF\},\{ HGL\}\}$&0\\\hline
			
			$C_{5}$&$\{\{BAF\},\{ GHL\}\}$&0\\\hline
			
			$C_{6}$&$\{\{AFB\},\{ GLH\}\}$&1\\\hline
			
			$C_{7}$&$\{\{BAF\},\{ HGL\}\}$&1\\\hline
			
			$C_{8}$&$\{\{AFB\},\{ HGL\}\}$&2\\\hline
			
			$C_{9}$&$\{\{BAF\},\{ GLH\}\}$&2\\\hline
		\end{tabular}\label{tab:3legWeb9} %
	\end{minipage} \\
	The $ R $ and $ D $ matrices are given by, 
	\begin{align}
		\begin{split}
			R=\frac{1}{6} \left(
			\begin{array}{ccccccccc}
				6 & -3 & -3 & -3 & -3 & 2 & 2 & 1 & 1 \\
				0 & 3 & 0 & 0 & -3 & -1 & 2 & -2 & 1 \\
				0 & 0 & 3 & -3 & 0 & -1 & 2 & 1 & -2 \\
				0 & 0 & -3 & 3 & 0 & 2 & -1 & -2 & 1 \\
				0 & -3 & 0 & 0 & 3 & 2 & -1 & 1 & -2 \\
				0 & 0 & 0 & 0 & 0 & 2 & 2 & -2 & -2 \\
				0 & 0 & 0 & 0 & 0 & 2 & 2 & -2 & -2 \\
				0 & 0 & 0 & 0 & 0 & -1 & -1 & 1 & 1 \\
				0 & 0 & 0 & 0 & 0 & -1 & -1 & 1 & 1 \\
			\end{array}
			\right),
		\end{split} 
		\mathcal{D}\,=\,\D{4}\,.
	\end{align}
	\item[\textbf{14}.] $\textbf{W}\,_{3,\text{II}}^{(2,1)}(3,3,1)$ \\
This is the second Cweb with same correlator and attachment content. It has nine diagrams, out which one is shown below. The table shows chosen order of all possible shuffles and their corresponding $ s $-factors. \\

	\begin{minipage}[c]{0.5\textwidth}%
		\begin{figure}[H]
			\vspace{-2mm}
			\includegraphics[width=5cm,height=5cm]{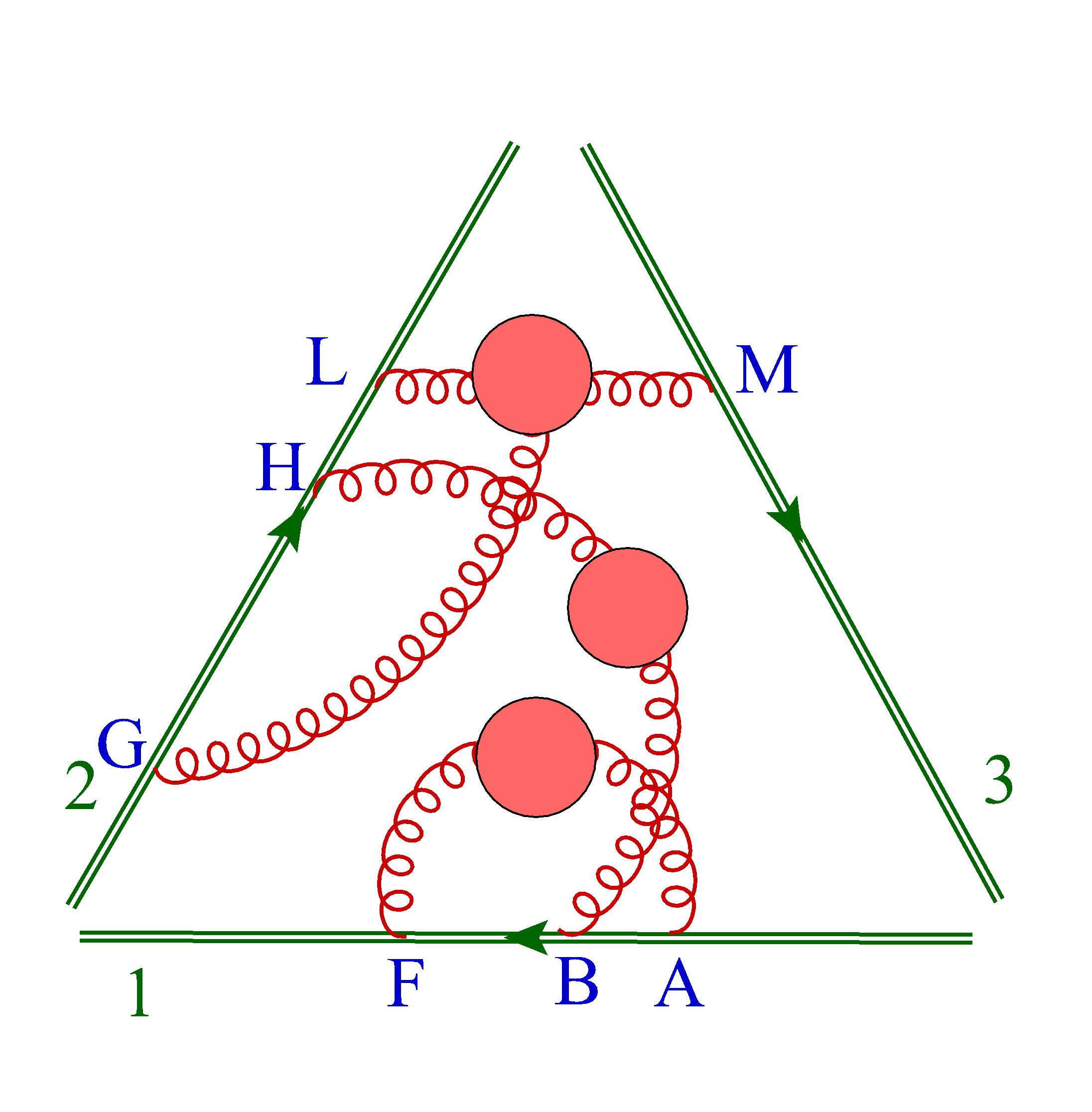} 
		\end{figure}
	\end{minipage}\hspace{-2cm} %
	\begin{minipage}[c]{0.46\textwidth}%
		%
		\begin{tabular}{|c|c|c|}
			\hline 
			\textbf{Diagrams}  & \textbf{Sequences}  & \textbf{s-factors}  \\ 
			\hline
			$C_{1}$&$\{\{ABF\},\{ GHL\}\}$&0\\\hline
			
			$C_{2}$&$\{\{AFB\},\{ GHL\}\}$&0\\\hline
			
			$C_{3}$&$\{\{ABF\},\{ HGL\}\}$&0\\\hline
			
			$C_{4}$&$\{\{ABF\},\{ GLH\}\}$&0\\\hline
			
			$C_{5}$&$\{\{BAF\},\{ GHL\}\}$&0\\\hline
			
			$C_{6}$&$\{\{AFB\},\{ HGL\}\}$&1\\\hline
			
			$C_{7}$&$\{\{AFB\},\{ GLH\}\}$&1\\\hline
			
			$C_{8}$&$\{\{BAF\},\{ HGL\}\}$&2\\\hline
			
			$C_{9}$&$\{\{BAF\},\{ GLH\}\}$&2\\\hline
		\end{tabular}\label{tab:3legWeb11} %
	\end{minipage} \\ \\
	The $ R $, and $ D $ matrices are given by, 
	\begin{align}
		\begin{split}
			R=\frac{1}{6} \left(
			\begin{array}{ccccccccc}
				6 & -3 & -3 & -3 & -3 & 2 & 2 & 1 & 1 \\
				0 & 3 & 0 & 0 & -3 & -1 & 2 & -2 & 1 \\
				0 & 0 & 3 & -3 & 0 & -1 & 2 & 1 & -2 \\
				0 & 0 & -3 & 3 & 0 & 2 & -1 & -2 & 1 \\
				0 & -3 & 0 & 0 & 3 & 2 & -1 & 1 & -2 \\
				0 & 0 & 0 & 0 & 0 & 2 & 2 & -2 & -2 \\
				0 & 0 & 0 & 0 & 0 & 2 & 2 & -2 & -2 \\
				0 & 0 & 0 & 0 & 0 & -1 & -1 & 1 & 1 \\
				0 & 0 & 0 & 0 & 0 & -1 & -1 & 1 & 1 \\
			\end{array}
			\right), \,
			\mathcal{D}\,=\,\D{4}
		\end{split}
	\end{align}

	\item[\textbf{15}.] $\textbf{W}\,_{3,\text{III}}^{(2,1)}(3,3,1)$ \\
	This is the third Cweb with same correlator and attachment content. It has nine diagrams, out which one is shown below. The table shows chosen order of all possible shuffles and their corresponding $ s $-factors. \\
	\begin{minipage}[c]{0.5\textwidth}%
		\begin{figure}[H]
			\vspace{-2mm}
			\includegraphics[width=5cm,height=5cm]{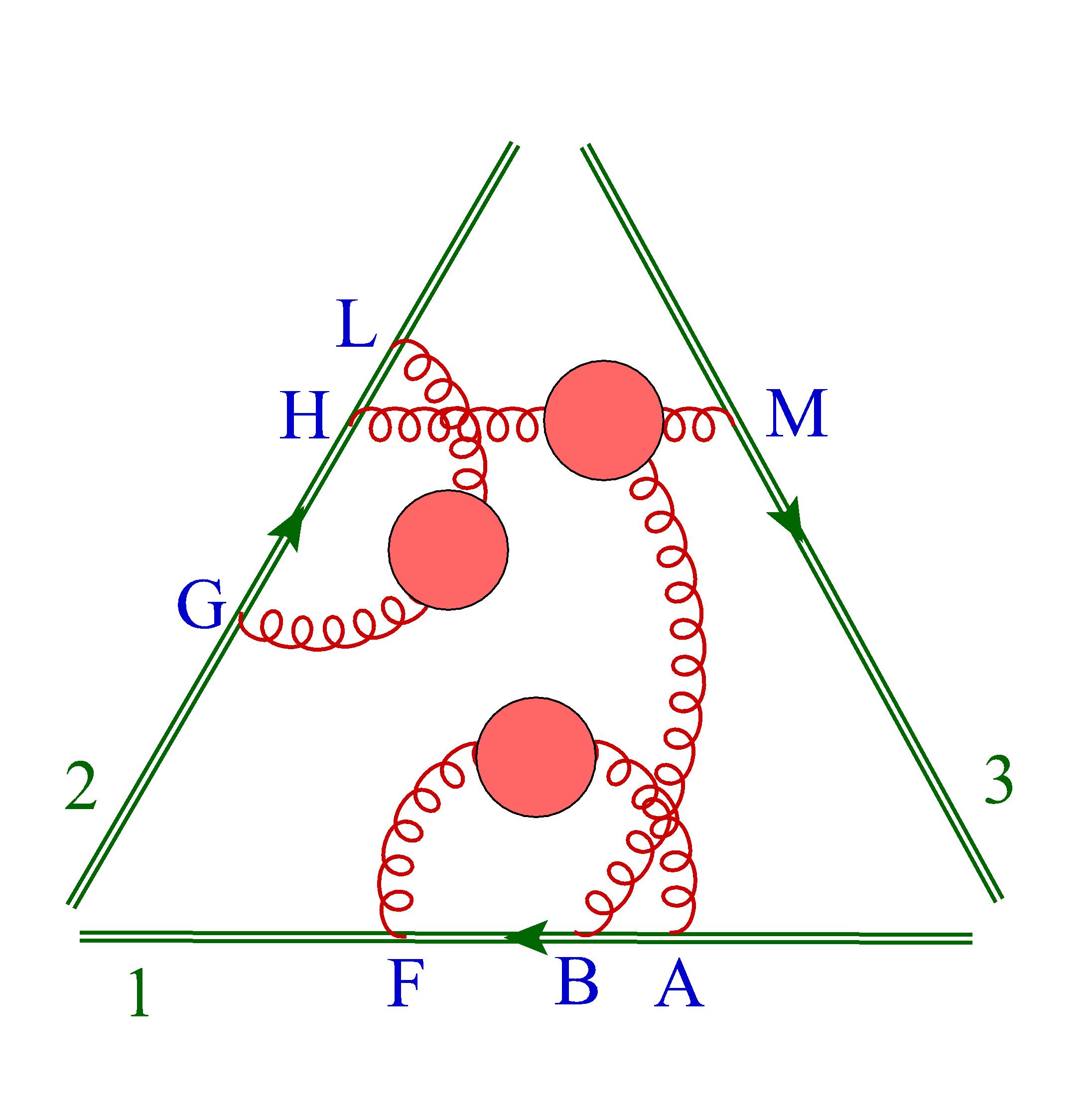} 
		\end{figure}
	\end{minipage}\hspace{-2cm} %
	\begin{minipage}[c]{0.46\textwidth}%
		\vspace{1cm}
		\begin{tabular}{|c|c|c|}
			\hline 
			\textbf{Diagrams}  & \textbf{Sequences}  & \textbf{s-factors}  \\ 
			\hline
			$C_{1}$&$\{\{ABF\},\{ GHL\}\}$&0\\\hline
			
			$C_{2}$&$\{\{AFB\},\{ GHL\}\}$&0\\\hline
			
			$C_{3}$&$\{\{ABF\},\{ GLH\}\}$&0\\\hline
			
			$C_{4}$&$\{\{ABF\},\{ HGL\}\}$&0\\\hline
			
			$C_{5}$&$\{\{BAF\},\{ GHL\}\}$&0\\\hline
			
			$C_{6}$&$\{\{AFB\},\{ HGL\}\}$&1\\\hline
			
			$C_{7}$&$\{\{BAF\},\{ GLH\}\}$&1\\\hline
			
			$C_{8}$&$\{\{AFB\},\{ GLH\}\}$&2\\\hline
			
			$C_{9}$&$\{\{BAF\},\{ HGL\}\}$&2\\\hline
		\end{tabular}\label{tab:4legWeb2} %
	\end{minipage} \\
	The mixing matrix and the diagonal matrix are given by,
	\begin{align}
		\begin{split}
			R=\frac{1}{6} \left(
			\begin{array}{ccccccccc}
				6 & -3 & -3 & -3 & -3 & 2 & 2 & 1 & 1 \\
				0 & 3 & 0 & 0 & -3 & -1 & 2 & -2 & 1 \\
				0 & 0 & 3 & -3 & 0 & 2 & -1 & -2 & 1 \\
				0 & 0 & -3 & 3 & 0 & -1 & 2 & 1 & -2 \\
				0 & -3 & 0 & 0 & 3 & 2 & -1 & 1 & -2 \\
				0 & 0 & 0 & 0 & 0 & 2 & 2 & -2 & -2 \\
				0 & 0 & 0 & 0 & 0 & 2 & 2 & -2 & -2 \\
				0 & 0 & 0 & 0 & 0 & -1 & -1 & 1 & 1 \\
				0 & 0 & 0 & 0 & 0 & -1 & -1 & 1 & 1 \\
			\end{array}
			\right),\,
			\mathcal{D}\,=\,\D{4}.
		\end{split}
	\end{align}

	\item[\textbf{16}.] $\textbf{W}\,_{3}^{(1,0,1)}(3,2,1)$ \\
	This Cweb has three diagrams, one which is displayed below. The mixing matrix for this particular Cweb agrees with the general form of any prime dimensional mixing matrices \cite{Agarwal:2021him}. The table shows the chosen order of shuffles and their corresponding $ s $-factors. \\
	\begin{minipage}[c]{0.5\textwidth}%
		\begin{figure}[H]
			\vspace{-2mm}
			\includegraphics[width=5cm,height=5cm]{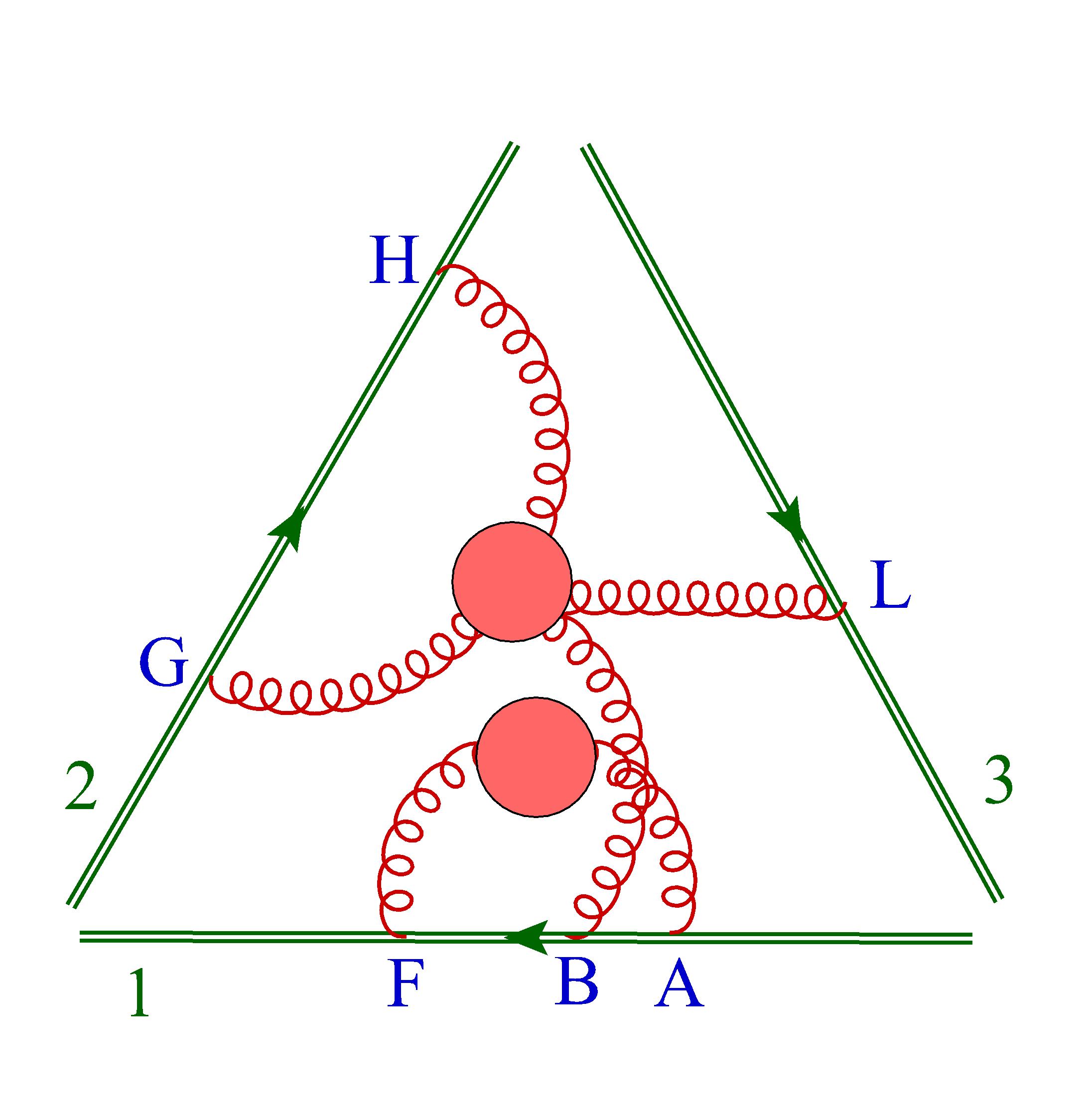} 
		\end{figure}
	\end{minipage}\hspace{-2cm} %
	\begin{minipage}[c]{0.46\textwidth}%
		%
		\begin{tabular}{|c|c|c|}
			\hline 
			\textbf{Diagrams}  & \textbf{Sequences}  & \textbf{s-factors}  \\ 
			\hline
			$C_{1}$&$\{\{ABF\}\}$&0\\\hline
			
			$C_{2}$&$\{\{AFB\}\}$&1\\\hline
			
			$C_{3}$&$\{\{FAB\}\}$&1\\\hline
		\end{tabular}\label{tab:3legsWeb13} %
	\end{minipage}
	\\
	The $ R $ and $ D $ matrices are given by, 
	\begin{align}
		\begin{split}
			R=\frac{1}{2} \left(
			\begin{array}{ccc}
				2 & -1 & -1 \\
				0 & 1 & -1 \\
				0 & -1 & 1 \\
			\end{array}
			\right),\,
			\mathcal{D}\,=\,\D{2}\,.
		\end{split}
	\end{align}
	\item[\textbf{17}.] $\textbf{W}\,_{3,\text{I}}^{(2,1)}(5,1,1)$ \\
	This is the first Cweb with same correlator and attachment content. It has fifteen diagrams, one of which is shown below. The table shows the chosen order of shuffle and their corresponding $ s $-factors.  \\
	\begin{minipage}[c]{0.5\textwidth}%
		\begin{figure}[H]
			\vspace{-2mm}
			\includegraphics[width=4cm,height=4cm]{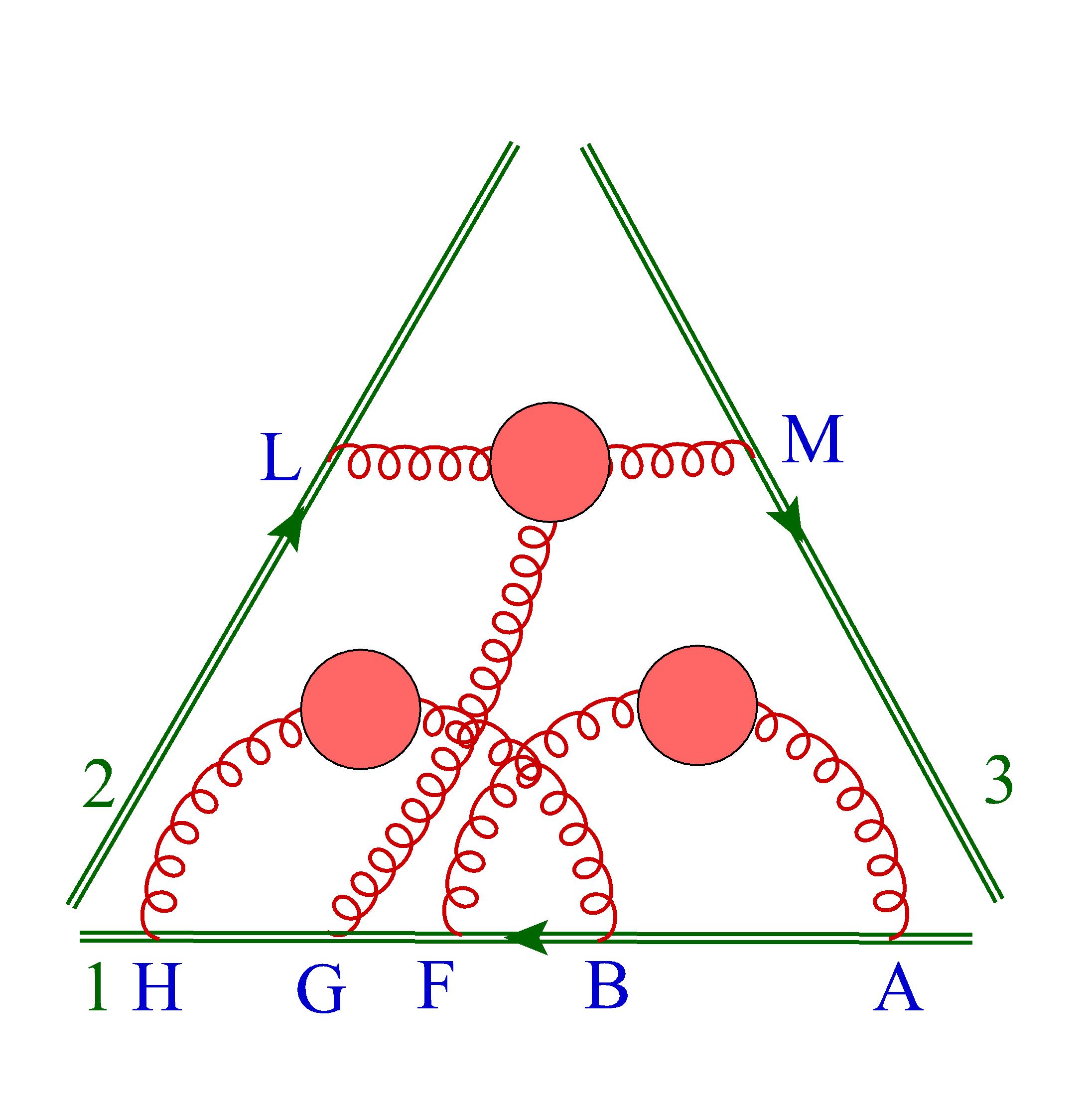} 
		\end{figure}
	\end{minipage}\hspace{-2cm} %
	\begin{minipage}[c]{0.46\textwidth}%
		\vspace{0.5cm}
		\begin{tabular}{|c|c|c|}
			\hline 
			\textbf{Diagrams}  & \textbf{Sequences}  & \textbf{s-factors}  \\ 
			\hline
			$C_{1}$&$\{\{ABFGH\}\}$&0\\\hline
			
			$C_{2}$&$\{\{ABHGF\}\}$&0\\\hline
			
			$C_{3}$&$\{\{ABGFH\}\}$&0\\\hline
			
			$C_{4}$&$\{\{ABGHF\}\}$&0\\\hline
			
			$C_{5}$&$\{\{AGBFH\}\}$&0\\\hline
			
			$C_{6}$&$\{\{AGBHF\}\}$&0\\\hline
			
			$C_{7}$&$\{\{AFBGH\}\}$&0\\\hline
			
			$C_{8}$&$\{\{ABFHG\}\}$&0\\\hline
			
			$C_{9}$&$\{\{ABHFG\}\}$&0\\\hline
			
			$C_{10}$&$\{\{AGFBH\}\}$&0\\\hline
			
			$C_{11}$&$\{\{GABFH\}\}$&0\\\hline
			
			$C_{12}$&$\{\{GABHF\}\}$&0\\\hline
			
			$C_{13}$&$\{\{AFBHG\}\}$&1\\\hline
			
			$C_{14}$&$\{\{AFGBH\}\}$&1\\\hline
			
			$C_{15}$&$\{\{GAFBH\}\}$&1\\\hline
		\end{tabular}\label{tab:3legWeb15} %
	\end{minipage} \\

	The mixing matrix and the diagonal matrix for this Cweb are given by, 
	\begin{align}
		\begin{split}
			R=&\frac{1}{12} \left(
			\begin{array}{ccccccccccccccc}
				12 & 0 & 0 & 0 & 0 & 0 & -6 & -6 & 0 & -6 & -6 & 0 & 2 & 2 & 8 \\
				0 & 12 & 0 & 0 & 0 & 0 & -6 & 0 & -6 & -6 & 0 & -6 & 2 & 2 & 8 \\
				0 & 0 & 12 & 0 & 0 & 0 & -12 & -6 & 0 & -12 & -6 & 0 & 8 & 8 & 8 \\
				0 & 0 & 0 & 12 & 0 & 0 & -12 & 0 & -6 & -12 & 0 & -6 & 8 & 8 & 8 \\
				0 & 0 & 0 & 0 & 12 & 0 & -6 & -6 & 0 & -6 & -6 & 0 & 8 & 2 & 2 \\
				0 & 0 & 0 & 0 & 0 & 12 & -6 & 0 & -6 & -6 & 0 & -6 & 8 & 2 & 2 \\
				0 & 0 & 0 & 0 & 0 & 0 & 6 & 0 & 0 & -6 & 0 & 0 & -4 & 2 & 2 \\
				0 & 0 & 0 & 0 & 0 & 0 & 0 & 6 & 0 & 0 & -6 & 0 & -4 & -4 & 8 \\
				0 & 0 & 0 & 0 & 0 & 0 & 0 & 0 & 6 & 0 & 0 & -6 & -4 & -4 & 8 \\
				0 & 0 & 0 & 0 & 0 & 0 & -6 & 0 & 0 & 6 & 0 & 0 & 2 & 2 & -4 \\
				0 & 0 & 0 & 0 & 0 & 0 & 0 & -6 & 0 & 0 & 6 & 0 & 8 & -4 & -4 \\
				0 & 0 & 0 & 0 & 0 & 0 & 0 & 0 & -6 & 0 & 0 & 6 & 8 & -4 & -4 \\
				0 & 0 & 0 & 0 & 0 & 0 & 0 & 0 & 0 & 0 & 0 & 0 & 2 & -4 & 2 \\
				0 & 0 & 0 & 0 & 0 & 0 & 0 & 0 & 0 & 0 & 0 & 0 & -4 & 8 & -4 \\
				0 & 0 & 0 & 0 & 0 & 0 & 0 & 0 & 0 & 0 & 0 & 0 & 2 & -4 & 2 \\
			\end{array}
			\right)\, \\
			\mathcal{D}\,=\,&\D{10}
		\end{split}
	\end{align}

	\item[\textbf{18}.] $\textbf{W}\,_{3,\text{II}}^{(2,1)}(5,1,1)$ \\
	This is the second Cweb with same correlator and attachment content. It has thirty diagrams, one of which is shown below. The table gives the chosen order of thirty shuffles and their corresponding $ s $-factors. We do not to present the mixing matrix fo this Cweb here due to its larger dimension, however it can be found in the ancillary file \textit{Boomerang.nb}. \\
	
	\begin{center}
			\includegraphics[width=5cm,height=5cm]{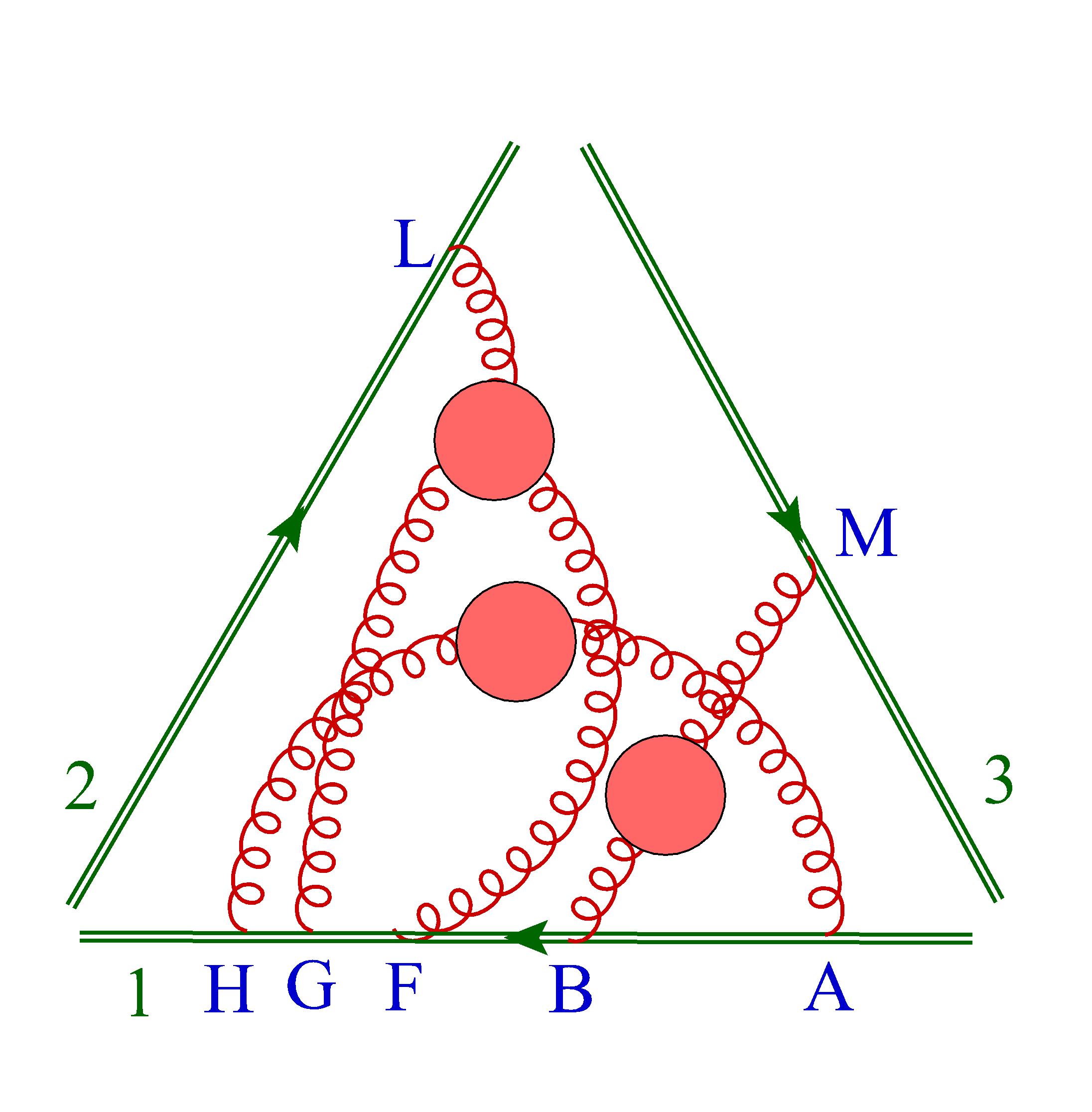} 
	\end{center}
			
	\begin{minipage}[c]{0.42\textwidth}%
		\vspace{2cm}
		\begin{tabular}{|c|c|c|}
			\hline 
			\textbf{Diagrams}  & \textbf{Sequences}  & \textbf{s-factors}  \\ 
			\hline
			$C_{1}$&$\{\{ABFG\}\}$&0\\\hline
			
			$C_{2}$&$\{\{ABFH\}\}$&0\\\hline
			
			$C_{3}$&$\{\{AFGB\}\}$&0\\\hline
			
			$C_{4}$&$\{\{AFBG\}\}$&0\\\hline
			
			$C_{5}$&$\{\{AFBH\}\}$&0\\\hline
			
			$C_{6}$&$\{\{AFHB\}\}$&0\\\hline
			
			$C_{7}$&$\{\{FAGB\}\}$&0\\\hline
			
			$C_{8}$&$\{\{FABG\}\}$&0\\\hline
			
			$C_{9}$&$\{\{FABH\}\}$&0\\\hline
			
			$C_{10}$&$\{\{FAHB\}\}$&0\\\hline
			
			$C_{11}$&$\{\{FBAG\}\}$&0\\\hline
			
			$C_{12}$&$\{\{FBAH\}\}$&0\\\hline
			
			$C_{13}$&$\{\{AGFB\}\}$&0\\\hline
			
			$C_{14}$&$\{\{ABGF\}\}$&0\\\hline
			
			$C_{15}$&$\{\{AFGH\}\}$&0\\\hline
		\end{tabular}\label{tab:3legWeb20} %
	\end{minipage}
\hspace{0.5cm}
\begin{minipage}[c]{0.42\textwidth}%
	\vspace{2cm}
	\begin{tabular}{|c|c|c|}
		\hline 
		\textbf{Diagrams}  & \textbf{Sequences}  & \textbf{s-factors}  \\ 
		\hline
		$C_{16}$&$\{\{AFHG\}\}$&0\\\hline
		
		$C_{17}$&$\{\{BAFG\}\}$&0\\\hline
		
		$C_{18}$&$\{\{BAFH\}\}$&0\\\hline
		
		$C_{19}$&$\{\{BFAG\}\}$&0\\\hline
		
		$C_{20}$&$\{\{BFAH\}\}$&0\\\hline
		
		$C_{21}$&$\{\{FAGH\}\}$&0\\\hline
		
		$C_{22}$&$\{\{FAHG\}\}$&0\\\hline
		
		$C_{23}$&$\{\{FBHA\}\}$&0\\\hline
		
		$C_{24}$&$\{\{FHAB\}\}$&0\\\hline
		
		$C_{25}$&$\{\{AGBF\}\}$&1\\\hline
		
		$C_{26}$&$\{\{AGFH\}\}$&1\\\hline
		
		$C_{27}$&$\{\{BAGF\}\}$&1\\\hline
		
		$C_{28}$&$\{\{BFHA\}\}$&1\\\hline
		
		$C_{29}$&$\{\{FHAG\}\}$&1\\\hline
		
		$C_{30}$&$\{\{FHBA\}\}$&1\\\hline
	\end{tabular} %
\end{minipage} \\

The diagonal matrix for this Cweb is given by, 

	\begin{align}
		\mathcal{D}\,=\,\D{20}
	\end{align}

	\item[\textbf{19}.] $\textbf{W}\,_{3}^{(1,0,1)}(4,1,1)$ \\
	This is a Cweb made out of one four-gluon correlator and a two-gluon correlator. It has six diagrams, one of them is displayed below. The table shows the chosen order of shuffles and their corresponding $ s $-factors. \\
	\begin{minipage}[c]{0.5\textwidth}%
		\begin{figure}[H]
			\vspace{-2mm}
			\includegraphics[width=5cm,height=5cm]{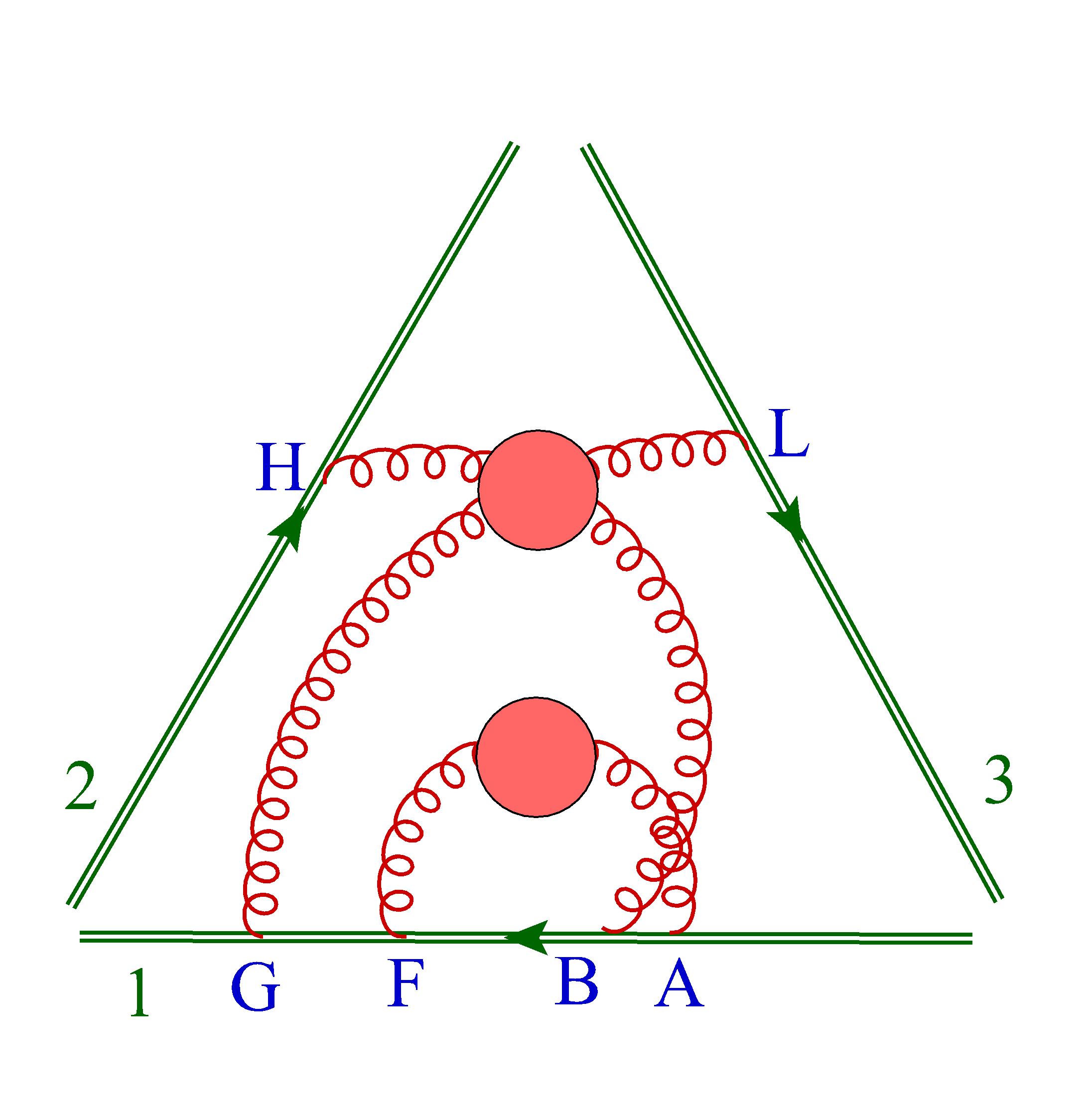} 
		\end{figure}
	\end{minipage}\hspace{-2cm} %
	\begin{minipage}[c]{0.46\textwidth}%
		\vspace{1cm}
		\begin{tabular}{|c|c|c|}
			\hline 
			\textbf{Diagrams}  & \textbf{Sequences}  & \textbf{s-factors}  \\ 
			\hline
			$C_{1}$&$\{\{ABFG\}\}$&0\\\hline
			
			$C_{2}$&$\{\{ABGF\}\}$&0\\\hline
			
			$C_{3}$&$\{\{BAFG\}\}$&0\\\hline
			
			$C_{4}$&$\{\{BAGF\}\}$&0\\\hline
			
			$C_{5}$&$\{\{BGAF\}\}$&1\\\hline
			
			$C_{6}$&$\{\{AFBG\}\}$&1\\\hline
		\end{tabular}\label{tab:3legWeb17} %
	\end{minipage}\\ \\
	The $ R $, and $ D $ matrices are given by, 
	\begin{align}
		\begin{split}
			R=\frac{1}{2} \left(
			\begin{array}{cccccc}
				2 & 0 & 0 & 0 & -1 & -1 \\
				0 & 2 & 0 & 0 & -1 & -1 \\
				0 & 0 & 2 & 0 & -1 & -1 \\
				0 & 0 & 0 & 2 & -1 & -1 \\
				0 & 0 & 0 & 0 & 1 & -1 \\
				0 & 0 & 0 & 0 & -1 & 1 \\
			\end{array}
			\right),\,
			\mathcal{D}\,=\,\D{5}
		\end{split}
		\label{eq:case-1-example}
	\end{align}
	
	\item[\textbf{20}.] $\textbf{W}\,_{3}^{(4)}(6,1,1)$ \\
	This is the largest Cweb, it has ninety diagrams. The table shows the chosen order of shuffle and their corresponding $ s $-factors.  We do not to present the mixing matrix fo this Cweb here due to its larger dimension, however it can be found in the ancillary file \textit{Boomerang.nb}.\\ 
\begin{center}
			\includegraphics[width=5cm,height=5cm]{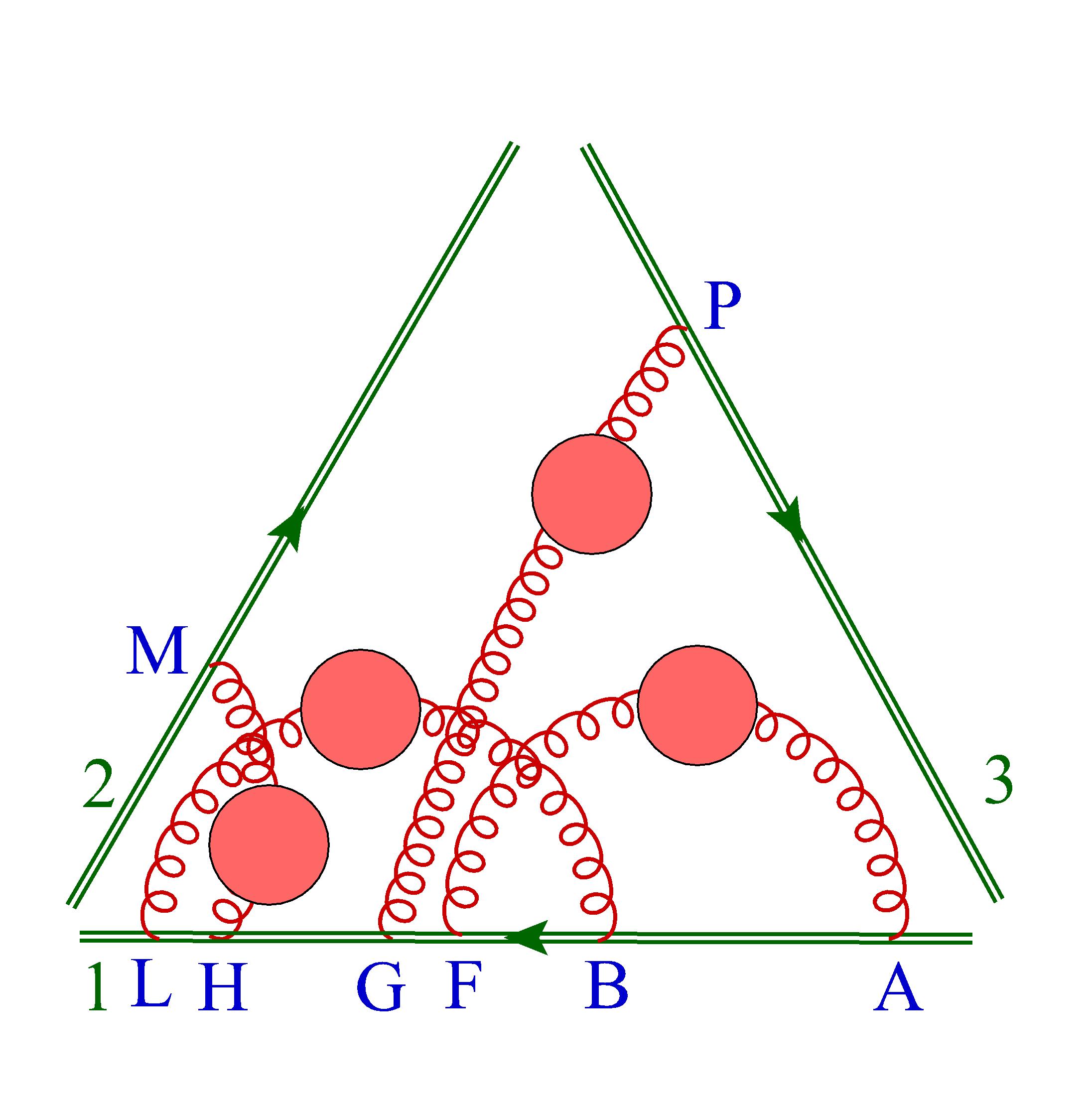} 

	\begin{minipage}[c]{0.46\textwidth}%
		%
		\begin{tabular}{|c|c|c|}
			\hline 
			\textbf{Diagrams}  & \textbf{Sequences}  & \textbf{s-factors}  \\ 
			\hline
			$C_{1}$&$\{\{ABFGHL\}\}$&0\\\hline
			
			$C_{2}$&$\{\{ABFHGL\}\}$&0\\\hline
			
			$C_{3}$&$\{\{ABLGHF\}\}$&0\\\hline
			
			$C_{4}$&$\{\{ABLHGF\}\}$&0\\\hline
			
			$C_{5}$&$\{\{ABGFHL\}\}$&0\\\hline
			
			$C_{6}$&$\{\{ABGLHF\}\}$&0\\\hline
			
			$C_{7}$&$\{\{ABGHFL\}\}$&0\\\hline
			
			$C_{8}$&$\{\{ABGHLF\}\}$&0\\\hline
			
			$C_{9}$&$\{\{ABHFGL\}\}$&0\\\hline
			
			$C_{10}$&$\{\{ABHLGF\}\}$&0\\\hline
			
			$C_{11}$&$\{\{ABHGFL\}\}$&0\\\hline
			
			$C_{12}$&$\{\{ABHGLF\}\}$&0\\\hline
			
			$C_{13}$&$\{\{AGBFHL\}\}$&0\\\hline
			
			$C_{14}$&$\{\{AGBLHF\}\}$&0\\\hline
			
			$C_{15}$&$\{\{AGBHFL\}\}$&0\\\hline
			
			$C_{16}$&$\{\{AGBHLF\}\}$&0\\\hline
			
			$C_{17}$&$\{\{AGHBFL\}\}$&0\\\hline
			
			$C_{18}$&$\{\{AGHBLF\}\}$&0\\\hline
			
			$C_{19}$&$\{\{AHBFGL\}\}$&0\\\hline
			
			$C_{20}$&$\{\{AHBLGF\}\}$&0\\\hline
			
			$C_{21}$&$\{\{AHBGFL\}\}$&0\\\hline
			
			$C_{22}$&$\{\{AHBGLF\}\}$&0\\\hline
			
			$C_{23}$&$\{\{AHGBLF\}\}$&0\\\hline
			
			$C_{24}$&$\{\{AFBGLH\}\}$&0\\\hline
			
			$C_{25}$&$\{\{AFBGHL\}\}$&0\\\hline
			
			$C_{26}$&$\{\{AFBHLG\}\}$&0\\\hline
			
			$C_{27}$&$\{\{AFBHGL\}\}$&0\\\hline
			
			$C_{28}$&$\{\{AFGBHL\}\}$&0\\\hline
			
			$C_{29}$&$\{\{AFHBGL\}\}$&0\\\hline
			
			$C_{30}$&$\{\{ABFLGH\}\}$&0\\\hline
		\end{tabular}\label{tab:4legWeb2} %
	\end{minipage} 
\end{center}

	\begin{minipage}[c]{0.46\textwidth}%
		%
		\begin{tabular}{|c|c|c|}
			\hline 
			\textbf{Diagrams}  & \textbf{Sequences}  & \textbf{s-factors}  \\ 
			\hline
			$C_{31}$&$\{\{ABFLHG\}\}$&0\\\hline
			
			$C_{32}$&$\{\{ABFGLH\}\}$&0\\\hline
			
			$C_{33}$&$\{\{ABFHLG\}\}$&0\\\hline
			
			$C_{34}$&$\{\{ABLFGH\}\}$&0\\\hline
			
			$C_{35}$&$\{\{ABLFHG\}\}$&0\\\hline
			
			$C_{36}$&$\{\{ABLGFH\}\}$&0\\\hline
			
			$C_{37}$&$\{\{ABLHFG\}\}$&0\\\hline
			
			$C_{38}$&$\{\{ABGFLH\}\}$&0\\\hline
			
			$C_{39}$&$\{\{ABGLFH\}\}$&0\\\hline
			
			$C_{40}$&$\{\{ABHFLG\}\}$&0\\\hline
			
			$C_{41}$&$\{\{ABHLFG\}\}$&0\\\hline
			
			$C_{42}$&$\{\{AGFBLH\}\}$&0\\\hline
			
			$C_{43}$&$\{\{AGFBHL\}\}$&0\\\hline
			
			$C_{44}$&$\{\{AGFHBL\}\}$&0\\\hline
			
			$C_{45}$&$\{\{AGBFLH\}\}$&0\\\hline
			
			$C_{46}$&$\{\{AGBLFH\}\}$&0\\\hline
			
			$C_{47}$&$\{\{AGHFBL\}\}$&0\\\hline
			
			$C_{48}$&$\{\{AHFBLG\}\}$&0\\\hline
			
			$C_{49}$&$\{\{AHFBGL\}\}$&0\\\hline
			
			$C_{50}$&$\{\{AHFGBL\}\}$&0\\\hline
			
			$C_{51}$&$\{\{AHBFLG\}\}$&0\\\hline
			
			$C_{52}$&$\{\{AHBLFG\}\}$&0\\\hline
			
			$C_{53}$&$\{\{AHGFBL\}\}$&0\\\hline
			
			$C_{54}$&$\{\{AHGBFL\}\}$&0\\\hline
			
			$C_{55}$&$\{\{GAFBHL\}\}$&0\\\hline
			
			$C_{56}$&$\{\{GABFLH\}\}$&0\\\hline
			
			$C_{57}$&$\{\{GABFHL\}\}$&0\\\hline
			
			$C_{58}$&$\{\{GABLFH\}\}$&0\\\hline
			
			$C_{59}$&$\{\{GABLHF\}\}$&0\\\hline
			
			$C_{60}$&$\{\{GABHFL\}\}$&0\\\hline
		\end{tabular}\label{tab:4legWeb2} %
	\end{minipage}\hspace{1.5cm}
	\begin{minipage}[c]{0.46\textwidth}%
		%
		\begin{tabular}{|c|c|c|}
			\hline 
			\textbf{Diagrams}  & \textbf{Sequences}  & \textbf{s-factors}  \\ 
			\hline
			$C_{61}$&$\{\{GABHLF\}\}$&0\\\hline
			
			$C_{62}$&$\{\{GAHFBL\}\}$&0\\\hline
			
			$C_{63}$&$\{\{GAHBFL\}\}$&0\\\hline
			
			$C_{64}$&$\{\{GAHBLF\}\}$&0\\\hline
			
			$C_{65}$&$\{\{GHABFL\}\}$&0\\\hline
			
			$C_{66}$&$\{\{GHABLF\}\}$&0\\\hline
			
			$C_{67}$&$\{\{HAFBGL\}\}$&0\\\hline
			
			$C_{68}$&$\{\{HABFLG\}\}$&0\\\hline
			
			$C_{69}$&$\{\{HABFGL\}\}$&0\\\hline
			
			$C_{70}$&$\{\{HABLFG\}\}$&0\\\hline
			
			$C_{71}$&$\{\{HABLGF\}\}$&0\\\hline
			
			$C_{72}$&$\{\{HABGFL\}\}$&0\\\hline
			
			$C_{73}$&$\{\{HABGLF\}\}$&0\\\hline
			
			$C_{74}$&$\{\{HAGFBL\}\}$&0\\\hline
			
			$C_{75}$&$\{\{HAGBFL\}\}$&0\\\hline
			
			$C_{76}$&$\{\{HAGBLF\}\}$&0\\\hline
			
			$C_{77}$&$\{\{HGABFL\}\}$&0\\\hline
			
			$C_{78}$&$\{\{HGABLF\}\}$&0\\\hline
			
			$C_{79}$&$\{\{AFBLGH\}\}$&1\\\hline
			
			$C_{80}$&$\{\{AFBLHG\}\}$&1\\\hline
			
			$C_{81}$&$\{\{AFGBLH\}\}$&1\\\hline
			
			$C_{82}$&$\{\{AFGHBL\}\}$&1\\\hline
			
			$C_{83}$&$\{\{AFHBLG\}\}$&1\\\hline
			
			$C_{84}$&$\{\{AFHGBL\}\}$&1\\\hline
			
			$C_{85}$&$\{\{GAFBLH\}\}$&1\\\hline
			
			$C_{86}$&$\{\{GAFHBL\}\}$&1\\\hline
			
			$C_{87}$&$\{\{GHAFBL\}\}$&1\\\hline
			
			$C_{88}$&$\{\{HAFBLG\}\}$&1\\\hline
			
			$C_{89}$&$\{\{HAFGBL\}\}$&1\\\hline
			
			$C_{90}$&$\{\{HGAFBL\}\}$&1\\\hline
		\end{tabular}\label{tab:4legWeb2} %
	\end{minipage}

Diagonalizing matrix is given as 
\begin{align}
	\mathcal{D}\,=\,\D{50}
\end{align}	
	
\end{itemize}

\bibliography{boom}
\bibliographystyle{bibstyle}
\end{document}